\documentclass{article}
\usepackage[a4paper,margin=0.75in]{geometry}
\usepackage{natbib}
\usepackage{graphicx}
\usepackage{subcaption}
\usepackage{enumerate}
\usepackage{caption}
\usepackage{amsmath}
\usepackage{bm}
\usepackage{amsfonts}
\usepackage{amssymb}
\usepackage{bbm}
\usepackage{listings}
\usepackage{wrapfig}
\usepackage{amsthm}
\usepackage{mathtools}
\usepackage{algorithm}
\usepackage{algpseudocode}

\DeclarePairedDelimiter\ceiling{\lceil}{\rceil}
\DeclarePairedDelimiter\flooring{\lfloor}{\rfloor}
\providecommand{\myceil}[1]{\left \lceil #1 \right \rceil }
\providecommand{\myfloor}[1]{\left \lfloor #1 \right \rfloor }

\newtheorem{Lemma}{Lemma}

\newtheorem{Thm}{Theorem}
\newtheorem{Proposition}{Proposition}
\usepackage{tikz}
\def\tick{\tikz\fill[scale=0.4](0,.35) -- (.25,0) -- (1,.7) -- (.25,.15) -- cycle;}

\makeatletter
\def\BState{\State\hskip-\ALG@thistlm}
\makeatother

\algnewcommand\algorithmicswitch{\textbf{switch}}
\algnewcommand\algorithmiccase{\textbf{case}}
\algnewcommand\algorithmicassert{\texttt{assert}}
\algnewcommand\Assert[1]{\State \algorithmicassert(#1)}

\algdef{SE}[SWITCH]{Switch}{EndSwitch}[1]{\algorithmicswitch\ #1\ \algorithmicdo}{\algorithmicend\ \algorithmicswitch}%
\algdef{SE}[CASE]{Case}{EndCase}[1]{\algorithmiccase\ #1}{\algorithmicend\ \algorithmiccase}%
\algtext*{EndSwitch}%
\algtext*{EndCase}%

\title{Subset Multivariate Collective And Point Anomaly Detection}
\author{Alexander Fisch, Idris Eckley, and Paul Fearnhead \\
	\small{Lancaster University, United Kingdom}}

\begin{document} 
 \maketitle	
\begin{abstract}
In recent years, there has been a growing interest in identifying anomalous structure within multivariate data streams. We consider the problem of detecting collective anomalies, corresponding to intervals where one or more of the data streams behaves anomalously. We first develop a test for a single collective anomaly that has power to simultaneously detect anomalies that are either rare, that is affecting few data streams, or common. We then show how to detect multiple anomalies in a way that is computationally efficient but avoids the approximations inherent in binary segmentation-like approaches. This approach, which we call MVCAPA, is shown to consistently estimate the number and location of the collective anomalies – a property that has not previously been shown for competing methods. MVCAPA can be made robust to point anomalies and can allow for the anomalies to be imperfectly aligned. We show the practical usefulness of allowing for imperfect alignments through a resulting increase in power to detect regions of copy number variation.\\

\end{abstract}
\noindent%


\section{Introduction}

In this article, we consider the challenge of estimating the location of both collective and point anomalies within a multivariate data sequence. The field of anomaly detection has attracted considerable attention in recent years, in part due to an increasing need to automatically process large volumes of data gathered without human intervention. See 
\cite{chandola2009anomaly} and  \cite{pimentel2014review} for comprehensive reviews of the area.

\cite{chandola2009anomaly} categorises anomalies into one of three categories: global, contextual, or collective. The first two of these categories are point anomalies, i.e. single observations which are anomalous with respect to the global, or  local, data context respectively. Conversely, a collective anomaly is defined as a sequence of observations which together form an anomalous pattern.

In this article, we focus on the following setting: we observe a multivariate time series $\textbf{x}_1,...,\textbf{x}_n \in \mathbb{R}^p$ corresponding to $n$ observations observed across $p$ different components. Each component of the series has a typical behaviour, interspersed by time windows where it behaves anomalously. In line with the definition in \cite{chandola2009anomaly}, we call the behaviour within such a time window a collective anomaly. Often the underlying cause of such a collective anomaly will affect more than one, but not necessarily all, of the components. Our aim is to accurately estimate the location of these collective anomalies within the multivariate series, potentially in the presence of point anomalies.
Examples of applications in which this class of anomalies is of interest include but are not limited to genetic data \citep{bardwell2017bayesian,jeng2012simultaneous} and brain imaging data \citep{Epidemic:fmri:Amoc,Epidemic:Aston}. In Genetics, it is of interest to detect regions of the genome containing an unusual copy number. Such genetic variations have been linked to a range of diseases including cancer \citep{diskin2009copy}. To detect these variations, a copy number log-ratio statistics is obtained for all locations along the genome. Segments in which the mean significantly deviates from the typical mean, 0, are deemed variations \citep{bardwell2017bayesian}. Jointly analysing the data of multiple individuals can allow for the detection of shared and even weaker variations \citep{jeng2012simultaneous}. In brain data analysis, sudden shocks can lead to certain parts of the brain exhibiting anomalous activity \citep{Epidemic:Aston} before returning to the baseline level.  

\begin{figure}
	\centering
	\begin{subfigure}[b]{0.5\linewidth}
		\centering
		\includegraphics[width=\linewidth]{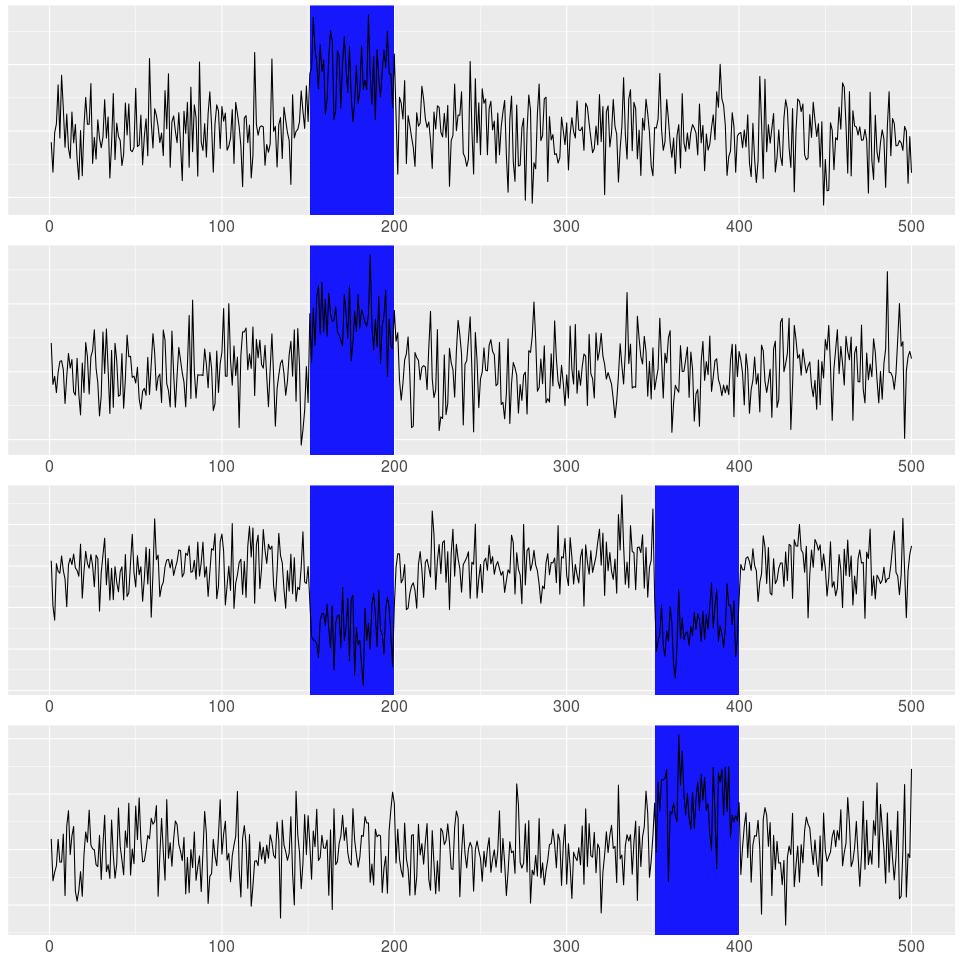} 
		\caption{No lags}
		\label{fig:no_lags}
	\end{subfigure}
	\begin{subfigure}[b]{0.5\linewidth}
		\centering
		\includegraphics[width=\linewidth]{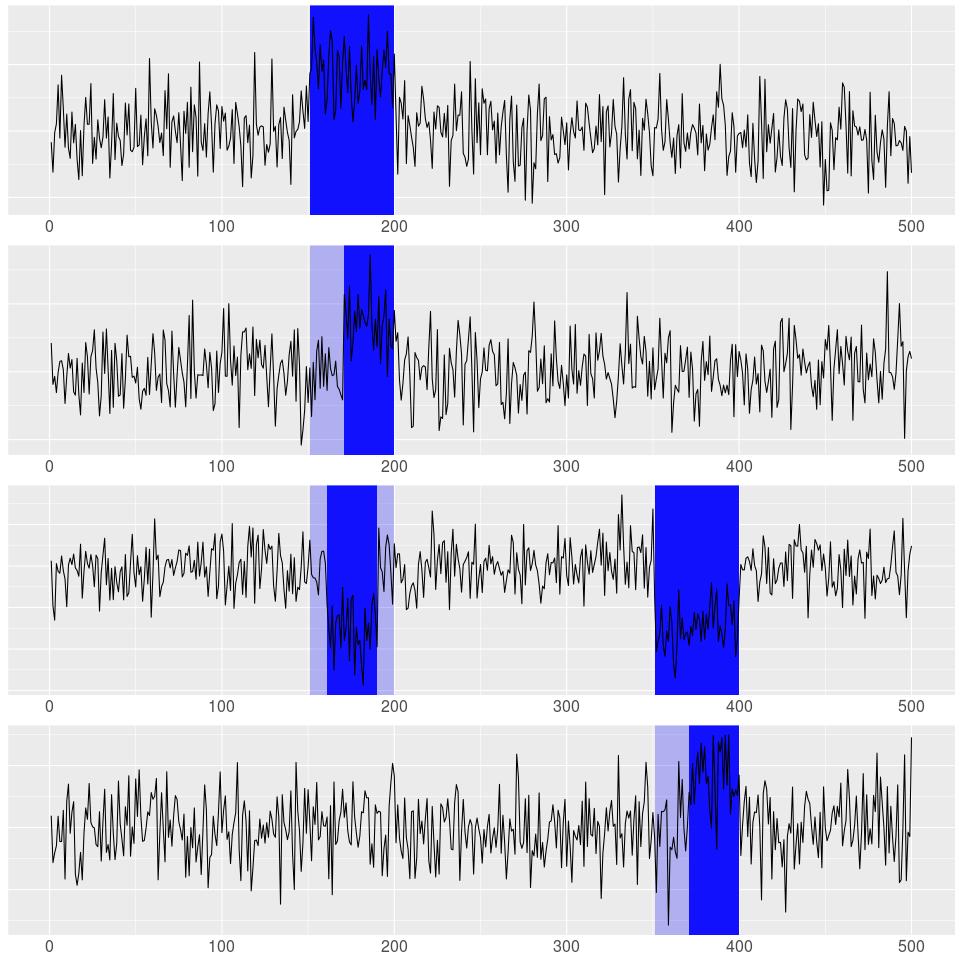} 
		\caption{Lags are indicated in light blue.}
		\label{fig:lags}
	\end{subfigure}
	\caption{An example time series with $K=2$ collective anomalies, highlighted in dark blue. Using the notation from Section \ref{sec:AMOC} these collective anomalies occur between times $s_1 = 150$ to $e_1 = 200$ and $s_2 = 350$ to $e_2 = 400$; and the affected components are $\textbf{J}_1 = \{1,2,3\}$ and $\textbf{J}_2 = \{3,4\}$.}
	\label{fig:example}
\end{figure}

Whilst it may be mathematically convenient to assume that anomalous structure occurs contemporaneously within the multivariate sequence, in practice one might expect some time delays (i.e.\ offsets or lags), as illustrated by Figure \ref{fig:example}. In this article we will consider two different scenarios for the alignment of related collective anomalies across different components. The first, idealised setting, is that concurrent collective anomalies perfectly align. That is we can segment our time series into windows of typical and anomalous behaviour. For each anomalous window the data from a subset of components will be collective anomalies. For some applications however, it is more realistic to assume that concurrent collective anomalies start and end at similar but not necessarily identical time points -- the second setting considered in this paper.

Current approaches aimed at detecting collective anomalies can broadly be divided into state space 
approaches and (epidemic) changepoint methods. State space models assume that a hidden state, which evolves following a Markov chain, determines whether the time series' behaviour is typical or anomalous.  Examples of state space models for anomaly detection can be found in \cite{bardwell2017bayesian} and \cite{smyth1994markov}. These models have the advantage of providing very interpretable output in the form of probabilities of certain segments being anomalous. However, they are often slow to fit and typically require prior parameters which can be difficult to specify.  

The epidemic changepoint model provides an alternative detection framework, built on an assumption that there is a typical behaviour from which the model deviates during certain windows. Early contributions in this area include the work of \cite{levin1985cusum}. Each epidemic changepoint can be modelled as two classical changepoints, one away from and one returning back to the typical distribution. Epidemic changepoints can therefore be inferred by using classical changepoint methods such as PELT \citep{killick2012optimal}, Binary Segmentation \citep{scott1974cluster}, or Wild Binary Segmentation \citep{fryzlewicz2014wild} if the data is univariate and Inspect \citep{wang2018high} or SMOP \citep{pickering2016changepoint} if the data is multivariate. We note, however, that this approach can lead to sub-optimal power, as it is unable to exploit the fact that the typical parameter is shared across the non-anomalous segments \citep{CAPApaper}. 

Many epidemic changepoint methods are based on the popular circular binary segmentation (CBS) by \cite{olshen2004circular}, an epidemic version of binary segmentation. A cost-function based approach for univariate data, CAPA, was introduced by \cite{CAPApaper}. However, within a multivariate setting, the theoretically detectable changes can potentially be very sparse, with a few components exhibiting strongly anomalous behaviour. Alternatively they might also be dense, i.e. with a large proportion of components exhibiting potentially very weak, anomalous behaviour \citep{cai2011optimal,jeng2012simultaneous}.
A range of CBS-derived methods for detecting epidemic changes has therefore been proposed such as Multisample Circular Binary Segmentation \citep{zhang2010detecting} for dense changes, LRS \citep{jeng2010optimal} for sparse changes, and higher criticism \citep{donoho2004higher} based methods like PASS \citep{jeng2012simultaneous} for both types of changes. 

This article makes three main contributions, the first of which is the derivation of a moment function based test for the presence of subset multivariate collective anomalies with finite sample guarantees on false positives and power against both dense and sparse changes. The second contribution is the introduction of a new algorithm \textbf{M}ulti-\textbf{V}ariate \textbf{C}ollective \textbf{A}nd \textbf{P}oint \textbf{A}nomalies (MVCAPA), capable of detecting both point anomalies and potentially lagged collective anomalies in a computationally efficient manner. Finally, the article provides finite sample consistency results for MVCAPA under certain settings which are independent of the number of collective anomalies $K$.

The article is organised as follows: We begin by introducing our modelling framework for (potentially lagged) multivariate collective anomalies in Section \ref{sec:AMOC} and define the penalised negative saving statistic to treat the restricted case of at most one single anomalous segment without lags. Section \ref{sec:Penalty}  introduces penalty regimes, prior to examining their power in Section \ref{sec:FiniteSamples}. We then turn to consider the problem of inferring multiple collective anomalies as well as point anomalies in Section \ref{sec:Multiple}, before discussing computational aspects in Section \ref{sec:Comp}, and establishing finite sample consistency results in Section \ref{sec:Consistency}. We extend penalised saving statistic to point anomalies in Section \ref{sec:Lags} and compare MVCAPA to other approaches in Section \ref{sec:Simulation}. We conclude the article by applying MVCAPA to CNV data in Section \ref{sec:Application}. All proofs can be found in the supplementary material at the end of this paper. MVCAPA has been implemented in the R package \texttt{anomaly} which is available from \texttt{https://github.com/Fisch-Alex/anomaly}.

\section{Model and Inference for a Single Collective Anomaly}

\subsection{Penalised Cost Approach}\label{sec:AMOC}

We begin by focussing on the case where collective anomalies are perfectly aligned. We consider a $p$-dimensional data source for which $n$ time-indexed observations are available. A general model for this situation is to assume that the observation $\textbf{x}_t^{(i)} \in \mathbb{R}$, where $1 \leq t \leq n$ and $1 \leq i \leq p$ index time and components respectively, has probability density function $f_{i}(x, \bm{\theta}^{(i)}(t))$ and that the parameter, $\bm{\theta}^{(i)}(t)$, depends on whether the observation is associated with a period of typical behaviour or an anomalous window. We let $\bm{\theta_0}^{(i)}$ denote the parameter associated with component $i$ during its typical behaviour. Let $K$ be the number of anomalous windows, with the $k$-th window starting at $s_k+1$ and ending at $e_k$ and affecting the subset of components denoted by the set $\textbf{J}_k$. We assume that anomalous windows do not overlap, so $e_k\leq s_{k+1}$ for $k=1,\ldots,K-1$. If collective anomalies of interest are assumed to be of length at least $l \geq 1$, we also impose $e_k - s_k \geq l$. Our model then assumes that the parameter associated with observation $\textbf{x}^{(i)}_t$ is
\begin{equation}\label{eq:Simple}
\bm{\theta}^{(i)}(t) = \begin{cases}
\bm{\theta}^{(i)}_1  & \text{if} \; s_1  < t \leq e_1   \; \text{and} \; i \in \textbf{J}_1, \\
&\vdots \\
\bm{\theta}^{(i)}_K  & \text{if} \; s_K  < t \leq e_K \; \text{and} \; i \in \textbf{J}_K,\\
\bm{\theta}^{(i)}_0  & \text{otherwise}.
\end{cases}
\end{equation}

We start by considering the case where there is at most one collective anomaly, i.e.\ where $K\leq1$, and introduce a test statistic to determine whether a collective anomaly is present and, if so, when it occurred and which components were affected. 
The methodology will be generalised to multiple collective anomalies in Section \ref{sec:Multiple}. Throughout we assume that the  parameter, $\bm{\theta}_0$, representing the sequence's baseline structure, is known. If this is not the case, an approximation can be obtained by estimating $\bm{\theta}_0$ robustly over the whole data, as in \cite{CAPApaper}.

Given the start and end of a window, $(s,e)$, and the set of components involved, $\textbf{J}$, we can calculate the log-likelihood ratio statistic for the collective anomaly. To simplify notation we introduce a cost,
\[
\mathcal{C}_i \left( \textbf{x}_{s+1:e}^{(i)} ,\bm{\theta}\right) = -2 \sum_{t=s+1}^e \log f (\textbf{x}_t^{(i)},\bm{\theta}),
\]
defined as minus twice the log-likelihood of data $\textbf{x}_{s+1:e}^{(i)}$ for parameter $\bm{\theta}$. We can then quantify the saving obtained by fitting component $i$ as anomalous for the window starting at $s$ and ending at $e$ as
\[
\mathcal{S}_{i}\left(s,e\right) =  \mathcal{C}_i \left( \textbf{x}_{(s+1):e}^{(i)} , \bm{\theta}_0^{(i)} \right) - \min_{\bm{\theta}}\left(\mathcal{C}_i \left( \textbf{x}_{(s+1):e}^{(i)} , \bm{\theta} \right)\right) .
\]
The log-likelihood ratio statistic is then 
$\sum_{i\in \textbf{J}} \mathcal{S}_{i}\left(s,e\right)$.
As the start or the end of the window, or the set of components affected, are unknown \textit{a priori}, it is natural to maximise the log-likelihood ratio statistic over the range of possible values for these quantities. However, in doing so we need to take account of the fact that different $ \textbf{J}$ will allow different numbers of components to be anomalous, and hence will allow maximising the log-likelihood, or equivalently minimising the cost, over differing numbers of parameters. This suggests penalising the log-likelihood ratio statistic differently, depending on the size of $ \textbf{J}$. That is we calculate
\begin{equation} \label{eq:test}
\max_{\textbf{J},s \leq e - l} \left[\sum_{i\in \textbf{J}} \mathcal{S}_{i}\left(s,e\right) - P(|\textbf{J}|)\right],
\end{equation}
where $P(.)$ is a suitable positive penalty function of the number of components that change, and $l$ is the minimum segment length.  We will detect a change if (\ref{eq:test}) is positive, and estimate its location and the set of components that are anomalous based on the values of $s$, $e$, and $\textbf{J}$ that give the maximum of (\ref{eq:test}). Whilst we have motivated this procedure based on a log-likelihood ratio statistic, the idea can be applied with more general cost functions. For example if we believe anomalies result in a change in mean then we could use cost functions based on quadratic loss. Suitable choices for the penalty function, $P(\cdot)$, will be discussed in Section \ref{sec:Penalty}.

To efficiently maximise (\ref{eq:test}),define positive constants $\alpha$, $\beta_{1:p}$ with $P(1)=\alpha+\beta_1$, and, for $i=2,\ldots,p$, $\beta_i=P(i)-P(i-1)$. So the $\beta_i$s are the first differences of our penalty function $P(\cdot)$.  Further, let the order statistics of $\mathcal{S}_{1}\left(s,e\right),\ldots,\mathcal{S}_{p}\left(s,e\right)$ be
\begin{equation*}
\mathcal{S}_{(1)}\left(s,e\right) \geq ... \geq \mathcal{S}_{(p)}\left(s,e\right),
\end{equation*}
and define the penalised saving statistic of the segment $\textbf{x}_{(s+1):e}$, 
\begin{equation*}
\mathcal{S}\left(s,e\right) = \max_{k}\left( \sum_{ i =1}^{k} \mathcal{S}_{(i)}\left(\textbf{x}_{s+1:e}\right) - \beta_i \right) - \alpha.
\end{equation*}
Then (\ref{eq:test}) is obtained by maximising $\mathcal{S}\left(s,e\right) $ over $s \leq e - l$. 

The choice of $\beta_1,...,\beta_p$ and $\alpha$ has a significant impact on the performance of the statistic. For example, if we set $\beta_i=0$ for $i=2,\ldots,p$, then our approach will assume all components are anomalous in any anomalous segment. Clearly, $\alpha$ and $\beta_1$ are only well specified up to their sum and $\alpha$ can be absorbed into $\beta_1$ without altering the properties of our statistic. However, not doing so can have computational advantages. Indeed, it removes the need of sorting when all the $\beta_i$s are identical, which can be appropriate in certain settings discussed in the next section.

\subsection{Choosing Appropriate Penalties}\label{sec:Penalty}

We now turn to the problem of choosing appropriate penalties for the methodology introduced in the previous section. Penalties are typically chosen in a way which controls false positives under the null hypothesis that no anomalous segments are present. However, the setting in which the penalty is to be used can also play an important role. In particular, it is typically required that under the null hypothesis
\begin{equation}\label{eq:Probboubd}
\mathbb{P}\left(\hat{K} = 0\right) \geq 1 - Ae^{-\psi(p,n)},
\end{equation}
where $A$ is a constant and $\psi:=\psi(p,n)$ increases with $n$ and/or $p$, depending on the setting. For example, in panel data \citep{bardwell2019most} or brain imaging data \citep{Epidemic:Aston}, the number of time points $n$ may be small but we may have data from a large number of variates, $p$. Setting $\psi(p,n) \propto \log(p)$ is therefore a natural choice so that the false positive probability tends to 0 as $p$ increases. In a streaming data context, the number of sampled components $p$ is typically fixed, while the number of observations $n$ increases. Setting $\psi(p,n) \propto \log(n)$ is then a natural choice. In an application where both $n$ and $p$ are large, or can increase, setting $\psi(p,n) \propto \log(n) + \log(p)$, would be a natural choice. 


Our approach to constructing appropriate penalty regimes is to construct a bound on the probability that $\sum_{ i =1}^{k} (\mathcal{S}_{(i)}(\textbf{x}_{s+1:e}) - \beta_i) - \alpha > 0$ under the null separately for all $k=1,\ldots,p$ and all windows $(s,e)$. We will derive explicit results for the case in which collective anomalies are characterised by an anomalous mean -- arguably one of the most common settings in applications -- and point to generalisations where appropriate. As mentioned in the previous section, we assume throughout this section that the parameter $\bm{\theta}_0$ of the typical distribution is known. 

\begin{figure}
	\centering
	\includegraphics[width=\linewidth]{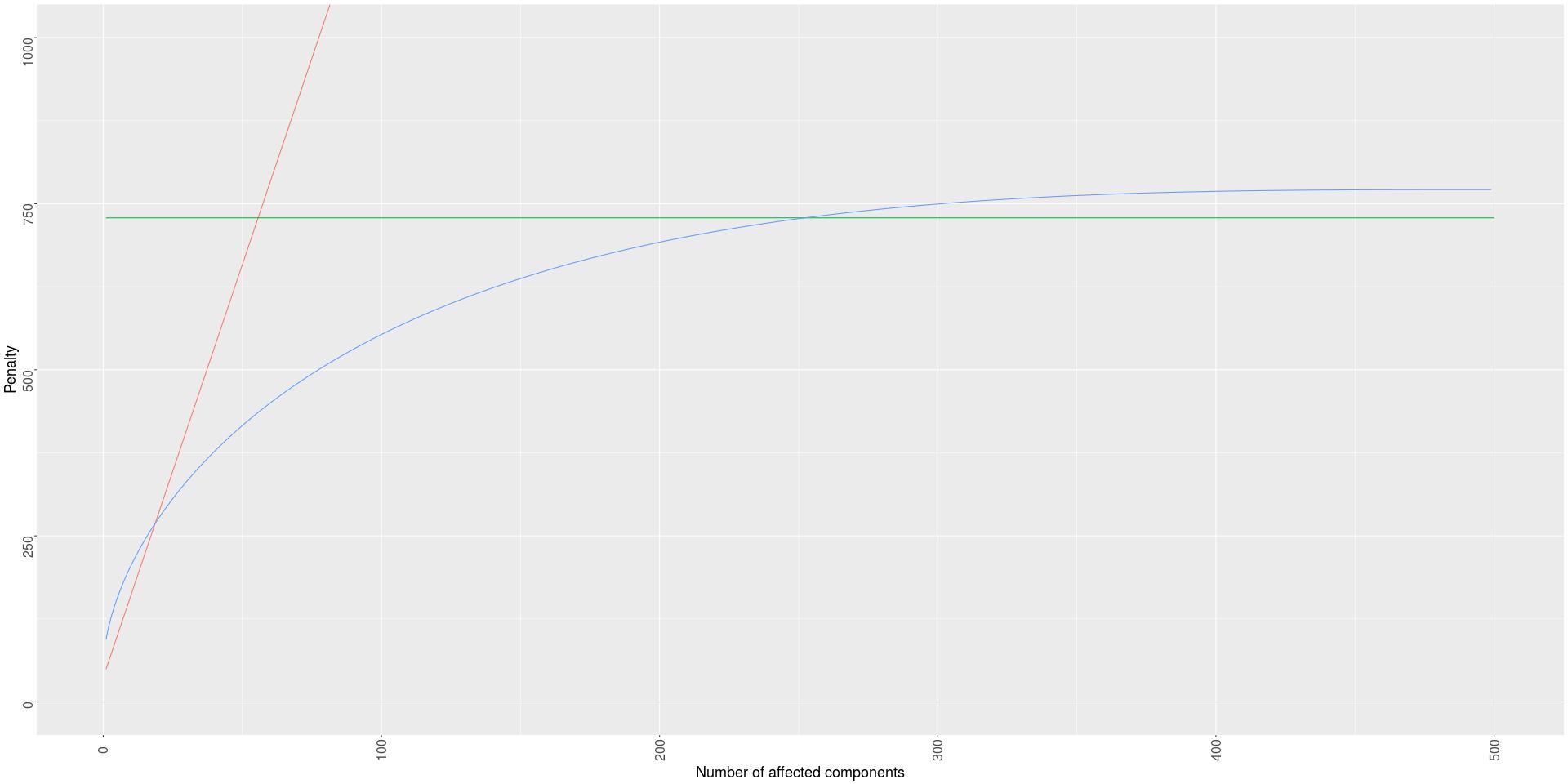}
	\caption{A comparison of the 3 penalty regimes for $p=500$ and $\psi=2\log(10000)$. Regime 1 is in green, regime 2 in red and regime 3 in blue.}
	\label{fig:penalties}
\end{figure} 

Assume without loss of generality that the typical mean is equal to 0 and the typical standard deviation is equal to 1. Under the the null model, which we denote $\mathcal{M}_0$, we therefore have that
\begin{equation*}
\textbf{x}_t^{(c)} \; \; \stackrel{i.i.d.}{\sim} \; \; \epsilon_{t,c} \; \; \; \;  \; \; \; 
\end{equation*}
where $\epsilon_{t,c}$ is Gaussian white noise. Denote the mean for data sequence $m$ within a window $(s,e)$ as $\bar{\textbf{x}}^{(m)}_{s+1:e}$. The cost function obtained if we model the data in this way is  the residual sum of squares. Under this cost function, for any $0\leq s<e\leq n$,
\begin{equation*}
\mathcal{S}_{m}\left(s,e\right) = \sum_{t=s+1}^e\left( \left(\mathbf{x}^{(m)}_t\right)^2- \left(\mathbf{x}^{(m)}_t- \bar{\textbf{x}}^{(m)}_{s+1:e}\right)^2 \right)=  (e-s)\left(\bar{\textbf{x}}^{(m)}_{(s+1):e} \right)^2.
\end{equation*}
Furthermore, under $\mathcal{M}_0$, this saving is $\chi^2_1$-distributed. 


We present three distinct penalty regimes that give valid choices for the penalties for this setting, each with different properties. All penalty regimes will be indexed by a parameter $\psi$ which corresponds to the exponent of the probability bound, as in \eqref{eq:Probboubd}.
A first strategy for defining penalties consists of just using a single global penalty.\\
\\
\textbf{Penalty Regime 1:} $\alpha = p + 2\sqrt{p\psi} +2\psi$ and $\bm{\beta}_m = 0$.
\\
\\
The following result based on standard tail bounds of the $\chi_p^2$ distribution shows that this penalty regime controls the overall type 1 error rate at the desired level
\begin{Proposition}\label{prop:FPcontrol_mean_penalty_1}
	Let $\textbf{x}$ follow $\mathcal{M}_0$ and let $\hat{K}$ denote the number of inferred collective anomalies under penalty regime 1. Then, there exists a constant $A_1$ such that $\mathbb{P}\{\hat{K} = 0\} \geq 1 - A_1n^2e^{-\psi}$.
\end{Proposition}

Under this penalty, we would infer that any detected anomaly region will affect all components. This is inappropriate, and is likely to lead to a lack of power, if we have anomalous regions that only affect a small number of components. For this type of anomalies, the following regime offers a good alternative as it typically penalises fitting collective anomalies with few components substantially less: \\
\\
\textbf{Penalty Regime 2:} $\alpha = 2(1+\epsilon)\psi$ and $\bm{\beta}_m = 2\log(p)$, for $1 \leq m \leq p$.
\\
\\
The following results based on tail bounds for the $\chi_1^2$-distribution shows that this penalty controls false positives:
\begin{Proposition}\label{prop:FPcontrol_mean_penalty_2}
	Let $\textbf{x}$ follow $\mathcal{M}_0$ and let $\hat{K}$ denote the number of inferred collective anomalies under penalty regime 2. Then, there exists a constant $A_2$ such that $\mathbb{P}\{\hat{K} = 0\} \geq 1 - A_2n^2e^{-\psi}$.
\end{Proposition}

Comparing penalty regime 2 with penalty regime 1 we see that it has a lower penalty for small $m$, but a much higher penalty for $m>>p/\log p$. As such it has higher power against collective anomalies affecting few components, but low power if the collective anomalies affect most components. Taking the point-wise minimum between both penalties therefore provides a balanced approach, providing power against both types of changes (see Figure \ref{fig:penalties}.) Moreover, this approach can be generalised to settings other than the change in mean case provided the distribution of the savings is sub-exponential under the null distribution. Indeed, the first penalty regime is derived from a Bernstein bound and the second one from an exponential Chernoff bound on the tail. 

For the special case considered here, in which the savings follow a $\chi^2_1$ distribution under the null hypothesis a third penalty regime can be derived:
\\
\\
\textbf{Penalty Regime 3:} $\sum_{i=1}^{m}\bm{\beta}_i = 2(\psi +\log(p)) + m + 2pa_mf(a_m) + 2\sqrt{(m + 2pa_mf(a_m))(\psi +\log(p))}$, where $f$ is the PDF of the $\chi^2_1$ distribution and $a_m$ is defined via the implicit equation $\mathbb{P}\left(\chi_1^2> a_m\right) = m/p$.
\\
\\
It satisfies the following proposition regarding false positive control
\begin{Proposition}\label{prop:FPcontrol_mean_penalty_3}
	Let $\textbf{x}$ follow $\mathcal{M}_0$ and let $\hat{K}$ denote the number of inferred collective anomalies under penalty regime 3. Then, there exists a constant $A_3$ such that $\mathbb{P}\{\hat{K} = 0\} \geq 1 - A_3n^2e^{-\psi}$.
\end{Proposition}
Moreover, as can be seen from Figure \ref{fig:penalties}, this penalty regime provides a good alternative to the other penalty regimes, especially in intermediate cases. It can be generalised to other distributions provided quantiles of $\mathcal{S}(s,e)$ can be estimated and exponential bounds for $(\mathcal{S}(s,e)-a)^+$ can be derived under the null hypothesis. 

In practice, to maximise power, we choose $\alpha, \beta_1, ..., \beta_p$ so that the resulting penalty function $P(k)$, is the point-wise minimum of the penalty functions $P_1(k)$, $P_2(k)$, and $P_3(k)$ resulting from using penalty regimes 1, 2, and 3 respectively. We call this the \textbf{composite regime}. It is a corollary from Propositions \ref{prop:FPcontrol_mean_penalty_1}, \ref{prop:FPcontrol_mean_penalty_2}, and \ref{prop:FPcontrol_mean_penalty_3}, that this composite penalty regime achieves $\mathbb{P}\{\hat{K} = 0\} \geq 1 - (A_1+A_2+A_3)n^2e^{-\psi}$ when $\textbf{x}$ follows $\mathcal{M}_0$. It should be noted that the $n^2$ term comes from a Bonferroni correction over all possible start and end points. Fixing a maximum segment length $M$ therefore reduces the $n^2$ to $nM$. 

Whilst the above propositions give guidance regarding the shape of the penalty function $P\left(\cdot\right)$ as well as a finite sample bound on the probability of false positives they do not give an exact false positive rate. A user specified rate can nevertheless be obtained by scaling the penalties $\beta_1,...\beta_p$ and $\alpha$ by a constant. This single constant can then be tuned using simulations on data exhibiting no anomalies. 

\subsection{Results on Power}\label{sec:FiniteSamples}

We now examine the power of the penalised saving statistic at detecting an anomalous segment $\textbf{x}_{s+1},...,\textbf{x}_{e}$ when using the thresholds defined in the previous section. In particular, we examine its behaviour under a fixed $n$ and large $p$ regime. It is well known \citep{jeng2012simultaneous,donoho2004higher,cai2011optimal}, that different regimes determining the detectability of collective anomalies apply in this setting depending on the proportion of affected components. We follow the asymptotic parametrisation of \cite{jeng2012simultaneous} and therefore assume that
\begin{equation}\label{eq:ANOM_PARAM}
\textbf{x}_t^{(i)} = v^{(i)} \mu  + \bm{\epsilon}_t^{(i)}, \; \; \; 
v^{(i)} \sim \begin{cases}
0 & \text{with prob.} \; \; 1 - p^{-\xi}, \\
1 & \text{with prob.} \; \; p^{-\xi},
\end{cases} \; \; \; \text{and} \; \; \; \bm{\epsilon}_t^{(i)} \sim N(0,1), \; \; \; \text{for} \; \; \; s < t \leq e.
\end{equation}

Typically \citep{jeng2012simultaneous}, changes are characterised as either sparse or dense. In a sparse change, only a few components are affected. Such changes can be detected based on the saving of those few components being larger than expected after accounting for multiple testing. The affected components therefore have to experience strong changes to be reliably detectable. On the other hand, a dense change is a change in which a large proportion of components exhibits anomalous behaviour. A well defined boundary between the two cases exist with $\xi \leq \frac{1}{2}$ corresponding to dense $\xi > \frac{1}{2}$ and corresponding to sparse changes \citep{jeng2012simultaneous,enikeeva2013high}. The changes affecting the individual components can consequently be weaker and still be detected by averaging over the affected components. Depending on the setting, the change in mean is parametrised by $r_p \in \mathbb{R}$ in the following manner:
\begin{equation*}
(e-s)\mu^2 = \begin{cases}
{2r_p\log(p)} &  \frac{1}{2} < \xi < 1,\\
p^{-2r_p} & 0 \leq \xi \leq \frac{1}{2}.
\end{cases}
\end{equation*}

Both \cite{jeng2012simultaneous} and \cite{cai2011optimal} derive detection boundaries for $r_p$, separating changes that are too weak to be detected from those changes strong enough to be detected. For the case in which the standard deviation in the anomalous segment is the same as the typical standard deviation, the detectability boundaries correspond to 
\begin{equation*}
\rho^- = \begin{cases}
\left(1 - \sqrt{1-\xi}\right)^2 &  \frac{3}{4} < \xi < 1,\\
\xi -  \frac{1}{2} & \frac{1}{2} < \xi \leq \frac{3}{4},
\end{cases}
\end{equation*}
for the sparse case ($\frac{1}{2} < \xi < 1$) and 
\begin{equation*}
\rho^+ = \left(\frac{1}{2} - \xi \right)
\end{equation*}
for the dense case ($0 \leq \xi \leq \frac{1}{2}$). The following proposition establishes that the penalised saving statistic has power against all sparse changes within the detection boundary, as well as power against most dense changes within the detection boundary
\begin{Proposition}\label{Prop:Power}
	Let the series $\textbf{x}_1, ... ,\textbf{x}_n$ contain an anomalous segment $\textbf{x}_{s+1},...,\textbf{x}_{e}$, which follows the model specified in equation \ref{eq:ANOM_PARAM}. Let $r_p > \rho^-$ if $\frac{1}{2} < \xi < 1$ or $r_p <\frac{1}{2} \rho^+$ if $0 \leq \xi \leq \frac{1}{2}$. Then the number of collective anomalies, $\hat{K}$, estimated by MVCAPA using the composite penalty regime with $\psi = 2 \log(n) + 2\log(\log(p))$ on the data $\textbf{x}_1, ... ,\textbf{x}_n$, satisfies
	\begin{equation*}
	\mathbb{P}\left(\hat{K} \neq 0 \right) \rightarrow 1 \;\;\; \; \text{as} \;\;\;  p \rightarrow \infty.
	\end{equation*}
\end{Proposition}

Whilst the above assumed $n$ to be fixed, the result also holds if $n=o(p)$. Moreover, rather than requiring $\mu_i$ to be 0, or a common value $\mu$, it is trivial to extend the result to the case where $\mu_1,...,\mu_p$ are i.i.d.\ random variables whose magnitude exceeds $\mu$ with probability $p^{-\xi}$. It is worth noticing that the third penalty regime is required to obtain optimal power against the intermediate sparse setting $\frac{1}{2} < \xi \leq \frac{3}{4}$.

\section{Inference for Multiple Anomalies}\label{sec:Multiple}

\subsection{Inference for $K$ Collective Anomalies and Point Anomalies}

We now turn to the problem of generalising the methodology introduced in Section \ref{sec:AMOC} to infer multiple collective anomalies. We will then borrow methodology from \cite{CAPApaper} to incorporate point anomalies within the inference. A natural way of extending the methodology introduced in Section \ref{sec:AMOC} to infer multiple collective anomalies in various ways, is to maximise the penalised saving jointly over the number and location of potentially multiple anomalous windows. That is we
infer $\hat{K}$, $\left(\hat{s}_1,\hat{e}_1,\hat{\bm{J}}_1\right)$,..., $\left(\hat{s}_{\hat{K}},\hat{e}_{\hat{K}}, \hat{\bm{J}}_{\hat{K}}\right)$ by directly maximising  
\begin{align}\label{eq:tobemaxed}
\sum_{k=1}^{\hat{K} }\mathcal{S}\left( \hat{s}_k,\hat{e}_k\right) ,
\end{align}
subject to $\hat{e}_k -\hat{s}_k  \geq l $ and $\hat{e}_{k} \leq \hat{s}_{k+1}$.



It is well know from the literature that many methods designed to detect changes, or collective anomalies, are vulnerable to point anomalies \citep{fearnhead2019changepoint,CAPApaper}. 
Distinguishing between point and collective anomalies only makes sense if they are different, that is collective anomalies are longer than length 1. Under such an assumption, our approach can be extended to model both point and collective anomalies.

Borrowing the approach of \cite{CAPApaper}, 
a point anomaly can be modelled as an epidemic changepoint of length 1 occurring during a segment of typical behaviour. Joint inference on collective and point anomalies can then be performed by maximising the penalised saving
\begin{align}\label{eq:quantitytobemaxedwithanomalies}
\sum_{k=1}^{\hat{K} }\mathcal{S}\left( \hat{s}_k,\hat{e}_k\right)  + \sum_{t \in O} \mathcal{S'}\left( \textbf{x}_t\right) ,
\end{align} 
with respect to $\hat{K}$, $\left(\hat{s}_1,\hat{e}_1,\hat{\bm{J}}_1\right)$,..., $\left(\hat{s}_{\hat{K}},\hat{e}_{\hat{K}}, \hat{\bm{J}}_{\hat{K}}\right)$, and the set of point anomalies $O$, subject to $\hat{e}_k - \hat{s}_k \geq l $, $\hat{e}_{k}<\hat{s}_{k+1}$ $ \left( \cup_i [s_i+1,e_i] \right) \cap O =\emptyset$. Here, $\mathcal{S'}\left( \textbf{x}_t\right)$ is the saving of introducing a point anomaly. 

For example, when collective anomalies are characterised by changes in mean
\begin{equation*}
\mathcal{S'}\left( \textbf{x}_t\right) = \sum_{i=1}^{p} \max \left(\left(\frac{\textbf{x}_t^{(i)} - {\bm{\mu}}_0^{(i)}}{{\bm{\sigma}}_0^{(i)}}\right)^2 - \beta', 0 \right).
\end{equation*}
The penalised savings $\mathcal{S'}\left(\cdot\right)$ and $\mathcal{S}\left(\cdot\right)$, as we assume point anomalies to be sparse. Suitable choices for $\beta'$ will be discussed in the next subsection.

\subsection{Penalties for Point Anomalies}\label{sec:PenaltiesPtAnomaly}

Penalties for point anomalies can be chosen with the aim of controlling false positives under the null hypothesis, that no collective or point anomalies are present. When collective anomalies are characterised by a change in mean the null hypothesis is identical to that in Section \ref{sec:Penalty}. The following proposition holds for any penalty $\beta'$:
\begin{Proposition}\label{prop:FPControlPtAnomalies}
	Let $\mathcal{M}_0$ hold and $\hat{O}$ denote the set of point anomalies inferred by MVCAPA using $\beta'$ as penalty for point anomalies. Then, there exists a constant $A$ such that
	\begin{equation*}
	\mathbb{P}\left( \hat{O} = \emptyset \right) \geq 1 - Anpe^{-\frac{1}{2}\beta'}.
	\end{equation*}
\end{Proposition}
This suggests setting $\beta' = 2\log(p) + 2\psi$, where $\psi$ is as in Section \ref{sec:Penalty}.

\section{Computation}\label{sec:Comp}

We now turn to the problem of maximising the penalised saving introduced in the previous section. 
The standard approach to extend a method for detecting an anomalous window to detecting multiple anomalous windows is through circular binary segmentation (CBS, \cite{olshen2004circular}) -- which repeatedly applies the method for detecting a single anomalous window or point anomaly. Such an approach is equivalent to using a greedy algorithm to approximately maximise the penalised saving and has computational cost of $O(Mn)$, where $M$ is the maximal length of collective anomalies and $n$ is the number of observations. Consequently, the runtime of CBS is $O(n^2)$ if no restriction is placed on the length of collective anomalies. We will show in this section that we can directly maximise the penalised saving by using a pruned dynamic programme. This enables us to jointly estimate the anomalous windows, at the same or at a lower computational cost than CBS.


\textbf{1. Only collective anomalies.}  The penalised saving defined in \eqref{eq:tobemaxed} can be maximised exactly using a dynamic programme. Indeed, defining $C(m)$ to be the largest penalised saving for all observations up to, and including, the time $m$, the following relationship holds:
\begin{align*}\label{eq:DP}
C(m) = \max \Bigg(C(m-1), \max_{0\leq t \leq m -l} \Bigg( C(t) + {\mathcal{S}}\left( t,m\right)  \Bigg) \Bigg).
\end{align*}
It should be noted that calculating $\mathcal{S}\left( t,m\right)$ is, on average, an $O(p\log(p))$ operation, since it requires sorting the savings made from introducing a change in each component. This sorting is not required when all $\bm{\beta}_i$ are identical. Setting all $\beta_i$ to the same value, as in penalty regime 1, therefore reduces the computational cost to $O(p)$. For a maximum segment length $M$, the computational cost of this dynamic programme approach scales like $O(Mn)$. If no maximum segment length is specified, it therefore scales quadratically in $n$. In this setting, it therefore achieves the same run-time as CBS in both cases.

\textbf{2. Collective and point anomalies.} The saving in (\ref{eq:quantitytobemaxedwithanomalies}) can be minimised exactly via a slight modification of the previous dynamic programme. Indeed, writing $C(m)$ for the largest penalised saving of all observations up to and including time $m$, the relationship
\begin{align*}
C(m) = \max \Bigg(C(m-1), \max_{0\leq t \leq m -l} \Bigg( C(t) + {\mathcal{S}}\left( t,m\right)  \Bigg), C(m-1) + {\mathcal{S'}}\left( \textbf{x}_t\right) \Bigg)
\end{align*}
holds for $C(0) = 0$. The previous observations regarding the computational complexity in $M$ and $n$ remain valid. 

\textbf{3. Pruning the dynamic programme.} Solving the whole dynamic programme if no maximum segment length is specified has a computational cost increasing quadratically in $n$. However, the solution space of the dynamic programme can be pruned in a fashion similar to \cite{killick2012optimal} and \cite{CAPApaper} to reduced this computational cost. Indeed, the following proposition holds:
\begin{Proposition}\label{prop:Pruning_easy}
	Let the costs $\mathcal{C}_i(,)$ be such that 
	\begin{equation*}
	\min_{\bm{\theta}} \left( \sum_{t=a+1}^{c} \mathcal{C}_i \left( \textbf{x}_t,\bm{\theta}\right) \right) \geq \min_{\bm{\theta}} \left( \sum_{t=a+1}^{b} \mathcal{C}_i \left( \textbf{x}_t,\bm{\theta}\right) \right) + \min_{\bm{\theta}} \left( \sum_{t=b+1}^{c} \mathcal{C}_i \left( \textbf{x}_t,\bm{\theta}\right) \right)
	\end{equation*}
	holds for all $\textbf{x}$ and $a,b,c$ such that $b-a \geq l$ and $c-b \geq l$. Then, if for some $t$ there exists an $m \geq t - l$ such that 
	\begin{equation*}
	C(m) - \alpha  - \sum_{1}^{p} \bm{\beta}_i > C(t) + {\mathcal{S}}\left( t,m\right),
	\end{equation*}
	then, for all $m' \geq m + l$,
	\begin{equation*}
	C(m') > C(t) + {\mathcal{S}}\left( t,m'\right).
	\end{equation*}
\end{Proposition}
A wide range of cost functions, such as the negative log-likelihood and the sum of squares satisfy the condition required by the above proposition. The proposition implies that if for some $t$ there exists an $m \geq t - l $ such that 
\begin{equation*}
C(m) - \alpha - \sum_{1}^{p} \bm{\beta}_i > C(t) + \mathcal{S}\left(t,m\right)  
\end{equation*}
holds, $t$ can be dropped as an option from the dynamic programme for all steps after step $m+l$, thus reducing the cost of the algorithm. As a result of this pruning we found the runtime of MVCAPA to be close to linear in $n$, when the number of collective anomalies increased linearly with $n$. 

\section{Accuracy of Detecting and Locating Multiple Collective Anomalies}\label{sec:Consistency}

Whilst we have shown our method has good properties when detecting a single anomalous window, it is natural to ask whether the extension to detecting multiple anomalous windows will be able to consistently infer the number of anomalous windows and accurately estimate their locations. Developing such results for the joint detection of sparse and dense collective anomalies is notoriously challenging, as can be seen from the fact that previous work on this problem \citep{jeng2012simultaneous} has not provided any such results. Another new feature of this proof is that the results allow for the number of anomalous segments $K$ to increase, whereas most results in the related changepoint literature (e.g.\ \cite{fryzlewicz2014wild}) assume $K$ to be fixed. 


Consider a multivariate sequence $\textbf{x}_1,...,\textbf{x}_n \in \mathbb{R}^p$, which is of the form $\textbf{x}_t = \bm{\mu}(t) + \bm{\eta}_t, $ where the mean $\bm{\mu}(t)$ follows a subset multivariate epidemic changepoint model with $K$ epidemic changepoints in mean. For simplicity, we assume that within an anomalous window all affected components experience the same change in mean, and that the noise process is i.i.d.\ Gaussian although the results extend to sub-Gaussian noise, i.e.\ 
\begin{equation}\label{eq:Model}
\bm{\mu}^{(i)}(t) = \begin{cases}
\bm{\mu}_1  & \text{if} \; s_1 < t \leq e_1  \; \text{and} \; i \in \textbf{J}_1, \\
&\vdots \\
\bm{\mu}_K  & \text{if} \; s_K < t \leq e_K, \; \text{and} \; i \in \textbf{J}_K,\\
0  & \text{otherwise},
\end{cases} \; \; \; \; \; \; \; \; \; \; \bm{\eta}_t \sim N(\bm{0}, \textbf{I}_p).
\end{equation}

Consider also the following choice of penalty, which is very similar to the the point-wise minimum between penalty regimes 1 and 2:
\begin{equation}\label{eq:MAINTHEOREMPENALTIES}
\sum_{i=1}^m \bm{\beta}_i = \begin{cases}
C\psi + Cm\log(p)  & \text{if} \; \; \; m \leq k^*, \\
p + C\psi + C\sqrt{p\psi}  & \text{if} \; \; \; m > k^*.
\end{cases}
\end{equation}
Here, $C$ is a constant, $\psi:=\psi(n,p)$ sets the rate of convergence and the threshold 
\begin{equation*}
k^* = p^{1/2} \frac{\psi}{\log(p)},
\end{equation*}
is defined as the threshold separating sparse changes from dense changes. 

Anomalous regions can be easier or harder to detect depending on the strength of the change in mean characterising them and the number of components they affect.
This intuition can be quantified by 

\begin{equation*}
\triangle_k^2 = \begin{cases}
\cfrac{\Huge\bm{\mu}_k^2}{\log(p) + \psi |J_k|^{-1}} \normalsize & \text{if} \; \; \; |J_k| \leq k^*, \\
\cfrac{\bm{\mu}_k^2}{\sqrt{p\psi}|J_k|^{-1} + \psi |J_k|^{-1} }  & \text{if} \; \; \; |J_k| > k^*,
\end{cases}
\end{equation*}
\normalsize
which we define to be the signal strength of the $k$th anomalous region. The following consistency result then holds

\begin{Thm}\label{THM:Main}
	There exists a global constant $C$ such that the inferred partition \\ $\tau = \{(\hat{s}_1,\hat{e}_1,\hat{\textbf{J}}_1),...,(\hat{s}_{\hat{K}},\hat{e}_{\hat{K}},\hat{\textbf{J}}_{\hat{K}}) \}$ obtained by applying MVCAPA using the penalty regime specified in (\ref{eq:MAINTHEOREMPENALTIES}) on data $\textbf{x}$ which follows the distribution specified in (\ref{eq:Model}) satisfies
	\begin{equation}
	\mathbb{P}\left( \hat{K} = K , \; \; \left| \hat{s}_k - s_k \right| < \frac{10C}{\triangle_k^2} , \; \;  \left| \hat{e}_k - e_k \right| < \frac{10C}{\triangle_k^2}  \right) > 1 - An^3e^{-\psi},
	\end{equation}
	provided that 
	\begin{align*}
	e_k-s_k \geq \frac{40C}{\triangle_k^2}, \; \; \; \; \; s_{k+1}-e_{k} \geq \frac{40C}{\triangle_k^2}, \; \; \; \; \; s_{k}-e_{k-1} \geq \frac{40C}{\triangle_k^2}
	\end{align*}
	holds for $k=1,...,K$.
\end{Thm}

The result is proved in the appendix. This finite sample result holds for a fixed $C$, which is independent of $n$, $p$, $K$, and/or $\triangle_{k}$. When $\psi=\log(p)$, the threshold $k^*$ is identical to that in \cite{jeng2012simultaneous}. However, if $\phi$ is chosen to increase with $\log(n)$, so will $k^*$. This formalises the intuition that when $n >> p$, all changes are in some sense sparse.

\section{Incorporating Lags}\label{sec:Lags}

So far, we have assumed that the collective anomalies were characterised by the model specified in  (\ref{eq:Simple}), which assumes all anomalous windows are perfectly aligned. In some applications, such as the vibrations recorded by seismographs at different locations, certain components will start exhibiting atypical behaviour later and/or return to the typical behaviour earlier. An example can be found in Figure \ref{fig:lags}. It is possible to extend the model in (\ref{eq:Simple}) to allow for this behaviour by allowing lags in the start or end of each window:  
\begin{equation}\label{eq:Superhard}
\bm{\theta}^{(i)}(t) = \begin{cases}
\bm{\theta}^{(i)}_1  & \text{if} \; s_1 + \textbf{d}_1^{(i)} < t \leq e_1 - \textbf{f}_1^{(i)}  \; \text{and} \; i \in \textbf{J}_1, \\
&\vdots \\
\bm{\theta}^{(i)}_K  & \text{if} \; s_K + \textbf{d}_K^{(i)} < t \leq e_K - \textbf{f}_K^{(i)} \; \text{and} \; i \in \textbf{J}_K,\\
\bm{\theta}^{(i)}_0  & \text{otherwise}.
\end{cases}
\end{equation}
Here the start and end lag of the $i$th component during the $k$th anomalous window are denoted, respectively, by $0 \leq \textbf{d}_k^{(i)} \leq w $ and $0 \leq \textbf{f}_k^{(i)} \leq w$, for some maximum lag-size, $w$, and satisfy $s_k + \textbf{d}_k^{(i)} < e_k - \textbf{f}_k^{(i)}$. The remaining notation is as before.

We can extend our penalised likelihood/penalised cost approach to this setting. We begin by extending the test statistic defined in Section \ref{sec:AMOC} and the inference procedure in Section \ref{sec:Multiple} to allow for lags of up  to $w$, before discussing modifications to the penalties. We conclude this section by introducing ways of making the method computationally efficient, leading to a computational cost increasing only linearly in $w$. 

\subsection{ Extending the Test Statistic} 

The statistic introduced in Section \ref{sec:AMOC} can easily be extended to incorporate lags. The only modification this requires is to re-define the saving $\mathcal{S}_{i}\left(s,e\right)$ to be \small
\begin{equation*}
\max_{\substack{0\leq \textbf{d}^{(i)},\textbf{f}^{(i)} \leq w \\ e-s -\textbf{d}^{(i)} - \textbf{f}^{(i)} \geq l }}\left[\mathcal{C}_i \left( \textbf{x}_{(s+1+\textbf{d}^{(i)}):(e-\textbf{f}^{(i)})}^{(i)} , \bm{\theta}_0 ^{(i)}\right) - \min_{\bm{\theta}}\left(\mathcal{C}_i \left( \textbf{x}_{(s+1+\textbf{d}^{(i)}):(e-\textbf{f}^{(i)})}^{(i)} , \bm{\theta} \right)\right) - \gamma \left(\mathbb{I}\left(\textbf{d}^{(i)}\neq 0 \lor \textbf{f}^{(i)}\neq 0\right)\right)\right],
\end{equation*} \normalsize
where $w$ is the maximal allowed lag, and $\gamma$ is a penalty for adding a lag. We then infer $O$, $\hat{K}$, $\left(\hat{s}_1,\hat{e}_1, \hat{\textbf{d}}_1,\hat{\textbf{f}}_1,\hat{\bm{J}}_1\right)$,..., $\left(\hat{s}_{\hat{K}},\hat{e}_{\hat{K}},\hat{\textbf{d}}_{\hat{K}},\hat{\textbf{f}}_{\hat{K}}, \hat{\bm{J}}_{\hat{K}}\right)$ by directly maximising the penalised saving 

\begin{align}\label{eq:quantitytobemaxedwithanomalies_with_lags}
\sum_{k=1}^{\hat{K} }\mathcal{S}\left( \hat{s}_k,\hat{e}_k\right) + \sum_{t \in O} \mathcal{S'}\left( \textbf{x}_t\right) ,
\end{align} 
with respect to $\hat{K}$, $\left(\hat{s}_1,\hat{e}_1, \hat{\textbf{d}}_1,\hat{\textbf{f}}_1,\hat{\bm{J}}_1\right)$,..., $\left(\hat{s}_{\hat{K}},\hat{e}_{\hat{K}},\hat{\textbf{d}}_K,\hat{\textbf{f}}_K, \hat{\bm{J}}_{\hat{K}}\right)$, and the set of point anomalies $O$, subject to $0 \leq \hat{\textbf{d}}_k,\hat{\textbf{f}}_k  \leq w$, $(\hat{e}_k - \hat{\textbf{f}}_k) - (\hat{s}_k + \hat{\textbf{d}}_k) \geq l $ and $\hat{e}_{k}<\hat{s}_{k+1}$. 

\subsection{Modifying the Penalties}

As discussed in Section \ref{sec:Penalty}, the main purpose of the penalties is to account for multiple testing. Introducing lags means searching over more possible start and end points, i.e.\ testing more hypotheses. Consequently, increased penalties are required. The following modified version of penalty regime 2 can be used to account for lags:
\\
\\
\textbf{Penalty Regime 2':} $\alpha = 2(1+\epsilon)\psi$ and $\bm{\beta}_m = 2(1+\epsilon)\log(p)+2(1+\epsilon)\log(w+1)$, for $1 \leq m \leq p$.
\\
\\
Here $\epsilon >0 $ is a (small) constant. The following proposition shows that penalty regime 2' controls false positives at the desired level.
\begin{Proposition}\label{prop:FPcontrol_mean_penalty_2'}
	Let $\textbf{x}$ follow $\mathcal{M}_0$ and let $\hat{K}$ denote the number of collective anomalies inferred by MVCAPA under regime 2'. Then, there exists a constant $A_2'$ such that \begin{equation*}
	\mathbb{P}\{\hat{K} = 0\} \geq 1 - n^2\exp\left(A_2'\left( \frac{1+\epsilon}{\epsilon} \right)^3\right)e^{-\psi}
	\end{equation*}
\end{Proposition}
An alternative to this penalty regime consists of using penalty regime 2, but setting the penalty $\gamma = 2(1+\epsilon)\log(w+1)$. 

Unlike penalty regime 2, which is based on a tail bound, regimes 1 and 3 are based on Bernstein-type mean bounds. However, the MGF of the maximum of correlated chi-squared distributions is not analytically tractable. Consequently we limited ourselves to extending regime 2.

\subsection{Computational Considerations} 
The dynamic programming approach described in Section  \ref{sec:Comp} can also be used to minimise the penalised negative saving in Equation (\ref{eq:quantitytobemaxedwithanomalies_with_lags}). Solving the dynamic programme requires the computation of $\mathcal{S}_{i}\left(t,m\right)$ 
for $1 \leq i \leq p$ for all permissible $t$ at each step of the dynamic programme. Computing these savings \textit{ex nihilo} every time leads to the computational cost of the dynamic programme to scale quadratically in $(w+1)$. 

However, it is possible to reduce the computational cost of including lags by storing the savings 
\begin{equation*}
\mathcal{C}_i\left(\textbf{x}_{(a+1):b}^{(i)},\bm{\theta}_0^{(i)}\right) -
\min_{\theta} \left[\mathcal{C}_i\left(\textbf{x}_{(a+1):b}^{(i)},\bm{\theta}\right)\right]
\end{equation*}
for $t - w \leq b \leq t$ and $0 \leq a \leq b - l$. These can then be updated in each step of the dynamic programme at a cost of at most $O(np)$. From these, it is possible to calculate all $\mathcal{S}_{i}\left(t,m\right)$ required for a step of the dynamic programme in just $O(np(w+1))$ comparisons. This reduces the computational cost of each step of the dynamic programme to $O(pn(w+1)+pn\log(p))$ or $O(pn(w+1))$, depending on the type of penalty used. Crucially, only the comparatively cheap operations of allocating memory and finding the maximum of two numbers now increase with $w+1$ and even this relationship is only linear.

\subsection{Pruning the Dynamic Programme} 
Even when lags are included in the model, the solution space of the dynamic programme can still be pruned in a fashion similar to \cite{killick2012optimal} and \cite{CAPApaper}. Indeed, the following generalisation of Proposition \ref{prop:Pruning_easy} holds:
\begin{Proposition}\label{prop:Pruning}
	Let the costs $\mathcal{C}_i(,)$ be such that 
	\begin{equation*}
	\min_{\bm{\theta}} \left( \sum_{t=a+1}^{c} \mathcal{C}_i \left( \textbf{x}_t,\bm{\theta}\right) \right) \geq \min_{\bm{\theta}} \left( \sum_{t=a+1}^{b} \mathcal{C}_i \left( \textbf{x}_t,\bm{\theta}\right) \right) + \min_{\bm{\theta}} \left( \sum_{t=b+1}^{c} \mathcal{C}_i \left( \textbf{x}_t,\bm{\theta}\right) \right)
	\end{equation*}
	holds for all $\textbf{x}$ and $a,b,c$ such that $b-a \geq l$ and $c-b \geq l$. Then, if for some $t$ there exists an $m \geq t - l - w$ such that 
	\begin{equation*}
	C(m) - \alpha - \sum_{1}^{p} \bm{\beta}_i > C(t) + {\mathcal{S}}\left( t,m\right) 
	\end{equation*}
	holds, 
	\begin{equation*}
	C(m') > C(t) + {\mathcal{S}}\left( t,m'\right)   
	\end{equation*}
	must also holds for all $m' \geq m + l +w$.
\end{Proposition}
\section{Simulation Study}\label{sec:Simulation}

We now compare the performance of MVCAPA to that of other popular methods. In particular, we compare ROC curves, precision, as well as the runtime with PASS \citep{jeng2012simultaneous} and Inspect \citep{wang2018high,wang2016inspectchangepoint}.  PASS  \citep{jeng2012simultaneous} uses higher criticism in conjunction with circular binary segmentation \citep{olshen2004circular} to detect subset multivariate epidemic changepoints. Code is available from the author's website. Inspect \citep{wang2018high} uses projections to find sparse classical changepoints. 

The comparison was carried out on simulated multivariate time series with $n=5000$ observations for $p$ components with i.i.d.\ $N(0,1)$ noise, for a range of values of $p$. To these, collective anomalies affecting $k$ components occurring at a geometric rate of 0.001 (leading to an average of about 5 collective anomalies per series) were added. The lengths of these collective anomalies are i.i.d.\ Poisson-distributed with mean 20. Within a collective anomaly, the start and end lags of each component are drawn uniformly from the set $\{0,...,w\}$, subject to their sum being less than the length of the collective anomaly. Note that $w=0$ implies the absence of lags. The means of the components during the collective anomaly are drawn from an $N(0,\sigma^2)$-distribution. In particular we considered the following cases, emulating different detectable regimes introduced in Section \ref{sec:FiniteSamples}. 
\begin{enumerate}
	\item The most sparse regime possible: a single component affected by a strong anomaly without lags, i.e.\ $\sigma = 2 \log(p)$, $w=0$, and $k=1$.
	\item The most dense regime possible: all components affected by weak anomalies without lags, i.e.\ $\sigma = p^{-1/4}$, $w=0$, and $k=p$. 
	\item A regime close to the boundary between sparse and dense changes, i.e.\ $k=2$ when $p=10$ and $k=6$ when $p=100$ with $\sigma = \log(p)$ and $w=0$.
	\item A regime close to the boundary between sparse and dense changes, but with lagged collective anomalies, i.e.\ the same as 3 but with $w=10$.
\end{enumerate}
This analysis was repeated with 5 point anomalies distributed $N(0,8\log(p))$. The $\log(p)$-scaling of the variance ensures that the point anomalies are anomalous even after correcting for multiple testing over the $p$ different components. 

\subsection{ROC Curves}

\begin{figure} 
	\begin{subfigure}[b]{0.5\linewidth}
		\centering
		\includegraphics[width=\linewidth]{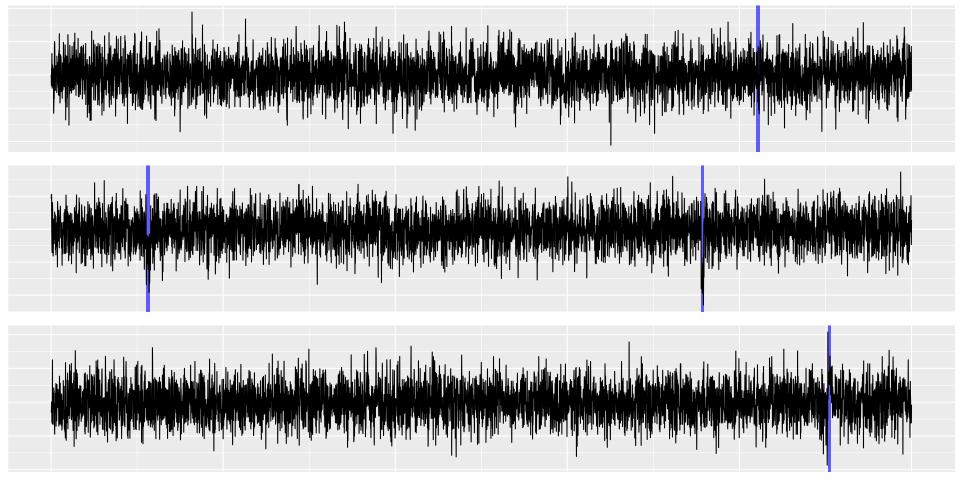} 
		\caption{Example}
		\label{fig:1_component_example} 
		\vspace{4ex}
	\end{subfigure} 
	\begin{subfigure}[b]{0.5\linewidth}
		\centering
		\includegraphics[width=\linewidth]{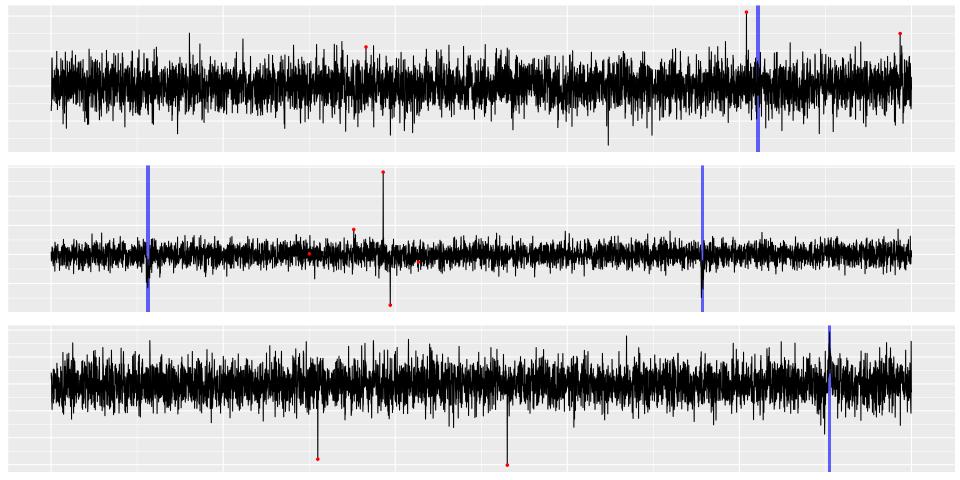} 
		\caption{Example with pt. anomalies}
		\label{fig:1_component_ANOM_example} 
		\vspace{4ex}
	\end{subfigure}
	\vspace{-20pt}
	\begin{subfigure}[b]{0.5\linewidth}
		\centering
		\includegraphics[width=0.9\linewidth]{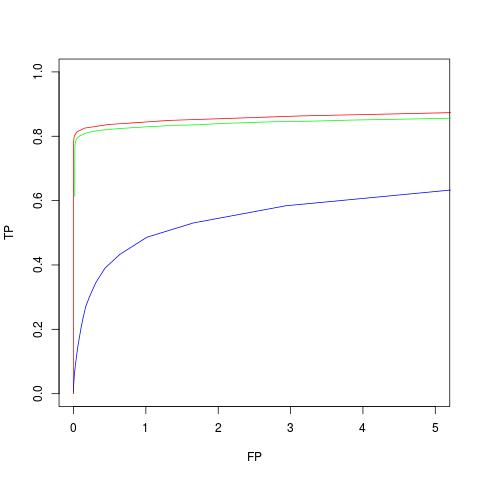} 
		\caption{p=10} 
		\label{fig:1_component_small} 
		\vspace{4ex}
	\end{subfigure} 
	\begin{subfigure}[b]{0.5\linewidth}
		\centering
		\includegraphics[width=0.9\linewidth]{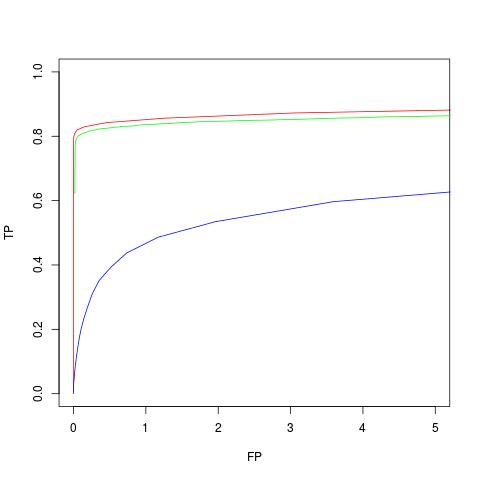} 
		\caption{p=10, with pt. anomalies} 
		\label{fig:1_component_ANOM_small} 
		\vspace{4ex}
	\end{subfigure} 
	\vspace{-20pt}
	\begin{subfigure}[b]{0.5\linewidth}
		\centering
		\includegraphics[width=0.9\linewidth]{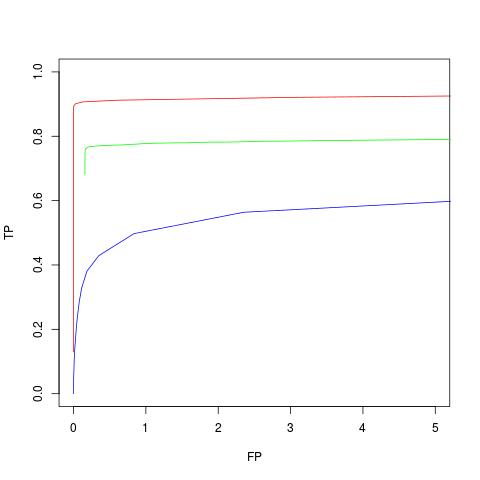} 
		\caption{p=100} 
		\label{fig:1_component_large} 
		\vspace{4ex}
	\end{subfigure}
	\begin{subfigure}[b]{0.5\linewidth}
		\centering
		\includegraphics[width=0.9\linewidth]{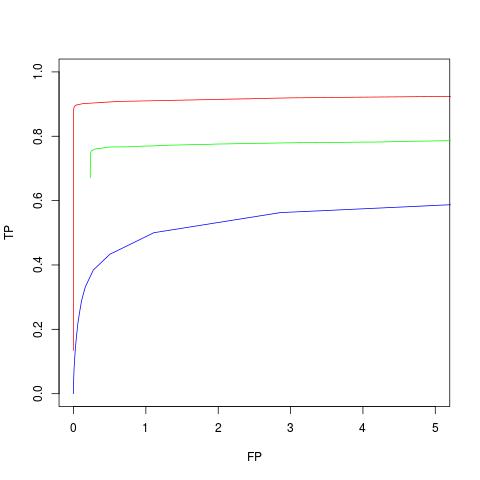} 
		\caption{p=100, with pt. anomalies} 
		\label{fig:1_component_ANOM_larg} 
		\vspace{4ex}
	\end{subfigure}
	\caption{Example series and ROC curves for setting 1. MVCAPA is in red, PASS in green, and Inspect in blue.}
	\label{fig:1_component} 
\end{figure}

\begin{figure} 
	\begin{subfigure}[b]{0.5\linewidth}
		\centering
		\includegraphics[width=\linewidth]{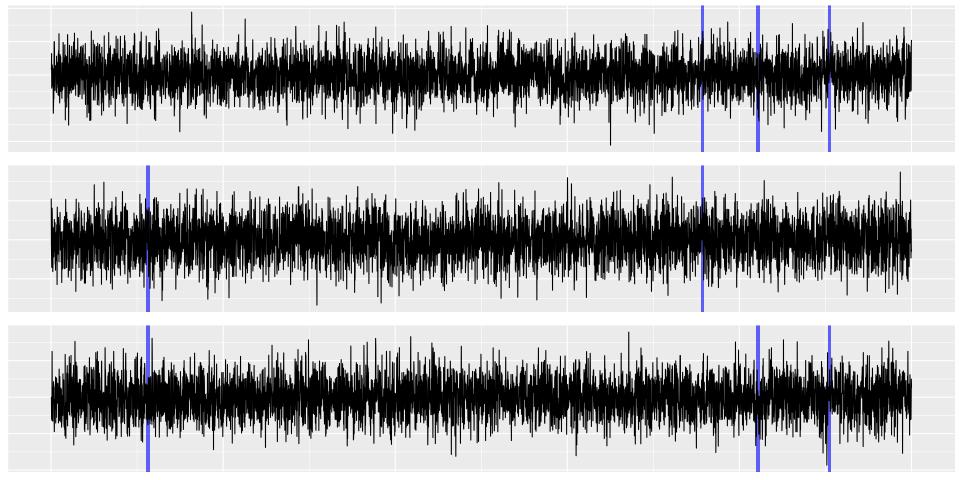} 
		\caption{Example}
		\label{fig:medium_component_example} 
		\vspace{4ex}
	\end{subfigure} 
	\begin{subfigure}[b]{0.5\linewidth}
		\centering
		\includegraphics[width=\linewidth]{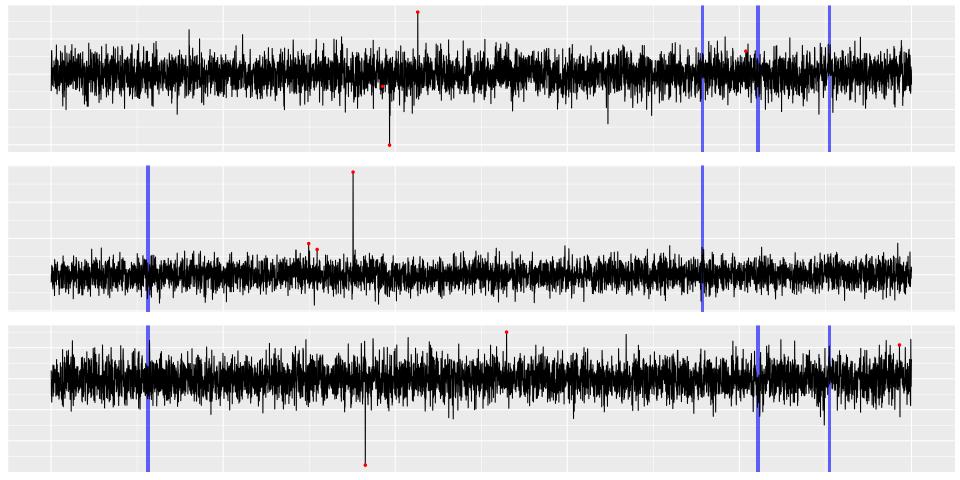} 
		\caption{Example, with pt. anomalies}
		\label{fig:medium_component_ANOM_example} 
		\vspace{4ex}
	\end{subfigure}
	\vspace{-20pt}
	\begin{subfigure}[b]{0.5\linewidth}
		\centering
		\includegraphics[width=0.9\linewidth]{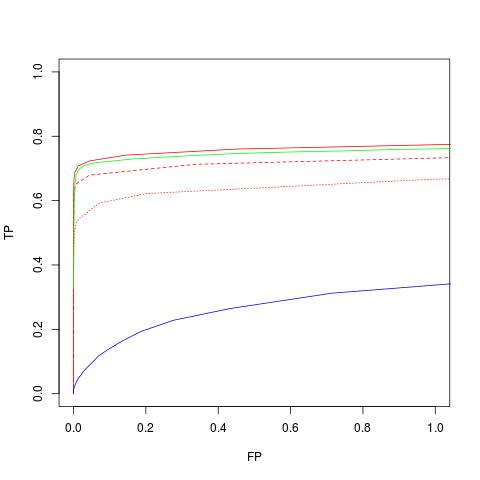} 
		\caption{p=10} 
		\label{fig:medium_component_small} 
		\vspace{4ex}
	\end{subfigure} 
	\begin{subfigure}[b]{0.5\linewidth}
		\centering
		\includegraphics[width=0.9\linewidth]{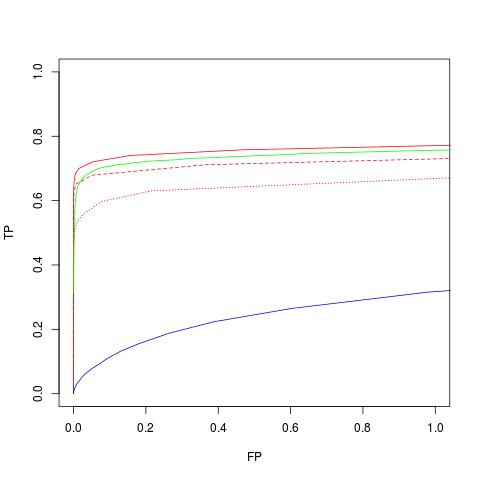} 
		\caption{p=10, with pt. anomalies} 
		\label{fig:medium_component_ANOM_small} 
		\vspace{4ex}
	\end{subfigure} 
	\vspace{-20pt}
	\begin{subfigure}[b]{0.5\linewidth}
		\centering
		\includegraphics[width=0.9\linewidth]{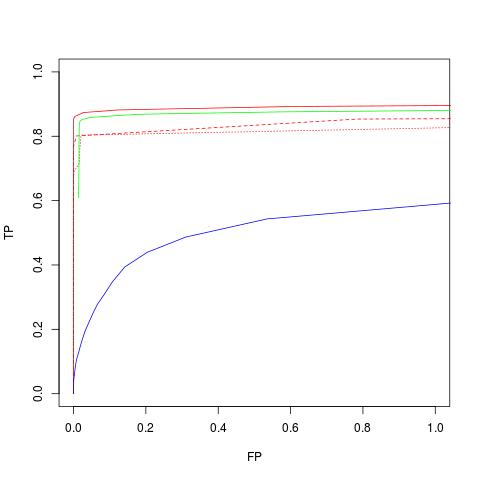} 
		\caption{p=100} 
		\label{fig:medium_component_large} 
		\vspace{4ex}
	\end{subfigure}
	\begin{subfigure}[b]{0.5\linewidth}
		\centering
		\includegraphics[width=0.9\linewidth]{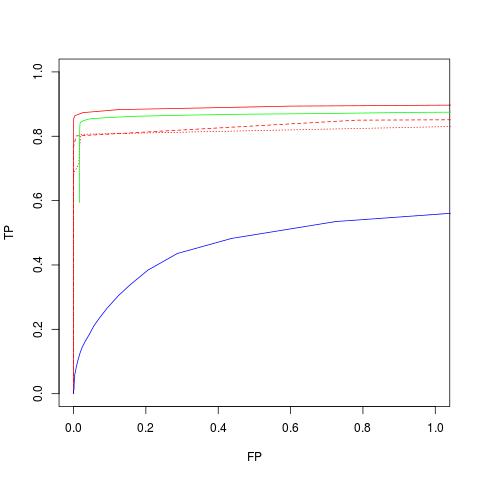} 
		\caption{p=100, with pt. anomalies} 
		\label{fig:medium_component_ANOM_larg} 
		\vspace{4ex}
	\end{subfigure}
	\caption{Example series and ROC curves for setting 3. MVCAPA is in red, PASS in green, and Inspect in blue. The solid red line corresponds to $w=0$, the dashed one to $w=10$ and the dotted one to $w=20$.}
	\label{fig:medium_component} 
\end{figure}

\begin{figure} 
	\begin{subfigure}[b]{0.5\linewidth}
		\centering
		\includegraphics[width=\linewidth]{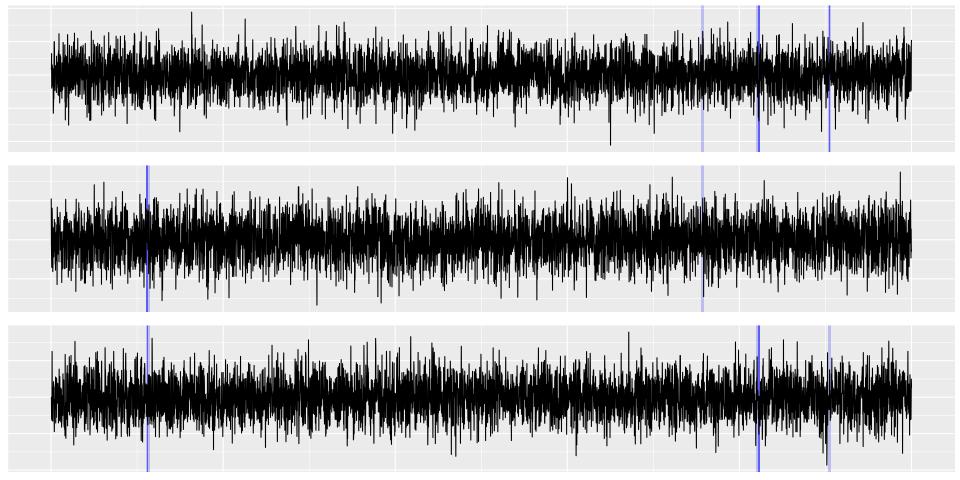} 
		\caption{Example}
		\label{fig:medium_component_LAG_example} 
		\vspace{4ex}
	\end{subfigure} 
	\begin{subfigure}[b]{0.5\linewidth}
		\centering
		\includegraphics[width=\linewidth]{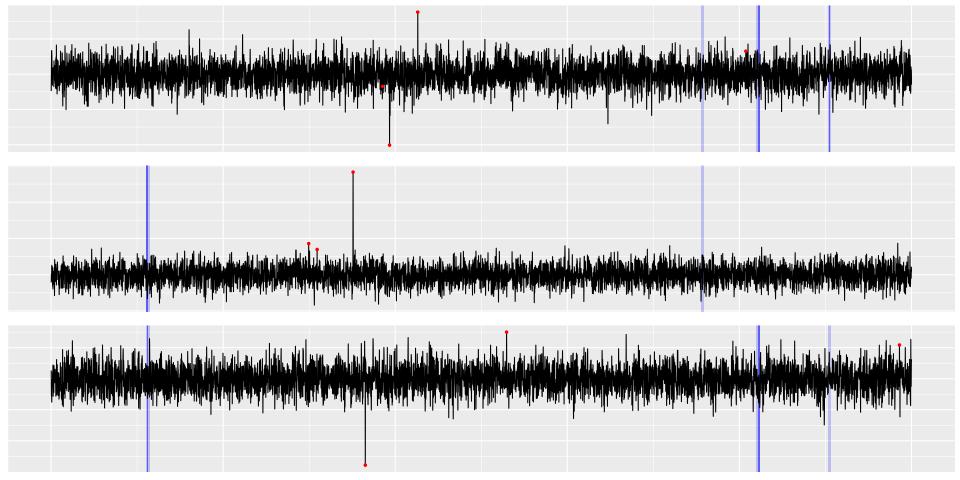} 
		\caption{Example, with pt. anomalies}
		\label{fig:medium_component_LAG_ANOM_example} 
		\vspace{4ex}
	\end{subfigure}
	\vspace{-20pt}
	\begin{subfigure}[b]{0.5\linewidth}
		\centering
		\includegraphics[width=0.9\linewidth]{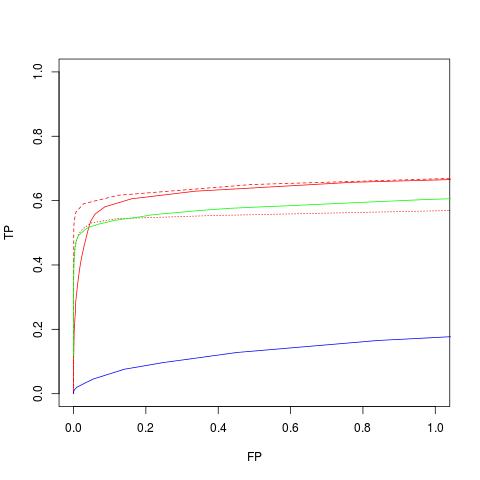} 
		\caption{p=10} 
		\label{fig:medium_component_LAG_small} 
		\vspace{4ex}
	\end{subfigure} 
	\begin{subfigure}[b]{0.5\linewidth}
		\centering
		\includegraphics[width=0.9\linewidth]{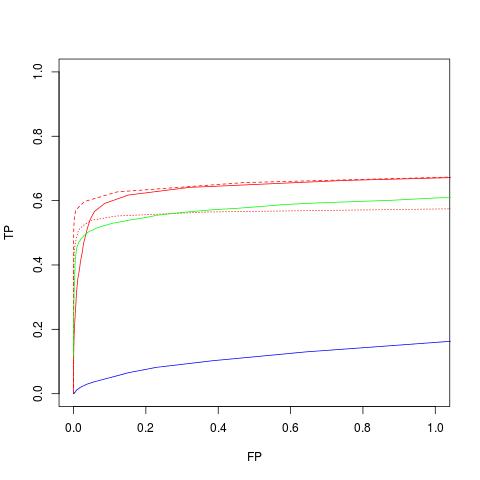} 
		\caption{p=10, with pt. anomalies} 
		\label{fig:medium_component_LAG_ANOM_small} 
		\vspace{4ex}
	\end{subfigure} 
	\vspace{-20pt}
	\begin{subfigure}[b]{0.5\linewidth}
		\centering
		\includegraphics[width=0.9\linewidth]{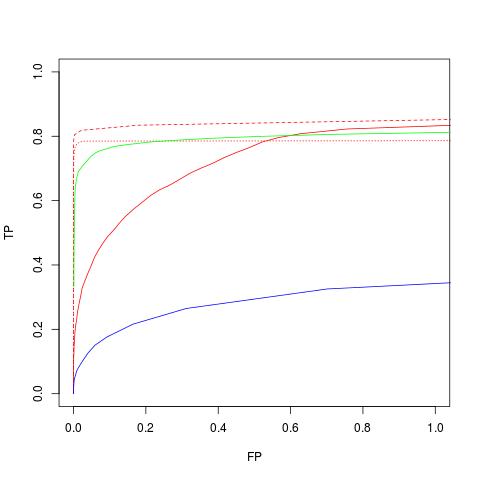} 
		\caption{p=100} 
		\label{fig:medium_component_LAG_large} 
		\vspace{4ex}
	\end{subfigure}
	\begin{subfigure}[b]{0.5\linewidth}
		\centering
		\includegraphics[width=0.9\linewidth]{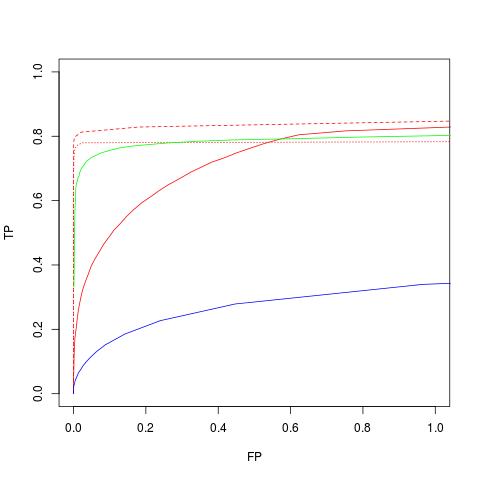} 
		\caption{p=100, with pt. anomalies} 
		\label{fig:medium_component_LAG_ANOM_larg} 
		\vspace{4ex}
	\end{subfigure}
	\caption{Example series and ROC curves for setting 4. MVCAPA is in red, PASS in green, and Inspect in blue. The solid red line corresponds to $w=0$, the dashed one to $w=10$ and the dotted one to $w=20$.}
	\label{fig:medium_component_LAG} 
\end{figure}

We obtained ROC curves by varying the threshold parameters of Inspect and PASS and by rescaling $\alpha,\beta',\beta_1,...,\beta_p$ for MVCAPA. The curves were obtained over 1000 simulated datasets. For MVCAPA, we typically set $w=0$, but also tried $w=10$ and $w=20$ for the third and fourth setting. We used and rescaled the composite penalty regime (Section \ref{sec:Penalty}) for $w=0$ and penalty regime 2' for $w>0$. We also set the maximum segment lengths for both MVCAPA and PASS to 100 and the minimum segment length of MVCAPA to 2. The $\alpha_0$ parameter of PASS, which excludes the $\alpha_0-1$ lowest $p$-values from the higher criticism statistic to obtain a better finite sample performance (see \cite{jeng2012simultaneous}) was set to $k$ or 5, whichever was the smallest. For MVCAPA and PASS, we considered a detected segment to be a true positive if its start and end point both lie within 20 observations of that of a true collective anomalies' start and end point respectively. For Inspect, we considered a detected change to be a true positive if it was within 20 observation of a true start or end point. When point anomalies were added to the data, we considered segments of length one returned by PASS to be point anomalies to make the comparison with MVCAPA fairer. 

The results  for three of settings considered can be found in Figures \ref{fig:1_component} to  \ref{fig:medium_component_LAG}. The qualitatively similar results for the second setting can be found in the supplementary material. We can see that Inspect usually does worst, especially when changes become dense, which is no surprise given the method was introduced to detect sparse changes. We additionally see that MVCAPA generally outperforms PASS. This advantage is particularly pronounced in the case in which exactly one component changes. This is a setting which PASS has difficulties dealing with due to the convergence properties of the higher criticism statistic at the lower tail \citep{jeng2012simultaneous}. PASS outperformed MVCAPA in the second setting for $p=10$, when it was assisted by a large value of $\alpha_0$, which considerably reduced the number candidate collective anomalies it had to consider. 

Figures \ref{fig:medium_component} and \ref{fig:medium_component_LAG}, show that MVCAPA performs best when the correct maximal lag is specified. They also demonstrate that specifying a lag and therefore overestimating the lag when no lag is present adversely affects performance of MVCAPA. However, when lags are present, over-estimating the maximal lag appears preferable to underestimating it. Finally, the comparison between Figures \ref{fig:medium_component_LAG_small}  and \ref{fig:medium_component_LAG_ANOM_small} shows that the performance of MVCAPA is hardly affected by the presence of point anomalies, unlike that of Inspect and, to a lesser extent, PASS, whose performance is adversely affected. 

\begin{figure} 
	\begin{center}
		\begin{tabular}{||c c c c ||c | c  | c  | c  |  c ||} 
			\hline
			\footnotesize
			\footnotesize Setting & \footnotesize p & \footnotesize Max.\ Lag & \footnotesize Pt. Anoms. & \footnotesize MVCAPA & \footnotesize MVCAPA, w=10 & \footnotesize MVCAPA, w=20 & \footnotesize Inspect & \footnotesize PASS   \\ [0.5ex] 
			\hline\hline
			\normalsize
			1   &   10      &  0   &  -       &  \textbf{0.09}& - & - & 0.64 & 0.31  \\ 
			1   &   100      &  0   & -       &  \textbf{0.02}& - & - & 0.40 & 0.62  \\ 
			1   &   10      &  0   &  \tick   &  \textbf{0.09}& - & - & 0.62 & 0.38  \\ 
			1   &   100      &  0   &  \tick  &  \textbf{0.03}& - & - & 0.40 & 0.67  \\ 
			2   &   10      &  0   &  -       &  \textbf{0.09} & - & - & 0.74 & 0.52  \\ 
			2   &   100      &  0   &  -      &  \textbf{0.01} & - & - & 0.71 & 0.54  \\ 
			2   &   10      &  0   &  \tick   &  \textbf{0.05} & - & - & 0.69 & 0.46  \\ 
			2   &   100      &  0   &  \tick  &  \textbf{0.01} & - & - & 0.67 & 0.51  \\
			3   &   10      &  0   &  -       &  \textbf{0.11} & 2.31 & 3.30 & 0.72 & 0.27  \\ 
			3   &   100      &  0   &  -      &  \textbf{0.01} & 3.43 & 3.83 & 0.53 & 0.29  \\ 
			3   &   10      &  0   &  \tick   &  \textbf{0.09} & 2.23 & 3.26 & 0.69 & 0.22  \\ 
			3   &   100      &  0   &  \tick  &  \textbf{0.01} & 3.35 & 3.82 & 0.53 & 0.23  \\ 
			4   &   10      &  10   &  -      &  0.63 & \textbf{0.46} & 1.09 & 0.80 & 2.53  \\ 
			4   &   100      &  10   &  -     &  1.27 & \textbf{0.18} & 1.57 & 0.61 & 3.64  \\ 
			4   &   10      &  10   &  \tick  &  0.72 & \textbf{0.51} & 1.22 & 0.83 & 2.60  \\ 
			4   &   100      &  10   &  \tick &  1.23 & \textbf{0.21} & 1.58 & 0.59 & 3.77  \\ 
			\hline
		\end{tabular}
		\caption{Precision of true positives detected by all methods measured in mean absolute distance for MVCAPA, PASS, and Inspect.} 
		\label{tble:Precision}
	\end{center}
\end{figure}

\subsection{Precision}

We compared the precision of the three methods by measuring the accuracy (in mean absolute distance) of true positives. Only true positives detected by all methods were taken into account to avoid selection bias. We used the default parameters for MVCAPA and PASS, whilst we set the threshold for Inspect to a value leading to comparable number of true and false positives. To ensure a suitable number of true positives for Inspect we doubled sigma in the second scenario. The results of this analysis can be found in Figure \ref{tble:Precision} and show that MVCAPA is usually the most precise approach, exhibiting a significant gain in accuracy against PASS. When comparing the influence of the user-specified maximal lag, we note that specifying he correct maximal lag gives the best performance. We also note that over- and under- estimating the lag have a similar effects on the accuracy.

\begin{figure} 
	\begin{subfigure}[b]{0.325\linewidth}
		\centering
		\includegraphics[width=0.9\linewidth]{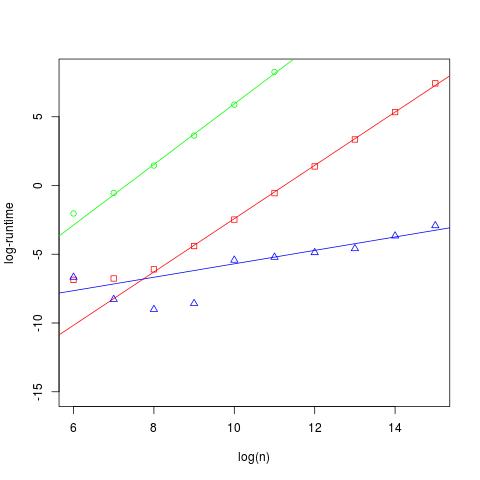} 
		\caption{No anomalies}
		\label{fig:No anomalies} 
		\vspace{4ex}
	\end{subfigure} 
	\begin{subfigure}[b]{0.325\linewidth}
		\centering
		\includegraphics[width=0.9\linewidth]{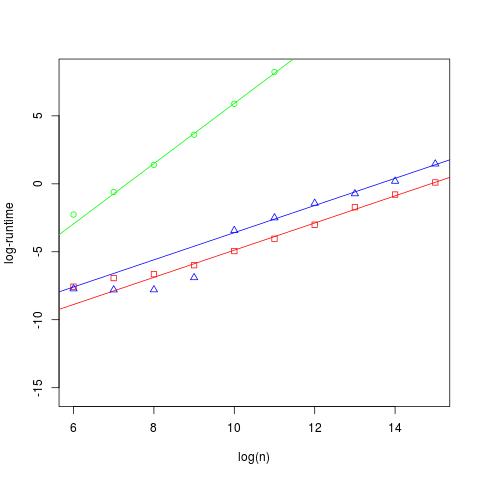} 
		\caption{Regular anomalies} 
		\label{fig:Regular anomalies} 
		\vspace{4ex}
	\end{subfigure} 
	\begin{subfigure}[b]{0.325\linewidth}
		\centering
		\includegraphics[width=0.9\linewidth]{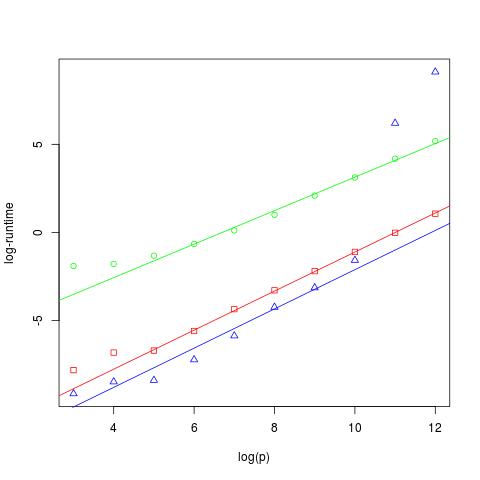} 
		\caption{Large p} 
		\label{fig:large p} 
		\vspace{4ex}
	\end{subfigure}
	\caption{A comparison of run-times for MVCAPA (red squared), PASS (green circles), and Inspect (blue triangles). Robust lines of best fit have been added. All logarithms are in base 2.}
	\label{Fig:Run-times} 
\end{figure}

\subsection{Runtime}

We compared the scaling of the runtime of MVCAPA, PASS, and Inspect in both the number of observations $n$, as well as the number of components $p$. To evaluate the scaling in $n$ we set $p=10$ and varied $n$ on data without any anomalies. We repeated this analysis with collective anomalies appearing (on average) every 100 observations. The results of these two analyses can be found in Figures \ref{fig:No anomalies} and  \ref{fig:Regular anomalies} respectively. We note that the slope of MVCAPA is very close to 2, in the anomaly-free setting and very close to 1 in the setting in which the number of anomalies increases linearly with the number of observations, suggesting quadratic and linear behaviour respectively, whilst the slopes of PASS and Inspect are close to 2 and 1 respectively in both cases.  

Turning to the scaling of the three methods in $p$, we set $n$ = 100 and varied $p$. The results of this analysis can be found in Figure \ref{fig:large p}. We note that the slopes of all methods are close to 1 suggesting linear behaviour. However, Inspect becomes very slow once $p$ exceeds a certain threshold.

\section{Application}\label{sec:Application}

We now apply MVCAPA to extract copy number variations (CNVs) from genetics data. The data consists of a log-likelihood ratio statistic evaluated along the genome which measures the likelihood of a CNV. A multivariate approach to detecting CNVs is attractive because they are often shared across individuals. By borrowing signal across individuals we should gain power for detecting CNVs which have a weak signal. However, as we will become apparent from our results, shared variations do not always align perfectly. 

In this section we re-use the design of \cite{bardwell2017bayesian} to compare MVCAPA with PASS. The only difference is that we set the maximum segment length for MVCAPA and PASS to 100, whilst \cite{bardwell2017bayesian} used 200. To investigate the potential benefit of allowing for lags, we repeated the experiment for MVCAPA both with $w=0$ (i.e. not allowing for lags) and $w=40$. Since $n>>p$ in this application, we used the sparse penalty setting for MVCAPA. 

The exact ground truth is unknown. Indeed, it is beyond the scope of this paper to differentiate between a false positive and a currently unknown CNV. We can nevertheless compare different methods by how accurately they detect known copy number variations for a given size. Like \cite{bardwell2017bayesian}, we used known CNVs from the HapMap project \citep{international2003international} as true positives and tuned the penalties and thresholds in such a way that 4\% of the genome was flagged up as anomalous. For MVCAPA this involved scaling the penalties $\alpha,\beta_1,...,\beta_p$ by a constant, as discussed in the final paragraph of Section \ref{sec:Penalty}.

The results of this analysis can be found in Figure \ref{tble:Accuracies_6}. These tables show that MVCAPA compares favourably with PASS. We can also see that allowing for lags generally led to a better performance of MVCAPA, thus suggesting non-perfect alignment of CNVs across individuals. Moreover, MVCAPA was very fast taking 5 seconds to analyse the longer genome on a standard laptop when we did not allow lags, and 10 seconds when we allowed for lags. The R implementation of PASS, on the other hand, took 17 minutes.

\begin{figure} 
	\begin{subfigure}[b]{\linewidth}
		\centering
		\begin{tabular}{| c |c c c|c c c|c c c|} 
	\hline
	\multicolumn{1}{|c|}{Truth} &  \multicolumn{3}{|c|}{PASS} & \multicolumn{3}{|c|}{MVCAPA ($w=40$)} & \multicolumn{3}{|c|}{MVCAPA ($w=0$)} \\ [0.5ex] 
	\hline 
	\footnotesize{Start}  &\footnotesize Rep 1 &\footnotesize Rep 2 &\footnotesize Rep 3 &\footnotesize Rep 1 &\footnotesize Rep 2 &\footnotesize Rep 3  &\footnotesize Rep 1 &\footnotesize Rep 2 &\footnotesize Rep 3  \\
	\hline
	\footnotesize{2619669}   & & \tick & &  &  &   &  &  &   \\
	\hline
	\footnotesize{2638575} &  & \tick &    &  &  &   &  &  &   \\
	\hline
	\footnotesize{21422575}  & \tick & \tick & \tick & \tick &\tick &\tick &\tick &\tick &\tick \\
	\hline
	\footnotesize{32165010} &\tick & \tick & \tick &\tick &\tick &\tick&\footnotesize \tick &\tick &\tick  \\
	\hline
	\footnotesize{34328205}&\tick & \tick  &  &\tick &\tick &\tick&\footnotesize \tick &\tick &\tick  \\
	\hline
	\footnotesize{54351338} &\tick & \tick & \tick &\tick &\tick &\tick& & & \\
	\hline
	\footnotesize{70644511}  & & & &\tick &\tick &\tick& \tick &\tick &\tick  \\ \hline
	\multicolumn{1}{c}{} & \multicolumn{9}{c}{}
	\normalsize
		\vspace{-10pt}
\end{tabular}
	\caption{Chromosome 16.}

\label{tble:Accuracies_16}
	\end{subfigure} 
	\begin{subfigure}[b]{\linewidth}
		\centering
			\vspace{15pt}
				\begin{tabular}{| c |c c c|c c c|c c c|} 
			\hline
			\multicolumn{1}{|c|}{Truth} & \multicolumn{3}{|c|}{PASS} & \multicolumn{3}{|c|}{MVCAPA ($w=40$)} & \multicolumn{3}{|c|}{MVCAPA ($w=0$)} \\ [0.5ex] 
			\hline 
			\footnotesize{Start}  &\footnotesize Rep 1 &\footnotesize Rep 2 &\footnotesize Rep 3 &\footnotesize Rep 1 &\footnotesize Rep 2 &\footnotesize Rep 3  &\footnotesize Rep 1 &\footnotesize Rep 2 &\footnotesize Rep 3  \\
			\hline
			\footnotesize{202314} &\tick & \tick & \tick& \tick & \tick & \tick  & \tick & \tick & \tick \\
			\hline
			\footnotesize{243582} &\tick & \tick & \tick & \tick & \tick & \tick  & \tick & \tick & \tick \\
			\hline
			\footnotesize{29945146} & \tick & & & \tick & \tick &   & \tick & \tick &   \\
			\hline
			\footnotesize{30569918} & &  &   &       & \tick &   &  &  &   \\
			\hline
			\footnotesize{31388628} & & &  & \tick & \tick &   & \tick & \tick &   \\
			\hline
			\footnotesize{31388628} & & &  & \tick & \tick &   &  & \tick &   \\
			\hline
			\footnotesize{32562531} & & & &  & \tick & \tick  &  & \tick &  \tick \\
			\hline
			\footnotesize{32605305} & &\tick & & \tick & \tick & \tick  & \tick & \tick &  \tick \\
			\hline
			\footnotesize{32717397} &\footnotesize \tick & &  & \tick & \tick & \tick  & \tick & \tick &  \tick \\
			\hline
			\footnotesize{74648424} & & &  & \tick & \tick &   & \tick & \tick &   \\
			\hline
			\footnotesize{77073620} & & & &  &  &   &  &  &   \\
			\hline
			\footnotesize{77155147} & & &  & \tick & \tick &   & \tick & \tick &   \\
			\hline
			\footnotesize{77496587}& \tick & & &  &  &   &  &  &   \\
			\hline
			\footnotesize{78936685} & &\tick &\tick & \tick & \tick & \tick  & \tick & \tick & \tick  \\
			\hline
			\footnotesize{103844990} &\tick & \tick & \tick & \tick  & \tick & \tick & \tick & \tick &  \tick \\
			\hline
			\footnotesize{126226035} & & &\tick  & \tick &  &   & \tick &  & \tick  \\
			\hline
			\footnotesize{139645437} & & & &  &  &   &  &  &  \\
			\hline
			\footnotesize{165647651} &\tick &  & \tick &  &  &   &  &  & \tick  \\
			\hline
			\multicolumn{1}{c}{} & \multicolumn{9}{c}{}
			\normalsize
		\end{tabular}
	\vspace{-10pt}
	\caption{Chromosome 6.}
	\end{subfigure}
		\caption{A comparison between PASS, MVCAPA without lags, and MVCAPA with a lag of up to 40 for two chromosomes. Successful detections are indicated by ticks. Note that chromosome 6 contains two different CNVs (of different lengths) begin at 31388628.} 
\label{tble:Accuracies_6}
\end{figure}

\clearpage

\section{Acknowledgements}

This work was supported by EPSRC grant numbers EP/N031938/1 (StatScale) and EP/L015692/1 (STOR-i). The authors also acknowledge British Telecommunications plc (BT) for financial support, David Yearling and Kjeld Jensen in BT Research \& Innovation for discussions. Thanks also to Lawrence Bardwell for providing us with data and code reproducing results from his paper.

\bibliographystyle{plain}
\bibliography{MVCAPA}

\newpage

\section{Supplementary Material}

\subsection{Proofs for Theorems and Propositions}	

\subsubsection{Proof of Proposition \ref{prop:FPcontrol_mean_penalty_1}}

Let $1\leq i\leq j \leq n$. The probability that the segment $(i,j)$ is not flagged up as anomalous is given by 
\begin{align*}
&\mathbb{P}\left( \sum_{ c \in S_m} (j-i+1)\left(\bar{\bm{\eta}}_{i:j}\right)^2 <  p + 2\psi + 2\sqrt{p\psi}, \, \; \; \, \forall S_m \subset \{1,...,p \}: |S_m| = m   , \;\;   1 \leq m \leq p\right) \\
& = \mathbb{P}\left( \sum_{ c =1}^{p} (j-i+1)\left(\bar{\bm{\eta}}_{i:j}\right)^2 <  p + 2\psi + 2\sqrt{p\psi}\right)  = \mathbb{P}\left( \chi^2_p <  p + 2\psi + 2\sqrt{p\psi}\right) \geq 1- e^{-\psi},
\end{align*}
where inequality follows from the bounds on the chi-squared distribution in \cite{laurent2000adaptive}. A Bonferroni correction over all possible pairs $i,j$ then finishes the proof. 

\subsubsection{Proof of Proposition \ref{prop:FPcontrol_mean_penalty_2}}

Let $1\leq i\leq j \leq n$. For this pair $(i,j)$ define
$Y_c = (j-i+1)\left(\bar{\bm{\eta}}_{i:j}^{(c)}\right) ^2$,
noting that they are all independent and $\chi^2_1$-distributed. The probability that this segment $i,j$ will not be considered anomalous is
\begin{align*}
&\mathbb{P}\left( \sum_{ c \in S_m} Y_c  <  2(1+\epsilon)\psi + 2m\log(p), \, \; \; \, \forall S_m \subset \{1,...,p \}: |S_m| = m  , \;\;  1 \leq m \leq p\right) \\
& \geq \mathbb{P}\left( \sum_{i=1}^p \left(Y_i - 2\log(p) \right)^+ < 2(1+\epsilon)\psi \right) \geq 1 - \mathbb{E}\left( e^{\lambda\left(Y_1 - 2\log(p) \right)^+} \right)^pe^{-2\lambda(1+\epsilon) \psi},
\end{align*}
for all $\lambda > 0$, where the second inequality corresponds to a Chernoff bound. Next set $\lambda = \frac{1}{2} \frac{1}{1+\epsilon}$ and note that for $p \geq 2$
\small
\begin{align*}
&\mathbb{E}\left( e^{\lambda\left(Y_1 - 2\log(p) \right)^+} \right)^p e^{-\psi} = \left[ 1 + \int_{1}^{\infty} \mathbb{P} \left( Y_1 > 2 \log(p) + \frac{1}{\lambda} \log(x) \right) dx      \right]^p e^{-\psi} \\
&= \left[ 1 + \int_{1}^{\infty} \mathbb{P} \left( \chi_1^2 > 2 \log(p) + \frac{1}{\lambda} \log(x) \right) dx      \right]^p e^{-\psi}  
\leq \left[ 1 + \frac{1}{p}\sqrt{\frac{1}{\pi \log(p)}} \int_{1}^{\infty} x ^{-\frac{1}{2\lambda}} dx \right]^pe^{-\psi}  \\
&= \left[ 1 + \frac{1}{p}\sqrt{\frac{1}{\pi \log(p)}} \frac{1}{1-2\lambda} \right]^pe^{-\psi} 
\leq 
\exp \left( \sqrt{\frac{1}{\pi \log(p)}} \frac{1+\epsilon}{\epsilon}\right)e^{-\psi}
\leq 
\exp \left( \sqrt{\frac{1}{\pi \log(2)}} \frac{1+\epsilon}{\epsilon}\right)e^{-\psi}
:= Ae^{-\psi},
\end{align*}
\normalsize
where the first inequality exploits well known tail bounds of the normal distribution. Similarly, for $p=1$: 
\begin{align*}
&\mathbb{E}\left( e^{\lambda\left(Y_1 - 2\log(p) \right)^+} \right)^p e^{-\psi} = \mathbb{E}\left( e^{\lambda Y_1} \right)e^{-\psi} = \sqrt{\frac{1}{1-2\lambda}}e^{-\psi} = \sqrt{\frac{1+\epsilon}{\epsilon}}e^{-\psi}
\end{align*}
A Bonferroni correction over all possible pairs $i,j$ then finishes the proof. 

\subsubsection{Proof of Proposition \ref{prop:FPcontrol_mean_penalty_3}}

Let $1\leq i\leq j \leq n$. For this pair $(i,j)$ define
$Y_c = (j-i+1)\left(\bar{\bm{\eta}}_{i:j}^{(c)}\right) ^2$,
noting that they are all independent and $\chi^2_1$-distributed. Next, define their order statistic $Y_{(1)} \geq  ... \geq Y_{(p)}$ The probability that the segment $(i,j)$ is not flagged up as anomalous is given by \small
\begin{align*}
&\mathbb{P}\left( \sum_{ c=1}^m Y_{(c)} < 2(\psi +\log(p)) + m + 2pa_mf(a_m) + 2\sqrt{(m + 2pa_mf(a_m))(\psi +\log(p))}  , \;\;\;  1 \leq m \leq p\right) \\
& \geq 1 - \sum_{m=1}^p\mathbb{P}\left( \sum_{ c=1}^m \left(Y_{(c)} - a_m\right) > 2(\psi +\log(p)) + m(1-a_m) + 2pa_mf(a_m) + 2\sqrt{(m + 2pa_mf(a_m))(\psi +\log(p))}\right)\\
& \geq 1 -  \sum_{m=1}^p\mathbb{P}\left( \sum_{ c=1}^m \left(Y_{(c)} - a_m\right)^{+} > 2(\psi +\log(p)) + m(1-a_m) + 2pa_mf(a_m) + 2\sqrt{(m + 2pa_mf(a_m))(\psi +\log(p))}\right)\\
& \geq 1 - \sum_{m=1}^p\mathbb{P}\left( \sum_{c=1}^p \left(Y_{c} - a_m\right)^{+} < 2(\psi +\log(p)) + m(1-a_m) + 2pa_mf(a_m) + 2\sqrt{(m + 2pa_mf(a_m))(\psi +\log(p))}\right).
\end{align*} \normalsize
We will now use the following lemma, which shows that $\left(Y_{c} - a_m\right)^+$ is sub-gamma. 
\begin{Lemma}\label{lemma:Subgamma}
	Let $x \sim \chi_1^2$. Then $(x - a)^+ -\left[ 2af(a) + (1 - a)\mathbb{P}\left(\chi_1^2 > a \right) \right]$ is sub-gamma with scale parameter 2 and variance $V = 4af(a) + 2\mathbb{P}\left(\chi_1^2 > a \right)$. 
\end{Lemma}
Using Lemma \ref{lemma:Subgamma} and the bounds on sub-gamma random-variables in \cite{boucheron2012concentration}, we have that 
\scriptsize
\begin{align*}
&\sum_{m=1}^p\mathbb{P}\left( \sum_{ c=1}^p \left(Y_{c} - a_m\right)^{+} < 2(\psi +\log(p)) + m(1-a_m) + 2pa_mf(a_m) + 2\sqrt{(m + 2pa_mf(a_m))(\psi +\log(p))}\right) \\
&\scriptsize
=\sum_{m=1}^p\mathbb{P}\left( \sum_{ c=1}^p \left(\left(Y_{c} - a_m\right)^{+} - (1-a_m)\mathbb{P}\left(\chi_1^2 > a_m \right) + 2a_mf(a_m)\right)< 2(\psi +\log(p)) + 2\sqrt{p\left(\mathbb{P}\left(\chi_1^2 > a_m \right) + 2a_mf(a_m)\right)(\psi +\log(p))}\right) \\
\footnotesize
&\leq \sum_{m=1}^p p^{-1}e^{-\psi} = e^{-\psi}. 
\end{align*}
\normalsize
A Bonferroni correction over all possible pairs $i,j$ then finishes the proof. 

\subsubsection{Proof of Proposition \ref{Prop:Power}}

\textbf{Proof of Proposition \ref{Prop:Power}}: We will show that the penalised saving for the true anomalous segment is positive with probability converging to 1 as $p$ increases. By the definition of signal strength, the distribution of the true anomalous segment's penalised saving does not depend on the length, $s-e$, of the segment. Thus, we assume, without loss of generality, that $e=s+1$ and treat the cases $0 < \xi \leq \frac{1}{2}$, $\frac{1}{2} < \xi < \frac{3}{4}$, and $\frac{3}{4} < \xi < 1$, separately. We write $X_i:=\textbf{x}_e^{(i)}$, for $ 1 \leq i \leq p$.\\

\textbf{Case 1:} $0 < \xi \leq \frac{1}{2}$. Remember that the composite penalty used is the minimum between regimes 1, 2, and 3. It is therefore sufficient to show that the saving will exceed the penalty specified by one of these three regimes (regime 1 in this case) at some point. By definition, $X_i = \epsilon_i + v_i\mu$, where $\epsilon_1,...,\epsilon_p$ are i.i.d.\ $N(0,1)$ and $v_1,...,v_p$ are i.i.d. $Ber(p^{-\xi})$. Therefore \footnotesize
\begin{align*}
&\mathbb{P}\left(\exists m: \sum_{ i=1}^m X_{(i)} \geq \alpha +\sum_{ i=1}^m\bm{\beta}_i\right) \geq \mathbb{P}\left( \sum_{i=1}^p X_i > p + 2\sqrt{p\psi} + 2\psi \right) = \mathbb{P}\left( \sum_{i=1}^p \epsilon_i^2 +\sum_{ i = 1} v_i \left(2\mu \epsilon_i +  \mu^2\right) > p + 2\sqrt{p\psi} + 2\psi \right) \\
& \geq 1 - \mathbb{P}\left( \sum_{i=1}^p \epsilon_i^2  < p - 2\sqrt{p\psi}\right) - \mathbb{P}\left( \sum_{ i = 1} v_i \left(2\mu \epsilon_i +  \mu^2\right) < 4\sqrt{p\psi} + 2\psi \right) \\ 
&\geq 1 - e^{-\psi} - \mathbb{P}\left( N\left(\mu^2 \left(\sum_{ i = 1} v_i\right),4\mu^2\left(\sum_{ i = 1} v_i\right)\right) < 4\sqrt{p\psi} + 2\psi \right) 
\end{align*}
\normalsize
Furthermore,
\small 
\begin{equation*}
\mathbb{P}\left( N\left(\mu^2 \left(\sum_{ i = 1} v_i\right),4\mu^2\left(\sum_{ i = 1} v_i\right)\right) > 4\sqrt{p\psi} + 2\psi \right) > \mathbb{P}\left( N\left(k\mu^2 ,4k\mu^2\right) > 4\sqrt{p\psi} + 2\psi \right)\mathbb{P}\left( \sum_{ i = 1} v_i> k\right),
\end{equation*}
\normalsize
for all k such that $1 \leq k \leq p$. We therefore only have to show that there exists some sequence of integers $k_p$ such that the right hand side converges to 1 as $p \rightarrow \infty$. Note that Hoeffding's inequality implies that
\begin{equation*}
\mathbb{P}\left( \sum_{ i = 1} v_i> p^{1-\xi} - p^{\frac{1}{2}-\frac{1}{2}\xi}\sqrt{\log(p)} \right) \; \rightarrow \; 1 \; \; \; \; as \; \; \; \; p \; \rightarrow \; \infty
\end{equation*}
and therefore 
\begin{equation*}
\mathbb{P}\left( \sum_{ i = 1} v_i> \frac{1}{2}p^{1-\xi}\right) \; \rightarrow \; 1 \; \; \; \; as \; \; \; \; p \; \rightarrow \; \infty.
\end{equation*}
Setting $k_p=\ceiling{\frac{1}{2}p^{1-\xi}}$, it is therefore sufficient to show that 
\begin{equation*}
\mathbb{P}\left( N\left(0 , 1\right) > \frac{4\sqrt{p\psi} + 2\psi - \frac{1}{2}\mu^2p^{1-\xi} }{\sqrt{2\mu^2p^{1-\xi}}}\right) = 
\mathbb{P}\left( N\left(0 , 1\right) > \frac{4\sqrt{p\psi} + 2\psi - \frac{1}{2}p^{1-\xi-2r_p} }{\sqrt{2\mu^2p^{1-\xi-2r_p}}}\right)
\end{equation*}
converges to 1 as $p$ tends to infinity. This is the case if $r_p < \frac{1}{2}(\frac{1}{2}-\xi)$, which finishes the proof. \\

\textbf{Case 2:} $\frac{3}{4} < \xi < 1$. By an argument similar to that made for case 1, it is sufficient to show that the saving will exceed the penalty specified by regime 2. We have that: 
\begin{align*}
&\mathbb{P}\left(\exists m: \sum_{ i=1}^m X_{(i)} \geq \alpha +\sum_{ i=1}^m\bm{\beta}_i\right) \geq \mathbb{P}\left( X_{(1)} > 2\psi + 2\log(p) \right) = 1 - \left(1 - \mathbb{P} \left(X_1 > 2\psi + 2\log(p) \right)\right)^p \end{align*}
By definition, $X_1 = (\mu v_1  + \epsilon_1 )^2$, where  $\epsilon_1 \sim N(0,1)$ and $v_1\sim Ber(p^{-\xi})$. We can therefore bound the above by
\begin{align*}
&1 - \left(1 - \mathbb{P} \left(X_1 > 2\psi + 2\log(p),v_1=1 \right)\right)^p =   1 - \left(1 - p^{-\xi} \mathbb{P} \left(   \left(\epsilon_1 + \sqrt{2r_p\log(p)}\right)^2 > 2\psi + 2\log(p)  \right)\right)^p \\ 
& > 1 - \left(1 - p^{-\xi} \mathbb{P} \left( N(0,1) > \sqrt{2\psi + 2\log(p)} - \sqrt{2r_p\log(p)} \right)\right)^p \\ 
&\geq 1 - \exp \left(-p^{1-\xi}\mathbb{P} \left(N(0,1) > \sqrt{2\psi + 2\log(p)} - \sqrt{2r_p\log(p)} \right) \right),
\end{align*}
where the second inequality follows from the fact that $1-x \leq e^{-x}$. We consider separately the cases $\sqrt{2\psi + 2\log(p)} - \sqrt{2r_p\log(p)} > 1 $ and $\sqrt{2\psi + 2\log(p)} - \sqrt{2r_p\log(p)} \leq 1$. In the latter case the above clearly converges to 1 as $p$ goes to infinity. In the former case we can use the lower tail bound $\mathbb{P} \left( N(0,1) > x \right) > \frac{1}{\sqrt{2\pi}}\frac{x}{x^2+1} \exp\left(-\frac{x^2}{2}\right)$, for $x > 0$ to bound the above by
\begin{equation*}
1 - \exp \left(-\frac{1}{\sqrt{2\pi}}p^{1-\xi} \frac{p^{\left(\sqrt{1+\frac{2\psi}{2\log(p)}}-\sqrt{r_p}\right)^2}}{ 1 + \left(\sqrt{2\psi + 2\log(p)} - \sqrt{2r_p\log(p)} \right)^2} \right).
\end{equation*}
Thus, for a fixed $r_p > \left(1-\sqrt{1-\xi}\right)^2$ this converges to 1, as $\psi/\log(p)$ converges to 0. \\

\textbf{Case 3:} $\frac{1}{2} < \xi < \frac{3}{4}$. By an argument similar to that made for case 1, it is sufficient to show that the saving will exceed the penalty specified by regime 3. We assume, without loss of generality, that $\mu > 0$. If $r_p\geq \frac{1}{4}$. Our approach is to define a threshold, $b$,  and a number of excesses, $\tilde{k}$, such that the number of savings in cost that exceed $b$ will be great than $\tilde{k}$ with probability going to $1$ as $p$ increases. We then show that the overall sum of the $\tilde{k}$ largest savings will be greater than the penalty for fitting $\tilde{k}$ components as anomalous.

We introduce the following new random variable: 
\begin{align*}
Y_i = \begin{cases}
\left[(\mu + \epsilon_i)^+\right] ^2 &  \text{if} \; \; \; v_i = 1\\
\epsilon_i^2 & \text{if} \; \; \; v_i = 0,
\end{cases}
\end{align*}
where $(x)^+$ denotes the positive part of $x$. Note that $Y_i \leq X_i$. We also introduce the following four technical lemmata
\begin{Lemma}\label{lemma:Hazardrate}
	Let $a > 0$ and let $Z \sim \chi^2_1$. Then, for all positive $x \in \mathbb{R}$ 
	\begin{equation*}
	\mathbb{P}\left(Y_i \geq a +x | Y_i \geq a \right) \geq \mathbb{P} \left( Z > a+x | Z \geq a\right).
	\end{equation*}
\end{Lemma}

\begin{Lemma}\label{lemma:LowerboundSubgaussian}
	Let $Z_i \stackrel{i.i.d.}{\sim} \chi^2_1$ for $1\leq i \leq k$ and $a>0$. Then for all $t \in R$ 
	\small
	\begin{equation*}
	\mathbb{P}\left(\sum_{i=1}^k(Z_i-a)|(Z_i>a) < k\mathbb{P}\left(\chi^2_1 >a\right)^{-1} \mathbb{E} \left( (Z-a)^+\right) - 2\sqrt{k\mathbb{P}\left(\chi^2_1 >a\right)^{-1}(\mathbb{P}\left(\chi^2_1 >a\right)+2af(a)) t } \right) < e^{-t}
	\end{equation*}
	\normalsize 
\end{Lemma}

\begin{Lemma}\label{lemma:Bounds_Regime3}
	Let $a_k$ be defined implicitly as $\mathbb{P}\left(\chi^2_1 > a_k\right) = \frac{k}{p}$ and let $f(\cdot)$ denote the probability density function of the $\chi^2_1$ distribution. Then
	\begin{equation*}
	p\mathbb{E} \left( (\chi_1^2-a_{k})^+\right) + k a_k = k + 2pa_kf(a_k) \leq 2k + 2k\log(p/k)
	\end{equation*}
\end{Lemma}

\begin{Lemma}\label{lemma:Mean_Increasing}
	For all $b>0$:
	\begin{equation*}
	\mathbb{E} \left((\chi_1^2 - b)^+|\chi_1^2 > b\right)  > 1.
	\end{equation*}
\end{Lemma}

Next write $b = 8r_p\log(p)$ and let $\tilde{k}$ be an integer such that both $p\mathbb{P}\left(\chi_1^2>b\right) \leq \tilde{k} \leq p$ and $a_{\tilde{k}} < b$. Note that since $r_p < \frac{1}{4}$, we have $b \leq 2\log(p)$ and such a $\tilde{k}$ is guaranteed to exist for sufficiently large values of $p$. For convenience, write $\tilde{\mu} = \mathbb{E} \left( (\chi_1^2-a_{\tilde{k}})^+\right)$. The following holds: 
\small
\begin{align*}
&\mathbb{P}\left(\exists m: \sum_{ i=1}^m X_{(i)} \geq \alpha +\sum_{ i=1}^m\bm{\beta}_i\right) \geq \mathbb{P}\left(\sum_{ i=1}^{\tilde{k}} Y_{(i)} \geq 2\psi +2\log(p) + \tilde{k}a_{\tilde{k}}+p\tilde{\mu} + 2 \sqrt{(\tilde{k}a_{\tilde{k}}+p\tilde{\mu})(\psi+2\log(p))} \right) \\
& \geq \mathbb{P}\left(\sum_{ i=1}^{\tilde{k}} Y_{(i)} \geq 2\psi +2\log(p) + \tilde{k}a_{\tilde{k}}+p\tilde{\mu} + 2 \sqrt{(\tilde{k}a_{\tilde{k}}+p\tilde{\mu})(\psi+2\log(p))} \Bigg| \sum_{ i =1}^p \mathbb{I} \left(Y_i > b \right)\geq \tilde{k} \right) \mathbb{P} \left(\sum_{ i =1}^p \mathbb{I} \left(Y_i > b \right)\geq \tilde{k} \right),
\end{align*}
\normalsize
where the first inequality follows from substituting the third penalty regime (using the equality from Lemma \ref{lemma:Bounds_Regime3}) and the second inequality follows from conditioning on the number of $Y_i$ exceeding $b$. Next note that
\begin{align*}
&\mathbb{P}\left(\sum_{ i=1}^{\tilde{k}} Y_{(i)} \geq 2\psi +2\log(p) + \tilde{k}a_{\tilde{k}}+p\tilde{\mu} + 2 \sqrt{(\tilde{k}a_{\tilde{k}}+p\tilde{\mu})(\psi+\log(p))} \Bigg| \sum_{ i =1}^p \mathbb{I} \left(Y_i > b \right)\geq \tilde{k} \right) \\ & \geq 
\mathbb{P}\left(\sum_{ i=1}^{\tilde{k}} Y_{(i)} \geq 2\psi +2\log(p) + \tilde{k}a_{\tilde{k}}+p\tilde{\mu} + 2 \sqrt{(\tilde{k}a_{\tilde{k}}+p\tilde{\mu})(\psi+\log(p))} \Bigg| \sum_{ i =1}^p \mathbb{I} \left(Y_i > b \right) = \tilde{k} \right) \\ 
&= \mathbb{P}\left(\sum_{ i=1}^{\tilde{k}}\left( Y_{i} - b \right)^+ \geq 2\psi +2\log(p) -\tilde{k}\left(b-a_{\tilde{k}}\right) + p\tilde{\mu} + 2 \sqrt{(\tilde{k}a_{\tilde{k}}+p\tilde{\mu})(\psi+\log(p))} \Bigg| Y_1,...,Y_{\tilde{k}} > b \right) 
\end{align*}
Let $Z_1, ..., Z_{\tilde{k}}$ be i.i.d.\ $\chi_1^2$ distributed. Lemma \ref{lemma:Hazardrate} then implies that the above exceeds 
\begin{equation*}
\mathbb{P}\left(\sum_{ i=1}^{\tilde{k}}\left( Z_{i} - b \right)^+ \geq 2\psi +2\log(p) -\tilde{k}\left(b-a_{\tilde{k}}\right) + p\tilde{\mu} + 2 \sqrt{(\tilde{k}a_{\tilde{k}}+p\tilde{\mu})(\psi+\log(p))} \Bigg| Z_1,...,Z_{\tilde{k}} > b \right).
\end{equation*}
Using the inequality in Lemma \ref{lemma:Bounds_Regime3} and the fact that $\psi < \log(p)$ for sufficiently large values of $p$ we can further bound the above by 
\begin{equation*}
\mathbb{P}\left(\sum_{ i=1}^{\tilde{k}}\left( Z_{i} - b \right)^+ \geq 4\log(p) + p \tilde{\mu} -\tilde{k}\left(b-a_{\tilde{k}}\right) + 2\sqrt{4(\tilde{k} + \tilde{k} \log(p/\tilde{k}) ) \log(p)} \Bigg| Z_1,...,Z_{\tilde{k}} > b \right).
\end{equation*}
Defining  $W_i = \left( Z_{i} - b \right) | (Z_i > b)$, we can further bound the above by
\begin{equation}\label{eq:tmp}
\mathbb{P}\left(\sum_{ i=1}^{\tilde{k}}W_i\geq p \tilde{\mu} -\tilde{k}\left(b-a_{\tilde{k}}\right) + 8\sqrt{\tilde{k}\log(p)^2}  \right), 
\end{equation}
provided $p$ is large enough. Next, note that since $b \geq a_{\tilde{k}}$
\begin{equation*}
\tilde{\mu} = \mathbb{E}\left[\left(\chi_1^2-a_{\tilde{k}}\right)^+\right] \leq \mathbb{E}\left[\left(\chi_1^2-b\right)^+\right] + \mathbb{P}\left(\chi_1^2>a_{\tilde{k}}\right)\left(b-a_{\tilde{k}}\right).
\end{equation*}
Consequently, we can bound (\ref{eq:tmp}) by 
\begin{align*}
&\mathbb{P}\left(\sum_{ i=1}^{\tilde{k}}W_i\geq p \mathbb{E}\left[\left(\chi_1^2 - b\right)^+\right] + 8 \sqrt{\tilde{k}\log(p)^2} \right)   \\
& = \mathbb{P}\left(\sum_{ i=1}^{\tilde{k}}\left[W_i - \mathbb{E}\left( \chi^2_1 - b | \chi^2_1 > b\right)\right] \geq \left(p\mathbb{P}\left(\chi^2_1 > b\right) - \tilde{k}\right) \mathbb{E}\left( \chi^2_1 - b | \chi^2_1 > b\right) + 8 \sqrt{\tilde{k}\log(p)^2} \right)  \\
& \geq 1 - \exp \left( - \frac{\left[\left((\tilde{k} - p\mathbb{P}\left(\chi_1^2 > b\right)) \mathbb{E} \left((\chi_1^2 - b)^+|\chi_1^2 > b\right) - 8\sqrt{\tilde{k}\log(p)^2} \right)^+\right]^2}{4\tilde{k} \mathbb{P}\left(\chi_1^2 > b\right)^{-1} (\mathbb{P}\left(\chi_1^2 > b\right) + 2bf(b))}  \right), 
\end{align*}
\normalsize
where the inequality follows from Lemma \ref{lemma:LowerboundSubgaussian}. Given Lemma \ref{lemma:Mean_Increasing} and the fact that Lemma \ref{lemma:Bounds_Regime3} implies that $\mathbb{P}\left(\chi_1^2 > b\right) + 2bf(b) < 2\mathbb{P}\left(\chi_1^2 > b\right)(1+\log(p))   $, we can further bound the above by 
\begin{equation*}
1 - \exp \left( - \frac{\left[\left((\tilde{k} - p\mathbb{P}\left(\chi_1^2 > b\right)) -8\sqrt{\tilde{k}}\log(p) \right)^+\right]^2}{8(\tilde{k}(1+\log(p)))}  \right).
\end{equation*} 
The arithmetic-mean-geometric-mean-inequality can be used to show that $((a-b)^+)^2 > \frac{\left((a)^+\right)^2}{2} - 4b^2$. The above quantity is therefore bounded by 
\begin{equation*}
1 - \exp \left( - \frac{1}{16(1+\log(p))}\left(\left[\frac{\tilde{k} - p\mathbb{P}\left(\chi_1^2 > b\right)}{\sqrt{\tilde{k}}}\right]^+\right)^2  +72 \log(p)\right).
\end{equation*} 
Note that if $\tilde{k} \geq p\mathbb{P}\left(\chi_1^2 > b\right) + p^{\frac{1}{2} - 2r_p + \delta}$ for some $\delta >0$, then 
\begin{equation*}
\frac{\tilde{k} - p\mathbb{P}\left(\chi_1^2 > b\right)}{\sqrt{\tilde{k}}} \geq \frac{ p^{\frac{1}{2} - 2r_p + \delta}}{\sqrt{p\mathbb{P}\left(\chi_1^2 > b\right) + p^{\frac{1}{2} - 2r_p + \delta}}} \geq \frac{ p^{\frac{1}{2} - 2r_p + \delta}}{\sqrt{p^{1-4r_p} + p^{\frac{1}{2} - 2r_p + \delta}}} \geq \frac{1}{2}p^{\frac{\delta}{2}},
\end{equation*}
with the first inequality following from the fact that the left-hand side is increasing in $\tilde{k}$, the second one following from the fact that $\mathbb{P}\left(\chi_1^2 > b\right)< \mathbb{P}\left(\chi_2^2 > b\right) = p^{-4r_p}$ and the last one following from the fact that $r_p < \frac{1}{4}$. 

Consequently, it is sufficient to show that there exists a $\delta>0$ such that
\begin{equation*}
\mathbb{P}\left( \sum_{i=1}^p \mathbb{I}\left(Y_i \geq b\right)> p\mathbb{P}\left(\chi_1^2 > b\right) + p^{\frac{1}{2} - 2r_p + \delta}\right) \rightarrow 1 \; \; \; \; \text{as} \; \; \; p \rightarrow \infty 
\end{equation*}
This can be seen from the fact that $\sum_{i=1}^p \mathbb{I}\left(Y_i > b\right)$ is $Bin(p,q)$-distributed with
\begin{equation*}
q = \mathbb{P}\left(Y_i > 8r_p\log(p)\right) = (1-p^{-\xi})\mathbb{P}\left(\chi^2_1>b\right) + p^{-\xi} \mathbb{P}\left( N(0,1) > \sqrt{2r_p\log(p)}\right).
\end{equation*}
Note that 
\begin{equation*}
q > \mathbb{P}\left(\chi^2_1>b\right) - p^{-\xi - 4r_p} +  p^{-\xi} \mathbb{P}\left( N(0,1) > \sqrt{2r_p\log(p)}\right),
\end{equation*}
since $\mathbb{P}\left(\chi_1^2 > b\right) < p^{-4r_p}$. Moreover, 
\begin{equation*}
q <  \mathbb{P}\left(\chi^2_1>b\right)  + p^{-\xi} \mathbb{P}\left( N(0,1) > \sqrt{2r_p\log(p)}\right) \leq p^{-4r_p} + p^{-\xi -r_p},
\end{equation*}
by standard tail bounds of the normal distribution and the definition of $b$. Standard Hoeffding bounds show that 
\begin{equation*}
\mathbb{P}\left(\sum_{i=1}^p \mathbb{I}\left(Y_i > b\right) > pq - \sqrt{pq\log(p)} \right) \rightarrow 1 \; \; \; \; as \; \; \; \; p \rightarrow \infty
\end{equation*}
Hence, 
\begin{equation*}
\mathbb{P}\left(\sum_{i=1}^p \mathbb{I}\left(Y_i > b\right) > p\mathbb{P}\left(\chi^2_1>b\right) +  p^{1-\xi} \mathbb{P}\left( N(0,1) > \sqrt{2r_p\log(p)}\right) - p^{1-\xi - 4r_p} - \sqrt{p(p^{-4r_p} + p^{-\xi -r_p})\log(p)} \right) 
\end{equation*}
converges to 1 as $ p \rightarrow \infty$. Note that 
\begin{equation*}
p^{1-\xi} \mathbb{P}\left( N(0,1) > \sqrt{2r_p\log(p)}\right) - p^{1-\xi - 4r_p} - \sqrt{p(p^{-4r_p} + p^{-\xi -r_p})\log(p)} > p^{\frac{1}{2} - 2r_p + \delta}
\end{equation*}
for all $\delta$ such that $r_p-\xi + \frac{1}{2}> \delta$, provided $p$ is large enough. This follows from the fact that 
\begin{equation*}
p^{1-\xi} \mathbb{P}\left( N(0,1) > \sqrt{2r_p\log(p)}\right) > \frac{1}{\sqrt{2\pi}}p^{1-\xi-r_p} \frac{\sqrt{2r_p\log(p)}}{1+2r_p\log(p)},
\end{equation*}
by standard tail bounds on the normal distribution. Since $0 < r_p$, the the above dominates $p^{1-\xi - 4r_p}$ as $p$ increases. Similarly, because $r_p-\xi + \frac{1}{2} > 0$, it dominates $\sqrt{p^{1-4r_p}}$, since, $r_p < \frac{1}{4}$ and $\xi < \frac{3}{4}$, $1-r_p-\xi > 0$ and the above therefore also dominates $\sqrt{p^{1-r_p-\xi}}$. Finally, if $\delta$ is such that $r_p-\xi + \frac{1}{2}> \delta$ it must also dominate $p^{\frac{1}{2} - 2r_p + \delta}$. This finishes the proof.

\subsubsection{Proof of Proposition \ref{prop:FPControlPtAnomalies}}

Standard tail bounds on the normal distribution give that
\begin{equation*}
\mathbb{P} \left(\left(\textbf{x}_t^{(i)}\right)^2  < 2\psi\right) \geq 1-Ae^{-\psi}
\end{equation*}
holds for a constant $A$ under the null hypothesis. A Bonferroni correction therefore gives  
$\mathbb{P} \left(\hat{O}=\emptyset\right) \geq 1-Anp\exp(-\frac{1}{2}\beta')$. \qed

\subsubsection{Proof of Propositions \ref{prop:Pruning_easy} and \ref{prop:Pruning}}

We give the proof of Proposition \ref{prop:Pruning}, as Proposition $\ref{prop:Pruning_easy}$ corresponds to as special case. We write
\begin{equation*}
{\mathcal{S}}\left( s,e,\textbf{d},\textbf{f},\textbf{J}\right) = \sum_{i \in \textbf{J}} \left( 
\mathcal{C}_i \left( \textbf{x}^{(i)}_{(s+1+\textbf{d}^{(i)}):(e-\textbf{f}^{(i)})},\bm{\theta}_0^{(i)}\right)
-
\min_{\bm{\theta}}\left[\mathcal{C}_i \left(\textbf{x}^{(i)}_{(s+1+\textbf{d}^{(i)}):(e-\textbf{f}^{(i)})},\bm{\theta}\right)\right]   \right) - \alpha - \sum_{ i=1 }^{|\textbf{J}|}\bm{\beta}_i
\end{equation*}
and note that 
\begin{equation*}
\mathcal{S}\left( t,m\right) = \max_{\textbf{d},\textbf{f},\textbf{J}: \; \; m - t - d - f \geq l}\left[{\mathcal{S}}\left( t,m,\textbf{d},\textbf{f},\textbf{J}\right) \right]
\end{equation*}
The proof of Proposition \ref{prop:Pruning} is then a corollary of the observation that for all $\textbf{d},\textbf{f} < w$
\footnotesize
\begin{align*}
&\mathcal{S}\left( t,m',\textbf{d},\textbf{f},\textbf{J}\right) 
=  \sum_{i \in \textbf{J}} \left( 
\mathcal{C}_i \left( \textbf{x}^{(i)}_{(t+1+\textbf{d}^{(i)}):(m'-\textbf{f}^{(i)})},\bm{\theta}_0^{(i)}\right)
-
\min_{\bm{\theta}}\left[\mathcal{C}_i \left(\textbf{x}^{(i)}_{(t+1+\textbf{d}^{(i)}):(m'-\textbf{f}^{(i)})},\bm{\theta}\right)\right]   \right) - \alpha - \sum_{ i=1 }^{|\textbf{J}|}\bm{\beta}_i \\
&\leq  \sum_{i \in \textbf{J}} \left( 
\mathcal{C}_i \left( \textbf{x}^{(i)}_{(t+1+\textbf{d}^{(i)}):(m'-\textbf{f}^{(i)})},\bm{\theta}_0^{(i)}\right)
-
\min_{\bm{\theta}}\left[\mathcal{C}_i \left(\textbf{x}^{(i)}_{(t+1+\textbf{d}^{(i)}):m},\bm{\theta}\right)\right]  
-\min_{\bm{\theta}}\left[\mathcal{C}_i \left(\textbf{x}^{(i)}_{(m+1):(m'-\textbf{f}^{(i)})},\bm{\theta}\right)\right]  \right)  - \alpha - 2\sum_{ i=1 }^{|\textbf{J}|}\bm{\beta}_i + \sum_{i = 1}^{p} \bm{\beta}_i \\
&= {\mathcal{S}}\left( t,m,\textbf{d},0,\textbf{J}\right) + {\mathcal{S}}\left( m,m',0,\textbf{f},\textbf{J}\right) + \sum_{i = 1}^{p} \bm{\beta}_i + \alpha, 
\end{align*}
\normalsize
since $m-t-w\geq l$ and $m'-m-w \geq l$. Indeed, the above inequality shows that
\begin{align*}
&C(t) + \max_{\textbf{d}\leq w,\textbf{f}\leq w,\textbf{J}: \; \; m - t - d - f \geq l}\left[{\mathcal{S}}\left( t,m',\textbf{d},\textbf{f},\textbf{J}\right) \right] \\
&\leq C(t) + \max_{\textbf{d}\leq w,\textbf{f}\leq w,\textbf{J}: \; \; m - t - d - f \geq l}\left[{\mathcal{S}}\left( t,m,\textbf{d},\textbf{0},\textbf{J}\right) + {\mathcal{S}}\left( m,m',\textbf{0},\textbf{f},\textbf{J}\right) \right]  + \alpha + \sum_{i = 1}^{p} \bm{\beta}_i \\
&\leq C(t) + \max_{\textbf{d}\leq w,\textbf{J}: \; \; m - t - d \geq l}\left[{\mathcal{S}}\left( t,m,\textbf{d},\textbf{0},\textbf{J}\right)\right] + \max_{\textbf{f}\leq w,\textbf{J}: \; \; m - t  - f \geq l} \left[{\mathcal{S}}\left( m,m',\textbf{0},\textbf{f},\textbf{J}\right) \right]  + \alpha + \sum_{i = 1}^{p} \bm{\beta}_i\\
&\leq C(t) + \max_{\textbf{d}\leq w,\textbf{f}\leq w,\textbf{J}: \; \; m - t - d - f \geq l}\left[{\mathcal{S}}\left( t,m,\textbf{d},\textbf{f},\textbf{J}\right)\right] + \max_{\textbf{d}\leq w,\textbf{f}\leq w,\textbf{J}: \; \; m - t - d - f \geq l} \left[{\mathcal{S}}\left( m,m',\textbf{d},\textbf{f},\textbf{J}\right) \right]  + \alpha + \sum_{i = 1}^{p} \bm{\beta}_i\\
&< C(m)+\max_{\textbf{d},\textbf{f},\textbf{J}: \; \; m - t - d - f \geq l} \left[{\mathcal{S}}\left( m,m',\textbf{d},\textbf{f},\textbf{J}\right) \right]   \leq C(m').
\end{align*}
This finishes the proof. 

\subsubsection{Proof of Proposition \ref{prop:FPcontrol_mean_penalty_2'}}

The proof of Proposition \ref{prop:FPcontrol_mean_penalty_2'} is similar to that of Proposition \ref{prop:FPcontrol_mean_penalty_2}. Let $1\leq i\leq j \leq n$. For this pair $(i,j)$ define: 
\begin{equation*}
Y_c^{(d,f)} = (j-f-d-i+1)\left(\bar{\bm{\eta}}_{(i+d):(j-f)}^{(c)}\right) ^2,
\end{equation*}
noting that they are all $\chi^2_1$ distributed. Then define $Y_c = \max \left( Y_c^{(d,f)}\right)$ and note that $Y_1,...,Y_p$ are independent. The probability that this segment $i,j$ will not be considered anomalous is
\begin{align*}
&\mathbb{P}\left( \sum_{ c \in S_m} Y_c  <  2(1+\epsilon)\psi + 2m(1+\epsilon)\log(p(w+1)), \, \; \; \, \forall S_m \subset \{1,...,p \}: |S_m| = m  , \;\;  1 \leq m \leq p\right) \\
& \geq \mathbb{P}\left( \sum_{i=1}^p \left(Y_i - 2(1+\epsilon)\log(p(w+1)) \right)^+ < 2(1+\epsilon)\psi \right) \geq 1 - \mathbb{E}\left( e^{\lambda\left(Y_1 - 2(1+\epsilon)\log(p(w+1)) \right)^+} \right)^pe^{-2(1+\epsilon)\lambda\psi},
\end{align*}
for all $\lambda > 0$, with the second inequality being a Chernoff bound. Now set $\lambda = \frac{1}{2} \frac{1}{1+\epsilon}$ and note that  \small
\begin{align}\label{eq:help}
&\mathbb{E}\left( e^{\lambda\left(Y_1 - 2(1+\epsilon)\log(p(w+1)) \right)^+} \right)^p e^{-\psi} = \left[ 1 + \int_{1}^{\infty} \mathbb{P} \left( Y_1 > 2(1+\epsilon) \log(p(w+1)) + \frac{1}{\lambda} \log(x) \right) dx      \right]^p e^{-\psi} 
\end{align} \normalsize
The following lemma holds
\begin{Lemma}\label{lemma:Martingalebound}
	Let $\eta_t \stackrel{i.i.d.}{\sim}N(0,1)$ for $i \leq t \leq j$. Then there exists a constant $\tilde{A}$ such that 
	\begin{equation*}
	\mathbb{P} \left( \max_{0\leq j,d \leq w: j-f-d-i \geq 0} \left((j-f-d-i+1)\left(\bar{\bm{\eta}}_{(i+d):(j-f)}^{(c)}\right) ^2 \right)> u \right) \leq \tilde{A}(w+1)\frac{1+\log(w+1)}{\log(b)}e^{-\frac{u}{2b}}.
	\end{equation*}
	for all $b \in \mathbb{R}$ such that $1 < b \leq 2$
\end{Lemma}
Using this lemma and setting $b = \sqrt{1+\epsilon}$, we can bound (\ref{eq:help}) by
\begin{align*}
&\left[ 1 + 2\tilde{A}(w+1)\frac{1+\log(w+1)}{\log(1+\epsilon)}\frac{1}{((w+1)p)^{\sqrt{1+\epsilon}}} \int_{1}^{\infty} x ^{-\frac{1}{2b\lambda}} dx      \right]^p e^{-\psi}  \\
&=\left[ 1 + \frac{1}{p^{\sqrt{1+\epsilon}}}\frac{2\tilde{A}}{\log(1+\epsilon)}\frac{1+\log(w+1)}{(w+1)^{\sqrt{1+\epsilon}-1}} \frac{1}{1-2\sqrt{1+\epsilon}\lambda}      \right]^p e^{-\psi}  \\
&\leq \exp\left[ \frac{1}{p^{\sqrt{1+\epsilon}-1}}\frac{2\tilde{A}}{\log(1+\epsilon)}\frac{1+\log(w+1)}{(w+1)^{\sqrt{1+\epsilon}-1}} \frac{\sqrt{1+\epsilon}}{\sqrt{1+\epsilon}-1}      \right] e^{-\psi},
\end{align*}
where the second inequality follows from the fact that $1-x \leq e^{-x}$. Note that
\begin{equation*}
\frac{1+\log(w+1)}{e^{(\sqrt{1+\epsilon}-1)\log(w+1)}} \leq  \frac{1+\log(w+1)}{1+(\sqrt{1+\epsilon}-1)\log(w+1)} \leq \max \left( 1 , \frac{1}{\sqrt{1+\epsilon}-1} \right) \leq \frac{\sqrt{1+\epsilon}}{\sqrt{1+\epsilon}-1},
\end{equation*} 
where the first inequality follows from the fact that $1+x \leq e^{x}$, the second inequality follows from the fact that the expression is monotone in $\log(w+1)$, and the third inequality follows from the fact that $max(a,b) \leq a +b $ for all positive $a,b$. The above result and the fact that $p \geq 1$ imply that the probability of false positives can further be bounded by
\begin{align*}
\exp\left[ \frac{2\tilde{A}}{\log(1+\epsilon)} \frac{1+\epsilon}{(\sqrt{1+\epsilon}-1)^2}      \right] e^{-\psi} = \exp\left[ \frac{2\tilde{A}}{\log(1+\epsilon)} \frac{(1+\epsilon)(\sqrt{1+\epsilon}+1)^2}{\epsilon^2}      \right] e^{-\psi} 
\leq  \exp\left[ 8\tilde{A} \frac{(1+\epsilon)^3}{\epsilon^3}      \right] e^{-\psi} ,
\end{align*}
where the inequality follows from the fact that $\log(1+\epsilon) \geq \frac{\epsilon}{1+\epsilon}$ and $1 \leq \sqrt{1+\epsilon}$. A Bonferroni correction over all possible pairs $i,j$ then finishes the proof. 

\subsubsection{Proof of Theorem \ref{THM:Main}}

We define the penalised cost of a segment $\textbf{x}_{i:j}$ under a partition $\tau = \{ \hat{\tau}_1, ... ,\hat{\tau}_{\hat{K}} \}$, where $\hat{\tau}_k = (\hat{s}_k,\hat{e}_k,\hat{\textbf{J}}_k)$ to be 
\begin{equation*}
\mathcal{C}\left(\textbf{x}_{i:j} , \hat{\tau} \right) = \sum_{k=1}^{\hat{K}} \left[ \mathcal{C}(\textbf{x}_{(\hat{s}_{k}+1):\hat{e}_k} ,\hat{\textbf{J}}_k)\right].
\end{equation*}
Here the penalised cost of introducing the $k$th anomalous window is 
\begin{equation*}
\mathcal{C}\left(\textbf{x}_{(s+1):e} , \{(s,e,\textbf{J})\} \right) = \mathcal{C}(\textbf{x}_{(s+1):e} ,\textbf{J}) := -\mathcal{S}(\textbf{x}_{(s+1):e} ,\textbf{J})  + \sum_{i=1}^{|\textbf{J}|}\bm{\beta_i}. := - (e-s)\sum_{i \in \textbf{J}} \mathcal{ C} \left(\bar{\textbf{x}}_{(s+1):e}^{(i)}\right)^2 + \sum_{i=1}^{|\textbf{J}|}\bm{\beta_i},
\end{equation*} \small
where $\mathcal{S}(\textbf{x}_{(s+1):e} ,\textbf{J})$, is defined as the saving made by fitting the segment $\textbf{x}_{(s+1):e}$ with $\textbf{J}$ and $\bar{\textbf{x}}_{(s+1):e}^{(i)} := (e-s)^{-1}\sum_{t=s+1}^e \textbf{x}_t^{i} $  is defined as the arithmetic mean of the $i$th component from time $t=s+1$ to $t=e$. It should be noted that minimising the penalised cost, is equivalent to maximising the penalised saving. We call the partition which minimises the penalised cost, $\mathcal{C}\left(\textbf{x}_{1:n} , \hat{\tau} \right)$, over all feasible partitions, $\hat{\tau}$, the optimal partition.

We also define the following event sets over all pairs $i,j$ such that $1 \leq i \leq j \leq n$
\begin{align*}
E_1 &:= \left\{ \sum_{ c \in S}(j-i+1)\left(\bar{\bm{\eta}}_{i:j}^{(c)}\right) ^2  < 2\psi +2|S|\log(p)  \; \; \; \forall S \subset \{1,...,p\} \right\} \\
E_2 &:= \left\{ \sum_{ c \in S}(j-i+1)\left(\bar{\bm{\eta}}_{i:j}^{(c)}\right) ^2  < p + 2\psi + 2\sqrt{p\psi}  \; \; \; \forall S \subset \{1,...,p\} \right\} \\
E_3 &:= \left\{ \sum_{ c = 1}^p(j-i+1)\left(\bar{\bm{\eta}}_{i:j}^{(c)} + \bar{\bm{\mu}}_{i:j}^{(c)} \right) ^2  > p - 2\sqrt{p\psi}\right\} \\
E_4 &:= \left\{ \left|\sum_{ c \in S}\sqrt{j-i+1}\bar{\bm{\eta}}_{i:j}^{(c)}\right| < \sqrt{2|S|\psi+2|S|^2\log(p)} \; \; \; \forall S \subset \{1,...,p\} \right\} \\
E_5 &:= \left\{ \sum_{ c \notin S }(j-i+1)\left(\bar{\bm{\eta}}_{i:j}^{(c)} + \bar{\bm{\mu}}_{i:j}^{(c)} \right) ^2  > p - 2\sqrt{p\psi} - 2\psi -2|S|\log(p) \; \; \; \forall S \subset \{1,...,p\}  \right\} \\
E_6 &:= \left\{ \left| \sum_{ c \in S }\left( \sum_{t=i}^j \left(\bm{\mu}_t^{(c)} - \bar{\bm{\mu}}_{i:j}^{(c)}  \right)\bm{\eta}_t^{(c)} \right) \right| \leq \sqrt{\sum_{ c \in S } \sum_{t=i}^j \left(\bm{\mu}_t^{(c)} - \bar{\bm{\mu}}_{i:j}^{(c)}  \right)^2}\sqrt{2\psi+2 \left| S \cap W_{i,j} \right| \log(p)} \; \; \; \forall S \subset \{1,...,p\}  \right\}\\
E_7 &:= \left\{ \sum_{ c \in S}\frac{(j-j')(j'-i+1)}{j-i+1}\left(\bar{\bm{\eta}}_{i:j'}^{(c)} - \bar{\bm{\eta}}_{(j'+1):j}^{(c)}\right) ^2  < 2\psi +2|S|\log(p)  \; \; \; \forall S \subset \{1,...,p\} \right\} \\
E_8 &:= \left\{ \sum_{ c \in S}\frac{(j-j')(j'-i+1)}{j-i+1}\left(\bar{\bm{\eta}}_{i:j'}^{(c)} - \bar{\bm{\eta}}_{(j'+1):j}^{(c)}\right) ^2   < p + 2\psi + 2\sqrt{p\psi}  \; \; \; \forall S \subset \{1,...,p\} \right\} \\
E_9 &:= \left\{ \sum_{ c = 1}^p\frac{(j-j')(j'-i+1)}{j-i+1}\left(\bar{\textbf{x}}_{i:j'}^{(c)} - \bar{\textbf{x}}_{(j'+1):j}^{(c)}\right) ^2   > p - 2\sqrt{p\psi}\right\} \\
E_{10} &:= \left\{ \left|\sum_{ c \in \textbf{J}_k}\sqrt{j-i+1}\bar{\bm{\eta}}_{i:j}^{(c)}\right| < \sqrt{2|\textbf{J}_k|\psi} \; \; \; \forall i,j: \exists k \in \{1,...,K\}: e_{k-1}<i, \; \; j \leq s_{k+1} \right\} \\
E_{11} &:= \left\{ \left|\sum_{ c \in \textbf{J}_k} \sqrt{\frac{(j-e_k)(e_k-i+1)}{j-i+1}}\left(\bar{\bm{\eta}}_{i:e_k}^{(c)} - \bar{\bm{\eta}}_{(e_k+1):j}^{(c)}\right)\right|  < \sqrt{2|\textbf{J}_k|\psi} \; \; \; \forall i,j: \exists k \in \{1,...,K\}: e_{k-1}<i, \; \; j \leq s_{k+1} \right\},
\end{align*}\normalsize
where the set $W_{i,j}$ of components with non constant mean in the interval $[i,j]$ is defined as
\begin{equation*}
W_{i,j} =  \left\{ c \in \{1,...,p\}: \exists t \in [i,j-1]: \bm{\mu}_t^{(c)} \neq \bm{\mu}_{t+1}^{(c)} \right\}.
\end{equation*}
The intuition behind these events is as follows: Events $E_1$ and $E_2$ bound the saving obtained from fitting an anomalous region on data belonging to the typical distribution and so ensure no false positives are fitted. Events $E_7$, $E_8$, $E_9$, and $E_{11}$ provide bounds on the additional un-penalised cost of splitting a fitted segment in two or merging two existing segments, assuring that anomalous regions are fitted by one rather than multiple adjacent segments. They are assisted by events $E_3$ and $E_5$ which bound the additional un-penalised cost incurred by fitting any given segment by a dense change, extending the result to showing the sub-optimality of a collective anomaly being fitted by multiple non-adjacent segments. Events $E_4$, $E_6$, and $E_{10}$ bound the interaction between the signal and the noise thus ensuring that anomalous regions are detected. For brevity, we denote $E= \cap E_i $ and note that it occurs with high probability. Indeed, the following Lemma holds: 
\begin{Lemma}\label{lemma:Eventsets}
	There exists a constant $A$ such that 
	\begin{equation*}
	\mathbb{P}\left(E\right)> 1 - An^3 e^{-\psi}
	\end{equation*}
\end{Lemma}
We now define the set of good partitions $\mathcal{B}_C$ to be 
\begin{equation}
\mathcal{B}_C = \left\{ \tau : |\tau| = K, \; \; |\hat{s}_k-s_k|\leq \frac{10C}{\triangle^2_k} \; \; |\hat{e}_k-e_k|\leq \frac{10C}{\triangle^2_k} \right\}.
\end{equation}
It is sufficient to prove the following proposition in order to prove Theorem \ref{THM:Main} 
\begin{Proposition}\label{Prop:MAIN}
	Let the assumptions of Theorem 1 hold. Given $E$ holds and $C$ exceeds a global constant, the partition $\tau_0$ minimising the penalised cost $\mathcal{C}\left(\textbf{x}_{1:n} , \tau \right)$ satisfies $\tau_0 \in \mathcal{B}_C$
\end{Proposition}

\noindent The main ideas of the proof of Proposition \ref{Prop:MAIN} are that given $E$: 
\begin{enumerate}[I]
	\item Each fitted anomalous segment overlaps with at most one true anomalous segment.
	\item Each fitted anomalous segment overlaps with at least one true anomalous region.
	\item Each true anomalous segment overlaps with at most one fitted anomalous region, i.e. there exists a bijection between fitted and true segments.
	\item Each fitted anomalous segment is close (in the sense of $\mathcal{B}_C$) to the true segment it fits. 
\end{enumerate}

We will prove these properties in the following order: First we will prove property II, then IV, then III, and then I. We will then use these to prove Proposition \ref{Prop:MAIN}. In the subsequent proofs we will use a certain number of technical Lemmata, all proved in the supplementary material. 

Throughout these proofs we will use the following two lemmata. The first one describes the increase in un-penalised cost incurred by splitting a fitted segment into two fitted segments and the second one bounds this increase in penalised cost for splitting fitted dense collective anomalies. 

\begin{Lemma}\label{lemma:splitsaving}
	Let $i \leq j' < j'+1 \leq j$. The following property is satisfied for all $\textbf{J}$ 
	\begin{equation*}
	\mathcal{S}\left(\textbf{x}_{i:j'},\textbf{J}\right) + \mathcal{S}\left(\textbf{x}_{(j'+1):j},\textbf{J}\right) = \mathcal{S}\left(\textbf{x}_{i:j},\textbf{J}\right) +\sum_{ c \in \textbf{J}}\left( \frac{(j'-i+1)(j-j')}{j-i+1}\left(\bar{\textbf{x}}_{i:j'}^{(c)} - \bar{\textbf{x}}_{(j'+1):j}^{(c)} \right)^2\right)
	\end{equation*}
\end{Lemma}

\begin{Lemma}\label{lemma:splitlemma}
	Let $i \leq j' < j'+1 \leq j$ The following holds given $E$ 
	\begin{equation*}
	\mathcal{C}\left(\textbf{x}_{i:j'},\textbf{1}\right) +\mathcal{C}\left(\textbf{x}_{(j'+1):j},\textbf{1}\right) \leq  \mathcal{C}\left(\textbf{x}_{i:j},\textbf{1}\right) + C\psi + C\sqrt{p\psi} + 2\sqrt{p\psi},
	\end{equation*}
	provided $C$ exceeds some global constant. 
\end{Lemma}

We will also use the following lemma which shows that merging two adjacent fitted collective anomalies which are both contained within a true anomalous segment reduces the penalised cost substantially. 
\begin{Lemma}\label{lemma:merging reduces cost}
	Let $i$, $j'$, and $j$ be such that there exists a $k$ such that $s_k < i \leq j' < j'+1 \leq j \leq e_k$. Then, 
	\begin{equation*}
	\mathcal{ C} \left(x_{i:j},\textbf{J}_k\right) \leq \mathcal{ C} \left(x_{i:j'},\textbf{J}_k\right) + \mathcal{ C} \left(x_{(j'+1):j},\textbf{J}_k\right) - \frac{79}{80}C \left( \psi+|\textbf{J}_k|\log(p)\right) 
	\end{equation*}
	and
	\begin{equation*}
	\mathcal{ C} \left(x_{i:j},\textbf{1}\right) \leq \mathcal{ C} \left(x_{i:j'},\textbf{1}\right) + \mathcal{ C} \left(x_{(j'+1):j},\textbf{1}\right) - \frac{79}{80}C \left( \psi+\sqrt{p\psi}\right) 
	\end{equation*}
	when $|\textbf{J}_k| \leq k^*$ and $|\textbf{J}_k| > k^*$ respectively , provided C exceeds some global constant and the event $E$ holds.
\end{Lemma}

The proof of part IV will mostly rely on the following three lemmata. The first one shows that fitting a true collective anomaly as anomalous reduces the penalised cost. The second and third one show that if a fitted sparse or dense collective anomaly contains a large number of observations both from a true anomalous segment and from a typical segment, then removing the typical data from the fitted anomaly reduces the penalised cost.

\begin{Lemma}\label{lemma:NonVSsome}
	Let $i$ and $j$ be such that there exists a $k$ such that $s_k < i \leq j \leq e_k$. Moreover assume that 
	\begin{equation*}
	j-i+1 \geq \frac{4C}{\triangle_k^2}.
	\end{equation*}
	Then given $E$ 
	\begin{equation*}
	\mathcal{C}\left(\textbf{x}_{i:j}, \textbf{J}_k \right) < 0 
	\end{equation*}
	holds if the $k$th anomalous window is sparse; i.e.\ if $|\textbf{J}_k| \leq k^*$; and 
	\begin{equation*}
	\mathcal{C}\left(\textbf{x}_{i:j}, \textbf{1} \right) < 0 
	\end{equation*}
	holds if the $k$th anomalous window is dense; i.e.\ if $|\textbf{J}_k| > k^*$;
	provided $C$ exceeds some global constant and the event $E$ holds.
\end{Lemma}

\begin{Lemma}\label{lemma:SplitCOSTSAVING}
	Let $i$ and $j$ be such that there exists a $k$ such that either of the following holds:
	\begin{enumerate}
		\item $s_k < i \leq j \leq s_{k+1}$ and 
		\begin{equation*}
		\min(e_k - i + 1, j - e_k) \geq \frac{10C}{\triangle_k^2}.
		\end{equation*}
		\item $e_{k-1} < i \leq j \leq e_{k}$ and 
		\begin{equation*}
		\min(s_k - i + 1, j - s_k) \geq \frac{10C}{\triangle_k^2}.
		\end{equation*}
	\end{enumerate}
	Then the corresponding holds given $E$ 
	\begin{enumerate}
		\item if the $k$th anomalous window is sparse; i.e.\ if $|\textbf{J}_k| \leq k^*$
		\begin{equation*}
		\mathcal{C}\left(x_{i:j},\textbf{J}_k\right) \geq 
		\mathcal{C}\left(x_{i:e_k},\textbf{J}_k\right) + 6C(\psi+\log(p))
		\end{equation*}
		if the $k$th anomalous window is dense; i.e.\ if $|\textbf{J}_k| > k^*$
		\begin{equation*}
		\mathcal{C}\left(x_{i:j},\textbf{1}\right) \geq
		\mathcal{C}\left(x_{i:e_k},\textbf{1}\right)  + 6C(\psi+\sqrt{p\psi}) 
		\end{equation*}
		\item if the $k$th anomalous window is sparse; i.e.\ if $|\textbf{J}_k| \leq k^*$
		\begin{equation*}
		\mathcal{C}\left(x_{i:j},\textbf{J}_k\right) \geq \mathcal{C}\left(x_{(s_k+1):j},\textbf{J}_k\right) + 6C(\psi+\log(p))
		\end{equation*}
		if the $k$th anomalous window is dense; i.e.\ if $|\textbf{J}_k| > k^*$
		\begin{equation*}
		\mathcal{C}\left(x_{i:j},\textbf{1}\right) \geq \mathcal{C}\left(x_{(s_k+1):j},\textbf{1}\right) + 6C(\psi+\sqrt{p\psi})
		\end{equation*}
	\end{enumerate}
	provided $C$ exceeds some global constant and the event $E$ holds.
\end{Lemma}

\begin{Lemma}\label{lemma:SplitCOSTSAVINGDENSE}
	Let $i$ and $j$ be such that there exists a $k$ such that the $k$th anomalous window is dense, $|\textbf{J}_k| > k^*$, and either of the following holds:
	\begin{enumerate}
		\item $s_k < i \leq j \leq s_{k+1}$ and 
		\begin{equation*}
		\min(e_k - i + 1, j - e_k) \geq \frac{10C}{\triangle_k^2}.
		\end{equation*}
		\item $e_{k-1} < i \leq j \leq e_{k}$ and 
		\begin{equation*}
		\min(s_k - i + 1, j - s_k) \geq \frac{10C}{\triangle_k^2}.
		\end{equation*}
	\end{enumerate}
	Then the corresponding holds for all $\textbf{J}$ given $E$ 
	\begin{enumerate}
		\item 
		\begin{equation*}
		\mathcal{C}\left(x_{i:j},\textbf{J}\right) \geq
		\mathcal{C}\left(x_{i:e_k},\textbf{1}\right)  + 4C(\psi+\sqrt{p\psi}) 
		\end{equation*}
		\item 
		\begin{equation*}
		\mathcal{C}\left(x_{i:j},\textbf{J}\right) \geq \mathcal{C}\left(x_{(s_k+1):j},\textbf{1}\right) + 4C(\psi+\sqrt{p\psi})
		\end{equation*}
	\end{enumerate}
	provided $C$ exceeds some global constant and the event $E$ holds.
\end{Lemma} 

For Part II, we will require the following six lemmata. The first one proves that merging two fitted collective anomalies contained within a truly anomalous segment reduces the overall penalised cost substantially, even if they are non-adjacent. The second one shows that if a fitted collective anomaly contains both typical and atypical data, then the atypical data can be removed from the fitted collective anomaly without increasing the penalised cost too much. The remaining Lemmata are mostly used to show that if a true anomaly has been fitted using the wrong set of components (i.e.\ fitting a sparse anomaly as a dense one, a dense anomaly as a sparse one, or a sparse anomaly as a sparse anomaly but not with the correct set of components), then it is possible to replace this fitted collective anomaly by one with the right components without increasing the overall penalised cost by too much. 

\begin{Lemma}\label{lemma:merging twice reduces cost}
	Let $i$, $j'$, and $j$ be such that there exists a $k$ such that $s_k < i \leq j' < j''+1 \leq j \leq e_k$. Then, 
	\begin{equation*}
	\mathcal{ C} \left(x_{i:j},\textbf{J}_k\right) \leq \mathcal{ C} \left(x_{i:j'},\textbf{J}_k\right) + \mathcal{ C} \left(x_{(j''+1):j},\textbf{J}_k\right) - \frac{19}{20}C \left( \psi+|\textbf{J}_k|\log(p)\right) 
	\end{equation*}
	and
	\begin{equation*}
	\mathcal{ C} \left(x_{i:j},\textbf{1}\right) \leq \mathcal{ C} \left(x_{i:j'},\textbf{1}\right) + \mathcal{ C} \left(x_{(j''+1):j},\textbf{1}\right) - \frac{19}{20}C \left( \psi+\sqrt{p\psi}\right) 
	\end{equation*}
	when $|\textbf{J}_k| \leq k^*$ and $|\textbf{J}_k| > k^*$ respectively, provided $C$ exceeds some global constant and the event $E$ holds.
\end{Lemma}

\begin{Lemma}\label{lemma:trimming(h)edges}
	Let $i,j$ be such that there exists a $k$ such that $[s_k+1,e_k] \cap [i,j] \neq \emptyset$, $e_{k-1} < i $, and $s_{k+1} \geq j$. Then,
	\begin{equation*}
	\mathcal{C}\left(\textbf{x}_{i':j'} , \textbf{J}\right) - \mathcal{C}\left(\textbf{x}_{i:j} , \textbf{J}\right) \leq 8\psi + 8|\textbf{J}|\log(p)
	\end{equation*}
	for $|\textbf{J}| \leq k^* $and 
	\begin{equation*}
	\mathcal{C}\left(\textbf{x}_{i':j'} , \textbf{1}\right) - \mathcal{C}\left(\textbf{x}_{i:j} , \textbf{1}\right) \leq 8\psi + 8\sqrt{p\psi}
	\end{equation*}	
	where $i' = \max(i,s_k+1)$ and $j' = \min(j,e_k)$ both hold given $E$.
\end{Lemma}

\begin{Lemma}\label{lemma:changeproperty}
	Let $E$ hold and $C$ exceed a global constant. Moreover, let $i$ and $j$ be such that there exists a $k$ such that $s_k < i \leq j \leq e_k$. Then
	\begin{equation*}
	\mathcal{S}(\textbf{x}_{i:j},\textbf{J}) \geq \alpha\left( C\psi + C|\textbf{J}|\log(p)\right)
	\end{equation*}
	for some $\alpha>0$ implies that
	\begin{equation*}
	\sqrt{|\textbf{J}|(j-i+1)\bm{\mu}_k^2} \geq \left( \sqrt{\alpha C} -\sqrt{2}\right)\sqrt{\psi+|\textbf{J}| \log(p)}
	\end{equation*}
	for any sparse $\textbf{J}$.
\end{Lemma}

\begin{Lemma}\label{lemma:Sparse_To_Sparse_CONDITION1}
	Let $i$ and $j$ be such that there exists a $k$ such that $s_k < i \leq j \leq e_k$. If the $k$th anomalous window is sparse; i.e.\ if $|\textbf{J}_k| \leq k^*$; and 
	\begin{equation*}
	\mathcal{S}\left( \textbf{x}_{i:j},\textbf{J}\right) \geq  \frac{19}{20}C\left(|\textbf{J}|\log(p) + \psi\right),
	\end{equation*} 
	then 
	\begin{equation*}
	\mathcal{C}\left( \textbf{x}_{i:j},\textbf{J}_k\right) - \mathcal{C}\left( \textbf{x}_{i:j},\textbf{J}\right) \leq \frac{1}{10}C|\textbf{J}_k|\log(p) + 2\psi
	\end{equation*}
	holds for all sparse $\textbf{J}$, i.e.\ $\textbf{J}$ satisfying $|\textbf{J}| \leq k^*$, if $C$ is larger than some global constant and the event $E$ holds.
\end{Lemma}

\begin{Lemma}\label{lemma:Sparse_To_Dense_CONDITION1}
	Let $i$ and $j$ be such that there exists a $k$ such that $s_k < i \leq j \leq e_k$. If the $k$th anomalous window is dense; i.e.\ if $|\textbf{J}_k| > k^*$; and 
	\begin{equation*}
	\mathcal{S}\left( \textbf{x}_{i:j},\textbf{J}\right) \geq  \frac{19}{20}C\left(|\textbf{J}|\log(p) + \psi\right),
	\end{equation*} 
	then 
	\begin{equation*}
	\mathcal{C}\left( \textbf{x}_{i:j},\textbf{1}\right) - \mathcal{C}\left( \textbf{x}_{i:j},\textbf{J}\right) \leq \frac{1}{10}C\sqrt{p\psi} + 2\psi
	\end{equation*}
	holds for all sparse $\textbf{J}$, i.e.\ $\textbf{J}$ satisfying $|\textbf{J}| \leq k^*$, if $C$ is larger than some global constant and the event $E$ holds.
\end{Lemma}

\begin{Lemma}\label{lemma:Dense_to_sparse}
	Let the event $E$ hold. Moreover, let $i$ and $j$ be such that there exists a $k$ such that $s_k < i \leq j \leq e_k$. Then, if the $k$th anomalous window is sparse; i.e.\ if $|\textbf{J}_k| \leq k^*$; 
	\begin{equation*}
	\mathcal{C}\left( \textbf{x}_{i:j},\textbf{J}_k\right) - \mathcal{C}\left( \textbf{x}_{i:j},\textbf{1}\right) \leq \frac{13}{20} C|\textbf{J}_k|\log(p) - \frac{6}{10} C\sqrt{p\psi} + 2\psi \leq \frac{1}{10} C|\textbf{J}_k|\log(p) - \frac{1}{20} C\sqrt{p\psi} + 2\psi
	\end{equation*}
	holds  if $C$ is larger than some global constant
\end{Lemma}

For Part I we will then require the following lemmata, which are again concerned with bounding the increase in penalised cost for replacing fitted segments with the wrong number of components by fitted segments with the right number of components. 

\begin{Lemma}\label{lemma:Sparse_To_Dense_CONDITION2}
	Let $i$ and $j$ be such that there exists a $k$ such that $s_k < i \leq j \leq e_k$. If the $k$th anomalous window is dense; i.e.\ if $|\textbf{J}_k| > k^*$; and 
	\begin{equation*}
	\mathcal{S}\left( \textbf{x}_{i:j},\textbf{J}\right) \geq  \frac{3}{10}C\left(|\textbf{J}|\log(p) + \psi\right),
	\end{equation*} 
	then 
	\begin{equation*}
	\mathcal{C}\left( \textbf{x}_{i:j},\textbf{1}\right) - \mathcal{C}\left( \textbf{x}_{i:j},\textbf{J}\right) \leq \frac{8}{10}C\sqrt{p\psi} - \frac{6}{10}C|\textbf{J}|\log(p) + 2\psi
	\end{equation*}
	holds for all sparse $\textbf{J}$, i.e.\ $\textbf{J}$ satisfying $|\textbf{J}| \leq k^*$, if $C$ is larger than some global constant and the event $E$ holds..
\end{Lemma}

\begin{Lemma}\label{lemma:Sparse_To_Sparse_CONDITION2}
	Let $i$ and $j$ be such that there exists a $k$ such that $s_k < i \leq j \leq e_k$. If the $k$th anomalous window is sparse; i.e.\ if $|\textbf{J}| \leq k^*$; and 
	\begin{equation*}
	\mathcal{S}\left( \textbf{x}_{i:j},\textbf{J}\right) \geq  \frac{3}{10}C\left(|\textbf{J}|\log(p) + \psi\right),
	\end{equation*} 
	then 
	\begin{equation*}
	\mathcal{C}\left( \textbf{x}_{i:j},\textbf{1}\right) - \mathcal{C}\left( \textbf{x}_{i:j},\textbf{J}\right) \leq \frac{8}{10}C|\textbf{J}_k|\log(p) - \frac{6}{10}C|\textbf{J}|\log(p) + 2\psi
	\end{equation*}
	holds for all sparse $\textbf{J}$, i.e.\ $\textbf{J}$ satisfying $|\textbf{J}| \leq k^*$, if $C$ is larger than some global constant and the event $E$ holds.
\end{Lemma}

\subsubsection{Property III}

We can prove that a fitted segment must overlap with at least one true anomalous segments:
\begin{Proposition}\label{Prop:NOFPs}
	Let the assumptions of Theorem 1 hold. Let $\tau$ be an optimal partition and $E$ hold. Then, $\forall (s,e,\textbf{J}) \in \tau \; \; \; \exists k: [s+1,e] \cap [s_k+1,e_k] \neq \emptyset$, provided $C > 2$.
\end{Proposition}
\textbf{Proof of Proposition \ref{Prop:NOFPs}}: By contradiction: If $(s,e,\textbf{J})$ overlaps with no true anomalous it can be shown that the partition $\tau \setminus (s,e,\textbf{J})$ has lower penalised cost than $\tau$, because of $E_1$ if $\textbf{J}$ is sparse and $E_4$ if $\textbf{J}$ is dense. \qed

\subsubsection{Property IV}

We now prove the following proposition, which shows that if each true anomalous region is fitted by exactly one segment, then the boundaries of that segment are close to the boundaries of the corresponding anomalous region. To this end, we define the set of partitions $\mathcal{T}_1$ as the set of all partitions fitting exactly $K$ anomalous segments in such a way that each fitted anomalous segment overlaps with exactly one true anomalous region and each true anomalous region overlaps with exactly one fitted anomalous segment. More formally, 
\begin{align*}
\mathcal{T}_1 = \{ \tau: |\tau| = K \; &\land \left(\forall (s,e,\textbf{J}) \in \tau \; \; \exists k: s_{k+1} \geq e \; \; \land \; \; e_{k-1} \leq s \; \; 
\land \; \; [s+1,e] \cap [s_k+1,e_k] \neq \emptyset\right) \\&\land (\forall k  \; \; \exists (s,e,\textbf{J}) \in \tau: [s+1,e] \cap [s_k+1,e_k] \neq \emptyset) \}. 
\end{align*}
The following proposition then holds:
\begin{Proposition}\label{Prop:WEAK}
	Let the assumptions of Theorem 1 hold. Given $E$, if a partition $\tau \in \mathcal{T}_1$ is optimal it must also satisfy $\tau \in \mathcal{B}_C$, if $C$ exceeds a global constant.
\end{Proposition}
\textbf{Proof of Proposition \ref{Prop:WEAK}}: Let $\tau$ be optimal. Consider the $k$th true anomalous segment $[s_k+1,e_k]$, which $\tau$ fits with the segment $(\hat{s}_k,\hat{e}_k,\hat{\textbf{J}})$. We begin by showing this fitted segment needs to cover most of the true anomalous region, because otherwise adding an additional segment to $\tau$ would reduce the penalised cost. 

Indeed, $\hat{s}_k \leq s_k + \frac{10C}{\triangle^2_k}$, as otherwise either the partition $\tau \cup (s_k,s_k+\myceil{\frac{10C}{\triangle^2_k}},\textbf{J}_k)$, if the $k$th anomalous segment is sparse, or the partition $\tau \cup (s_k,s_k+\myceil{\frac{10C}{\triangle^2_k}},\textbf{1})$, if the $k$th anomalous segment is dense, would have a lower overall penalised cost than $\tau$ by Lemma \ref{lemma:NonVSsome}, which would contradict the optimality of $\tau$. $\hat{e}_k \geq e_k - \frac{10C}{\triangle^2_k}$ holds by a similar argument. 

The next step consists of showing that $(\hat{s}_k,\hat{e}_k,\hat{\textbf{J}})$ does not extend too far beyond the $k$th anomalous region. Our approach consists of using Lemmata \ref{lemma:SplitCOSTSAVING} and \ref{lemma:SplitCOSTSAVINGDENSE} to show that if this were to happen we could replace $(\hat{s}_k,\hat{e}_k,\hat{\textbf{J}})$ by a different fitted segment in a way which reduces penalised cost. We will just show that $\hat{e}_k \leq e_k + \frac{10C}{\triangle^2_k}$, as a similar argument implies that $\hat{s}_k\geq s_k -\frac{10C}{\triangle^2_k}$.

We already know that  $\hat{s}_k \leq s_k + \frac{10C}{\triangle^2_k}$. Thus, if $\hat{e}_k > e_k + \frac{10C}{\triangle^2_k}$, the segment $[\hat{s}_k+1,\hat{e}_k]$ would contain at least $\myceil{\frac{10C}{\triangle^2_k}}$ observations both from the typical distribution and the $k$th anomalous window. It is possible to show that this partition can be replaced by splitting up $[\hat{s}_k+1,\hat{e}_k]$ in such a way that the penalised cost is reduced.  
\begin{itemize}
	\item If $\textbf{J}_k$ is dense, we can replace $(\hat{s}_k,\hat{e}_k,\hat{\textbf{J}})$ first with $(\hat{s}_k,e_k - \myfloor{\frac{10C}{\triangle^2_k}},\hat{\textbf{J}})$ and $(e_k - \myfloor{\frac{10C}{\triangle^2_k}},\hat{e}_k,\hat{\textbf{J}})$, increasing the penalised cost by no more than $C\psi+C|\textbf{J}|\log(p)$ if $\hat{\textbf{J}}$ is sparse and $C\psi+(C+2)\sqrt{p\psi}$ if $\hat{\textbf{J}} = \textbf{1}$ (By event $E_9$). Lemma \ref{lemma:SplitCOSTSAVINGDENSE} then shows that replacing $(e_k - \myfloor{\frac{10C}{\triangle^2_k}},\hat{e}_k,\hat{\textbf{J}})$ with $(e_k - \myfloor{\frac{10C}{\triangle^2_k}},e_k,\textbf{1})$ reduces the penalised cost by at least $4C\psi+4C|\hat{\textbf{J}}|\log(p)$ if $\hat{\textbf{J}}$ is sparse and $4C\psi+4C\sqrt{p\psi}$ when $\hat{\textbf{J}} = \textbf{1}$ respectively. Chaining these two transformations therefore leads to a reduction in penalised cost contradicting optimality of $\tau$.
	\item If $\textbf{J}_k$ is sparse, the cases $\hat{\textbf{J}} = \textbf{1}$, and $|\hat{\textbf{J}}| \leq k^*$ have to be considered separately. If $\hat{\textbf{J}} = \textbf{1}$,
	\begin{align*}
	&\mathcal{C}\left(\textbf{x}_{(\hat{s}_k+1):\hat{e}_k},\textbf{1} \right) = \mathcal{C}\left(\textbf{x}_{(\hat{s}_k+1):\hat{e}_k},\textbf{J}_k \right) + p +  C\sqrt{p\psi} - \sum_{ c \notin \textbf{J}_k } (\hat{e}_k - \hat{s}_k) \left(\bar{\bm{\eta}}_{(\hat{s}_k+1):\hat{e}_k}\right)^2 + C|\textbf{J}_k|\log(p) \\
	&\geq \mathcal{C}\left(\textbf{x}_{(\hat{s}_k+1):\left(e_k - \myceil{\frac{10C}{\triangle^2_k}} \right)} ,\textbf{J}_k \right) + \mathcal{C}\left(
	\textbf{x}_{\left(e_k - \myfloor{\frac{10C}{\triangle^2_k}} \right):\hat{e}_k}
	,\textbf{J}_k \right) - 2C|\textbf{J}_k|\log(p)  - (C+2)\psi + (C-2) \sqrt{p\psi},s
	\end{align*} 
	with the inequality following from $E_2$ and the fact that splitting a segment does not increase the un-penalised cost. Lemma \ref{lemma:SplitCOSTSAVING}, then shows that the above quantity exceeds
	\begin{equation*}
	\mathcal{C}\left(\textbf{x}_{(\hat{s}_k+1):\left(e_k - \myceil{\frac{10C}{\triangle^2_k}} \right)} ,\textbf{J}_k \right) + \mathcal{C}\left(
	\textbf{x}_{\left(e_k - \myfloor{\frac{10C}{\triangle^2_k}} \right):e_k}
	,\textbf{J}_k \right) +6C\psi + 4C|\textbf{J}_k|\log(p)  - (C+2)\psi + (C-2) \sqrt{p\psi},
	\end{equation*}
	which exceeds 
	\begin{equation*}
	\mathcal{C}\left(\textbf{x}_{(\hat{s}_k+1):\left(e_k - \myceil{\frac{10C}{\triangle^2_k}} \right)} ,\textbf{J}_k \right) + \mathcal{C}\left(
	\textbf{x}_{\left(e_k - \myfloor{\frac{10C}{\triangle^2_k}} \right):e_k}
	,\textbf{J}_k \right),
	\end{equation*}
	thus contradicting the optimality of $\tau$. Similarly, if $|\hat{\textbf{J}}| \leq k^*$, \footnotesize
	\begin{align*} 
	&\mathcal{C}\left(\textbf{x}_{(\hat{s}_k+1):\hat{e}_k},\hat{\textbf{J}} \right) \geq  \mathcal{C}\left(\textbf{x}_{(\hat{s}_k+1):\hat{e}_k},\textbf{J}_k \right) - \sum_{ c \in \hat{\textbf{J}} \setminus \textbf{J}_k } (\hat{e}_k - \hat{s}_k) \left(\bar{\bm{\eta}}_{(\hat{s}_k+1):\hat{e}_k}\right)^2 + C(|\hat{\textbf{J}}| - |\textbf{J}_k|)\log(p) \\
	&\geq\mathcal{C}\left(\textbf{x}_{(\hat{s}_k+1):\left(e_k - \myceil{\frac{10C}{\triangle^2_k}} \right)} ,\textbf{J}_k \right) + \mathcal{C}\left(
	\textbf{x}_{\left(e_k - \myfloor{\frac{10C}{\triangle^2_k}} \right):\hat{e}_k}
	,\textbf{J}_k \right) - 2C|\textbf{J}_k|\log(p)  - C\psi - 2\psi - 2|\hat{\textbf{J}} \setminus \textbf{J}_k |\log(p) + C|\hat{\textbf{J}}|\log(p) \\
	&\geq \mathcal{C}\left(\textbf{x}_{(\hat{s}_k+1):\left(e_k - \myceil{\frac{10C}{\triangle^2_k}} \right)} ,\textbf{J}_k \right) + \mathcal{C}\left(
	\textbf{x}_{\left(e_k - \myfloor{\frac{10C}{\triangle^2_k}} \right):e_k}
	,\textbf{J}_k \right) - 2C\psi - 2C|\textbf{J}_k|\log(p),
	\end{align*}\normalsize 
\end{itemize}
with the second inequality following from $E_1$ and the fact that splitting a segment does not increase the un-penalised cost. The third equality holds for large enough values of $C$. As before, Lemma \ref{lemma:SplitCOSTSAVING} shows that the above quantity exceeds
\begin{equation*}
\mathcal{C}\left(\textbf{x}_{(\hat{s}_k+1):\left(e_k - \myceil{\frac{10C}{\triangle^2_k}} \right)} ,\textbf{J}_k \right) + \mathcal{C}\left(
\textbf{x}_{\left(e_k - \myfloor{\frac{10C}{\triangle^2_k}} \right):e_k}
,\textbf{J}_k \right), 
\end{equation*} thus contradicting the optimality of $\tau$. \qed

\subsubsection{Property II}

We now prove that if all fitted segments of the optimal partition overlap with at most one true anomalous segment, then each true anomalous segment must overlap with exactly one fitted segment. To this end, we now define $\mathcal{T}_2$ as the set of partitions in which each fitted anomalous segment overlaps with exactly one truly anomalous region. More formally we define 
\begin{align*}
\mathcal{T}_2 = \{ \tau: \forall (s,e,\textbf{J}) \in \tau \; \; \exists k: s_{k+1} \geq e \; \; \land \; \; e_{k-1} \leq  s \; \; 
\land \; \; [s+1,e] \cap [s_k+1,e_k] \neq \emptyset \}. 
\end{align*}
and note that $\mathcal{T}_1 \subset \mathcal{T}_2$. The following proposition holds:
\begin{Proposition}\label{Prop:BitStronger}
	Let the assumptions of Theorem 1 hold. Given $E$, if a partition $\tau \in \mathcal{T}_2$ is optimal it must also satisfy $\tau \in \mathcal{T}_1$ if $C$ exceeds a global constant.
\end{Proposition}
\textbf{Proof of Proposition \ref{Prop:BitStronger}}: The proof has two parts: 
\begin{enumerate}
	\item We need to show that the optimality of $\tau$ implies that each true anomalous segment overlaps with at least one fitted segment in $\tau$. 
	\item We need to show that the optimality of $\tau$ implies that each true anomalous segment overlaps with at most one fitted segment in $\tau$.
\end{enumerate} 

We prove both statements by contradiction: First assume that $\tau$ is optimal but that there exists a $k$ such that $[s_k+1,e_k]$ is not covered at all by any fitted segment in $\tau$. Then by Lemma \ref{lemma:NonVSsome}, the partition $\tau \cup (s_k,e_k,\textbf{J}_k)$, if the $k$th change is sparse, or $\tau \cup (s_k,e_k,\textbf{1})$, if the $k$th change is dense, has a lower penalised cost than $\tau$, so contradicting its optimality. 

Now assume that there exists a $k$ such that $\tau$ contains two or more fitted segments overlapping with $[s_k+1,e_k]$. We will show that it is possible to merge any two fitted segments (called $(a,b,\textbf{J}_1), (c,d,\textbf{J}_2)$, where $c\geq b$ without loss of generality) in a way which reduces the total penalised cost thereby contradicting the optimality of $\tau$. In order to do so, we define $a' = \max(s_k,a)$ and $d' = \min(e_k,d)$.
The following two cases have to be considered separately, but share in the following idea: Merging  $(a,b,\textbf{J}_1), (c,d,\textbf{J}_2)$, into a single fitted segment increases the un-penalised cost by at most $O(\sqrt{C})$. At the same time merging reduces the penalty by $O(C)$. Hence, if $C$ is large enough, merging reduces the overall penalised cost.

\textbf{1. $\textbf{J}_k$ is dense} : We will show that replacing $(a,b,\textbf{J}_1), (c,d,\textbf{J}_2)$ with $(a',d',\textbf{1})$ reduces the penalised cost. Lemma \ref{lemma:merging twice reduces cost}, implies that it is sufficient to show that 
\begin{equation*}
\mathcal{ C}(\textbf{x}_{a':b},\textbf{1}) - \mathcal{ C}(\textbf{x}_{a:b},\textbf{J}_1) \leq \frac{5}{40}C \left(\psi+\sqrt{p\psi}\right) \; \; \; \; \; \text{and} \; \; \; \; \; \mathcal{ C}(\textbf{x}_{c:d'},\textbf{1}) - \mathcal{ C}(\textbf{x}_{c:d},\textbf{J}_2) \leq \frac{5}{40}C \left(\psi+\sqrt{p\psi}\right).
\end{equation*}
We limit ourselves to proving the first statement, as the second one can be proven via a symmetrical argument. If $\textbf{J}_1=\textbf{1}$, the statement follows directly from Lemma \ref{lemma:trimming(h)edges}. If $|\textbf{J}_1| \leq k^*$, we first note that Lemma \ref{lemma:trimming(h)edges} implies that 
\begin{equation} \label{eq:Dummy1}
\mathcal{ C}(x_{a':b},\textbf{J}_1) \leq \mathcal{ C}(\textbf{x}_{a:b},\textbf{J}_1) + 8\psi + 8|\textbf{J}_1|\log(p) \leq \mathcal{ C}(\textbf{x}_{a:b},\textbf{J}_1) + 8\psi + 8\sqrt{p\psi}
\end{equation}
By optimality of $\tau$, $\mathcal{ C}(\textbf{x}_{a:b},\textbf{J}_1)<0$, must hold. This implies that
\begin{equation*}
\mathcal{S}(x_{a':b},\textbf{J}_1) \geq \frac{19}{20}C\left( \psi + |\textbf{J}|\log(p) \right).
\end{equation*}
Consequently, Lemma \ref{lemma:Sparse_To_Dense_CONDITION1} shows that 
\begin{equation} \label{eq:Dummy2}
\mathcal{ C}(x_{a':b},\textbf{1}) \leq \mathcal{ C}(x_{a':b},\textbf{J}_1) + \frac{1}{10}C\left(\psi+\sqrt{p\psi}\right).
\end{equation}
Combining (\ref{eq:Dummy1}) and (\ref{eq:Dummy2}) finishes the proof. \qed

\textbf{2. $\textbf{J}_k$ is sparse} : We will show that replacing $(a,b,\textbf{J}_1), (c,d,\textbf{J}_2)$ with $(a',d',\textbf{J}_k)$ reduces the penalised cost. Lemma \ref{lemma:merging twice reduces cost}, implies that it is sufficient to show that 
\begin{equation*}
\mathcal{ C}(\textbf{x}_{a':b},\textbf{J}_k) - \mathcal{ C}(\textbf{x}_{a:b},\textbf{J}_1) \leq \frac{5}{40}C \left(\psi+|\textbf{J}_k|\log(p)\right) \; \; \; \; \; \text{and} \; \; \; \; \; \mathcal{ C}(\textbf{x}_{c:d'},\textbf{J}_k) - \mathcal{ C}(\textbf{x}_{c:d},\textbf{J}_2) \leq \frac{5}{40}C \left(\psi+|\textbf{J}_k|\log(p)\right).
\end{equation*}
These proofs for both statements are symmetrical. We therefore only prove the first one. As before we begin by considering the case $\textbf{J}_1 = \textbf{1}$. We have 
\begin{align*}
\mathcal{ C}(\textbf{x}_{a':b},\textbf{J}_k) &= \mathcal{ C}(\textbf{x}_{a:b},\textbf{1}) + \left(\mathcal{ C}(\textbf{x}_{a':b},\textbf{1}) -  \mathcal{ C}(\textbf{x}_{a:b},\textbf{1}) \right) + \left(\mathcal{ C}(\textbf{x}_{a':b},\textbf{J}_k) -  \mathcal{ C}(\textbf{x}_{a':b},\textbf{1}) \right) \\
&\leq \mathcal{ C}(\textbf{x}_{a:b},\textbf{1}) + \left(8\psi+8\sqrt{p\psi}\right) + \left(2\psi + \frac{1}{10}C|\textbf{J}_k|\log(p) - \frac{1}{20} C \sqrt{p\psi}\right) \\
&\leq \mathcal{ C}(\textbf{x}_{a:b},\textbf{1}) + 10\psi + \frac{1}{10}C|\textbf{J}_k|\log(p),
\end{align*}
where the first inequality follows from Lemmata \ref{lemma:trimming(h)edges} and \ref{lemma:Dense_to_sparse}, while the second inequality golds if $C$ exceeds a fixed constant. Turning to the case in which $|\textbf{J}_1| \leq  k^*$, we note that the same strategy of proof used for the case in which $\textbf{J}_k$ is dense can be reapplied, the only difference being that Lemma \ref{lemma:Sparse_To_Sparse_CONDITION1} has to be used instead of Lemma \ref{lemma:Sparse_To_Dense_CONDITION1}. \qed

\subsubsection{Property I}

We will now prove that an optimal partition can not contain a fitted segment overlapping with more than one true anomalous segment. We formalise this in the following Proposition: 
\begin{Proposition}\label{Prop:PropertyI}
	Let the assumptions of Theorem 1 hold. Let $\tau$ be an optimal partition. Then, $\tau \in \mathcal{T}_2$, given that the event $E$ holds and that the constant $C$ exceeds a global constant.
\end{Proposition}
Note that this result trivially holds when $K=1$. In order to prove this proposition, we will use a variation of Proposition \ref{Prop:BitStronger}. For this we introduce the set of fitted sparse segments, which either begin or end at the start of a true anomalous segment and only contain a small fraction of the true anomalous segment 
\begin{equation*}
\mathcal{A}_1 = \left\{ (s,e,\textbf{J}) : |\textbf{J}| < k^* \; \land \; \exists k: \left(s=s_k \; \land \; e \leq s_k + \frac{10C}{\triangle_k^2}\right) \lor \left(e=e_k \; \land \; s \geq e_k - \frac{10C}{\triangle_k^2}\right)  \right\},
\end{equation*}
as well as its analogue for dense changes
\begin{equation*}
\mathcal{A}_2 = \left\{ (s,e,\textbf{1}) : \exists k: \left(s=s_k \; \land \; e \leq s_k + \frac{10C}{\triangle_k^2}\right) \lor \left(e=e_k \; \land \; s \geq e_k - \frac{10C}{\triangle_k^2}\right)  \right\}.
\end{equation*}
The following two propositions can then be proven 
\begin{Proposition}\label{Prop:Helper}
	Let the assumptions of Theorem 1 hold. Let $\tau' \in \mathcal{T}_2$ and $E$ hold true. Then there exists another partition $\tau'' \in \mathcal{T}_2$ such that 
	\begin{equation*}
	\mathcal{C}\left(\textbf{x}_{1:n},\tau''\right) \leq \mathcal{C}\left(\textbf{x}_{1:n},\tau'\right) - \frac{6}{10}\left( \sum _{(s,e,\textbf{J}) \in \tau' \cap \mathcal{A}_1 }  \left( C\psi + C|\textbf{J}|\log(p)\right)  + \sum _{(s,e,\textbf{1}) \in \tau' \cap \mathcal{A}_2 }\left( C\psi + C\sqrt{p\psi}\right) \right),
	\end{equation*}
	if $C$ exceeds a global constant.
\end{Proposition}
\begin{Proposition}\label{Prop:FINAL_PROP}
	Let the assumptions of Theorem 1 hold. Let $\tau$ be an optimal partition and $E$ hold true. Then, there exists a partition $\tau' \in \mathcal{T}_2$ such that
	\begin{equation*}
	\mathcal{C}\left(\textbf{x}_{1:n},\tau'\right) \leq \mathcal{C}\left(\textbf{x}_{1:n},\tau\right) + \frac{11}{20}\left( \sum _{(s,e,\textbf{J}) \in \tau \cap \mathcal{A}_1 }  \left( C\psi + C|\textbf{J}|\log(p)\right)  + \sum _{(s,e,\textbf{1}) \in \tau \cap \mathcal{A}_2 }\left( C\psi + C\sqrt{p\psi}\right) \right),
	\end{equation*}
	with equality if and only if $\tau \in \mathcal{T}_2$, if $C$ exceeds a global constant.
\end{Proposition}

Note that Proposition \ref{Prop:Helper} does not assume that $\tau'$ is optimal. Using these two propositions it is easy to derive the following: 

\textbf{Proof of Proposition \ref{Prop:PropertyI}}: Assume that the optimal partition $\tau $ is such that $\tau \notin \mathcal{T}_2$. Then, by Proposition \ref{Prop:FINAL_PROP}  there exists a partition $\tau' \in \mathcal{T}_2$ such that
\begin{equation*}
\mathcal{C}\left(\textbf{x}_{1:n},\tau\right) > \mathcal{C}\left(\textbf{x}_{1:n},\tau'\right) - \frac{11}{20}\left( \sum _{(s,e,\textbf{J}) \in \tau \cap \mathcal{A}_1 }  \left( C\psi + C|\textbf{J}|\log(p)\right)  + \sum _{(s,e,\textbf{1}) \in \tau \cap \mathcal{A}_2 }\left( C\psi + C\sqrt{p\psi}\right) \right),
\end{equation*}
Moreover, Proposition \ref{Prop:Helper} implies that there exists another partition $\tau'' \in \mathcal{T}_2$ such that
\begin{equation*}
\mathcal{C}\left(\textbf{x}_{1:n},\tau'\right) \geq \mathcal{C}\left(\textbf{x}_{1:n},\tau''\right) +  \frac{6}{10}\left( \sum _{(s,e,\textbf{J}) \in \tau' \cap \mathcal{A}_1 }  \left( C\psi + C|\textbf{J}|\log(p)\right)  + \sum _{(s,e,\textbf{1}) \in \tau' \cap \mathcal{A}_2 }\left( C\psi + C\sqrt{p\psi}\right) \right),
\end{equation*}
Consequently, 
\begin{equation*}
\mathcal{C}\left(\textbf{x}_{1:n},\tau\right) > \mathcal{C}\left(\textbf{x}_{1:n},\tau''\right), 
\end{equation*}
which contradicts the optimality of $\tau$. \qed 

\textbf{Proof of Proposition \ref{Prop:Helper}}: Proposition \ref{Prop:BitStronger} shows that fitting an anomalous region with two segments, or with one very short segment leaving most of the anomalous region uncovered is sub-optimal. This proposition goes further by showing it is suboptimal by at least $O(\frac{6}{10}C)$. Crucially, this is larger than $O(\frac{1}{2}C)$ and will help us break up fitted segments spanning multiple anomalous regions. The proof of this Proposition is similar in flavour to the proof of the second part of Proposition \ref{Prop:BitStronger}. The main idea is that there are at most two fitted partitions $\in \tau' \cap \left( \mathcal{A}_1 \cup \mathcal{A}_2 \right)$ overlapping with the $k$th true anomalous region. These partitions therefore leave at least $\frac{20C}{\triangle_k^2}$ of the $k$th anomalous region uncovered. Therefore, if no other segment in $\tau'$ overlaps with the $k$th anomalous region, one can be added without increasing the penalised cost. It can then be merged with the fitted partitions in $\in \tau' \cap \left( \mathcal{A}_1 \cup \mathcal{A}_2 \right)$ and overlap with the $k$th true anomalous region. This yields a new partition still in $\mathcal{T}_2$ with the claimed reduction in penalised cost.

Since $\tau' \in \mathcal{T}_2$, we can consider each of the $K$ true anomalous regions separately. We define the set of fitted segments in $\tau'$ which overlap with the $k$th anomalous region to be
\begin{equation*}
\tau'_k = \left\{(s,e,\textbf{J}) \in \tau' : [s+1,e] \cap [s_k+1,e_k] \neq \emptyset  \right\}.
\end{equation*}
Proving the full result is therefore equivalent to proving the existence of a $\tau''_k$ which yields the required reduction in penalised cost. The following 3 cases are possible:

\begin{enumerate}
	\item $ |\tau'_k \cap \left(\mathcal{A}_1 \cup \mathcal{A}_2 \right)| = 0$, which happens when $\tau'$ does not contain a short fitted segment at either the beginning or the end of the $k$th anomalous region. No further transformation is required in this case, i.e.\ $\tau''_k = \tau'_k$
	\item $ |\tau'_k \cap \left(\mathcal{A}_1 \cup \mathcal{A}_2 \right)| = 1$.
	\item $ |\tau'_k \cap \left(\mathcal{A}_1 \cup \mathcal{A}_2 \right)| = 2$.
\end{enumerate}

We will only explicitly describe the transformation for the second case, as applying it twice yields a transformation for the third case. Without loss of generality we further assume that $\tau'_k \cap \left(\mathcal{A}_1 \cup \mathcal{A}_2 \right) = (s,e_k,\textbf{J})$, i.e.\ that the short fitted segment lies at the end of the $k$th anomalous window. A first special case can be treated very quickly. If $|\textbf{J}| \leq k^*$ and $\mathcal{ C}\left(\textbf{x}_{(s+1):e_k},\textbf{J}\right) \geq \frac{6}{10}C(\psi+|\textbf{J}|\log(p))$, removing $(s,e_k,\textbf{J})$ from $\tau'_k$ is sufficient. If $|\textbf{J}| \leq k^*$ and $\mathcal{ C}\left(\textbf{x}_{(s+1):e_k},\textbf{J}\right) < \frac{6}{10}C(\psi+|\textbf{J}|\log(p))$, we nevertheless have 
\begin{equation*}
\mathcal{ C}\left(\textbf{x}_{(s+1):e_k},\textbf{J}_k\right) \leq \mathcal{ C}\left(\textbf{x}_{(s+1):e_k},\textbf{J}\right) + \frac{8}{10}C|\textbf{J}_k|\log(p) - \frac{6}{10}C|\textbf{J}|\log(p) + 2\psi
\end{equation*}
if $\textbf{J}_k$ is sparse and 
\begin{equation*}
\mathcal{ C}\left(\textbf{x}_{(s+1):e_k},\textbf{1}\right) \leq \mathcal{ C}\left(\textbf{x}_{(s+1):e_k},\textbf{J}\right) + \frac{8}{10}C\sqrt{p\psi} - \frac{6}{10}C|\textbf{J}|\log(p) + 2\psi
\end{equation*}
if $\textbf{J}_k$ is dense, by Lemmata \ref{lemma:Sparse_To_Sparse_CONDITION2} and \ref{lemma:Sparse_To_Dense_CONDITION2} respectively. Similarly, if $\textbf{J} = \textbf{1}$ we have that
\begin{equation*}
\mathcal{ C}\left(\textbf{x}_{(s+1):e_k},\textbf{J}_k\right) \leq \mathcal{ C}\left(\textbf{x}_{(s+1):e_k},\textbf{J}\right) + \frac{8}{10}C|\textbf{J}_k|\log(p) - \frac{6}{10}C\sqrt{p\psi}  + 2\psi
\end{equation*}
if $\textbf{J}_k$ is sparse as a direct consequence of Lemma \ref{lemma:Dense_to_sparse} and, trivially,
\begin{equation*}
\mathcal{ C}\left(\textbf{x}_{(s+1):e_k},\textbf{1}\right) \leq \mathcal{ C}\left(\textbf{x}_{(s+1):e_k},\textbf{J}\right) + \frac{8}{10}C\sqrt{p\psi} - \frac{6}{10}C\sqrt{p\psi} + 2\psi
\end{equation*}
if $\textbf{J}_k$ is dense. 

Consequently, if the next fitted change in $\tau'_k$ to the left of $(s,e,\textbf{J})$ is of the form $(\tilde{s},\tilde{e}   ,\textbf{J}_k)$, if $\textbf{J}_k$ is sparse or $(\tilde{s},\tilde{e},\textbf{1})$ if $\textbf{J}_k$ is dense, for some $\tilde{s} \geq s_k$, Lemma \ref{lemma:merging twice reduces cost} shows that the required reduction in penalised cost can be obtained by merging these two fitted segments. If there is no other fitted change in $\tau'_k$, or if the next fitted segment in $\tau'_k$ to the left of $(s,e,\textbf{J})$ is $(\tilde{s},\tilde{e}   ,\textbf{J})$, where $\tilde{e}$ satisfies $s - \tilde{e} \geq \frac{10C}{\triangle_{k}^2}$, Lemma \ref{lemma:NonVSsome} implies that adding $(s - \myceil{\frac{10C}{\triangle_{k}^2}},s   ,\textbf{J}_k)$, if $\textbf{J}_k$ is sparse or $(s -\myceil{ \frac{10C}{\triangle_{k}^2}},s,\textbf{1})$ if $\textbf{J}_k$ is dense, does not increase the penalised cost. Lemma \ref{lemma:merging twice reduces cost} can then be applied as before to show that merging this new fitted segment with $(s,e,\textbf{J})$ yields a new partition exhibiting the required reduction in penalised cost. 

Hence, in order to finish proving the result we only need to show that any $(\tilde{s},\tilde{e},\textbf{J}) \in \tau_k'$ can either be removed without increasing the penalised cost or replaced by $(\max(\tilde{s},s_k),\tilde{e},\textbf{J}_k)$ in a way which increases the penalised cost by at most $\frac{5}{40}C(|\textbf{J}_k|\log(p)+\psi)$ if $\textbf{J}_k$ is  sparse or $(\max(\tilde{s},s_k),\tilde{e},\textbf{1})$ in a way which increases the penalised cost by at most $\frac{5}{40}C\left(\sqrt{p\psi} + \psi \right) $ if $\textbf{J}_k$ is dense. This however, was already shown in the proof of Proposition \ref{Prop:BitStronger}. This finishes the proof. \qed

\textbf{Proof of Proposition \ref{Prop:FINAL_PROP}:} If $\tau \in \mathcal{T}_2$, the result trivially holds. In order to prove the result when $\tau' \notin \mathcal{T}_2$, we consider all possible fitted segments $(s,e,\textbf{J}) \in \tau \setminus \mathcal{T}_2$ which overlap with at least two anomalous regions and show that 
\begin{enumerate}
	\item No such segment can overlap a true fitted dense change, the $k'$th say, by more than $\frac{10C}{\triangle_k'^2}$ as this would contradict the optimality of $\tau$.
	\item All other fitted segments, overlapping with at least two anomalous regions, including, potentially, a certain number of sparse changes by more that $\frac{10C}{\triangle}$ can be replaced by fitted segments each overlapping with exactly one true anomalous segment in a way which strictly bounds the increase in penalised cost as stipulated by the proposition.
\end{enumerate}

\textbf{1)} First of all we can show that the optimality of $\tau$ implies that no partition $(s,e,\textbf{J})\in \tau \setminus \mathcal{T}_2 $ can overlap a dense change (the $k'$th change say) by more than $\frac{10C}{\triangle_k'^2}$. Otherwise, the interval $[s+1,e]$ would also contain at least $\frac{10C}{\triangle_k'^2}$ observations belonging to the typical distribution. We could therefore split it up into three segments (increasing the penalised cost by at most $2C\psi+2C|\textbf{J}|\log(p)$ or $2C\psi+2(C+2)\sqrt{p\psi}$), one of which containing exactly $\myceil{\frac{10C}{\triangle_k'^2}}$ of observations belonging to the typical distribution and $\myceil{\frac{10C}{\triangle_k'^2}}$ of observations belonging to the $k'$th anomalous window. Lemma \ref{lemma:SplitCOSTSAVINGDENSE} shows that such a segment can be replaced in a way which reduces the penalised cost by at least $4C\psi+4C\sqrt{p\psi}$. Overall, we would thus obtain a new partition with a lower penalised cost than $\tau$ contradicting the optimality of $\tau$. 

\textbf{2)} Consider now, a segment $(s,e,\textbf{J})\in \tau \setminus \mathcal{T}_2 $ not overlapping with any dense changes by more than $\frac{10C}{\triangle_k^2}$. For this segment define the set of true anomalous segments it overlaps by more than $\frac{10C}{\triangle_k^2}$ to be 
\begin{equation*}
\mathcal{D}_{e,s} := \left\{k: \left| [s+1,e] \cap [s_{k}+1,e_k+1] \right| \geq \frac{10C}{\triangle_k^2} \right\}.
\end{equation*}
and note that $|\textbf{J}_k|$ is sparse if $k \in \mathcal{D}_{s,e}$ for some $(s,e,\textbf{J})\in \tau \setminus \mathcal{T}_2 $. We have to consider the following 4 scenarios
\begin{enumerate}
	\item The beginning of the fitted segment $(s,e,\textbf{J})\in \tau \setminus \mathcal{T}_2 $ overlaps with a true anomalous region $[s_{k'}+1,e_{k'}]$, but does so by less than $\frac{10C}{\triangle_{k'}^2}$. i.e. $\exists k': e_{k'} - \frac{10C}{\triangle_{k'}^2} \leq s+1 \leq e_{k'}$.
	\item The end of the fitted segment $(s,e,\textbf{J})\in \tau \setminus \mathcal{T}_2 $ overlaps with a true anomalous region $[s_{k''}+1,e_{k''}]$, but does so by less than $\frac{10C}{\triangle_{k''}^2}$. i.e. $\exists k'': s_{k''} +1 + \frac{10C}{\triangle_{k''}^2} \geq e \geq s_{k''}+1$.
	\item Both apply
	\item None of 1 and 2 apply. Note that this allows for the beginning and or the end of $(s,e,\textbf{J})\in \tau \setminus \mathcal{T}_2 $ to lie in a truly anomalous region provided the overlap with that region exceeds the critical threshold  of $\frac{10C}{\triangle^2}$.
\end{enumerate}
We then replace $(s,e,\textbf{J})$ in $\tau$ to obtain a new partition $\tilde{\tau}$.  depending on the cases above we define $\tilde{\tau}$ to be
\begin{enumerate}
	\item \begin{equation*}
	\left(\tau \setminus \{(s,e,\textbf{J})\}\right) \cup \{(s,e_{k'},\textbf{J})\} \cup \left(\bigcup\limits_{k \in \mathcal{D}_{e,s}} \{\left(s_k,e_k,\textbf{J}_k \right)\} \right)
	\end{equation*}
	\item \begin{equation*}
	\left(\tau \setminus \{(s,e,\textbf{J})\}\right) \cup \left(\bigcup\limits_{k \in \mathcal{D}_{e,s}} \{\left(s_k,e_k,\textbf{J}_k \right)\}\right) \cup \{ (s_{k''},e,\textbf{J}) \}
	\end{equation*}
	\item \begin{equation*}
	\left(\tau \setminus \{(s,e,\textbf{J})\}\right) \cup \{(s,e_{k'},\textbf{J})\} \cup \left(\bigcup\limits_{k \in \mathcal{D}_{e,s}} \{\left(s_k,e_k,\textbf{J}_k \right)\} \right)\cup \{ (s_{k''},e,\textbf{J}) \}
	\end{equation*}
	\item \begin{equation*}
	\left(\tau \setminus \{(s,e,\textbf{J})\}\right) \cup\bigcup\limits_{k \in \mathcal{D}_{e,s}} \{\left(s_k,e_k,\textbf{J}_k \right)\}
	\end{equation*}
\end{enumerate}
depending on which case applies. The main effect of this transformation is the same across all cases: It results in all true anomalous regions contained in $(s,e,\textbf{J})$ to be fitted separately and according to the ground truth. Only the number of fitted segments belonging to $\mathcal{A}_1$ and/or $\mathcal{A}_2$ depends on the case.
Since applying this transformation for all $(s,e,\textbf{J}) \in \tau \setminus \mathcal{T}_2$ leads to a new partition $\tau'$ which is contained in $\mathcal{T}_2$, it is sufficient to prove that each transformation individually increases the penalised cost by strictly less than
\begin{enumerate}
	\item $\frac{11}{20}C\left(\psi + |\textbf{J}|\log(p)\right)$ if $\textbf{J}$ is sparse or $\frac{11}{20}C\left(\psi + \sqrt{p\psi}\right)$ if $\textbf{J}$ is dense.
	\item $\frac{11}{20}C\left(\psi + |\textbf{J}|\log(p)\right)$ if $\textbf{J}$ is sparse or $\frac{11}{20}C\left(\psi + \sqrt{p\psi}\right)$ if $\textbf{J}$ is dense.
	\item $\frac{22}{20}C\left(\psi + |\textbf{J}|\log(p)\right)$ if $\textbf{J}$ is sparse or $\frac{22}{20}C\left(\psi + \sqrt{p\psi}\right)$ if $\textbf{J}$ is dense.
	\item $0$
\end{enumerate}
depending on the case in order to prove the proposition. The fourth case follows directly from the following Lemma: 

\begin{Lemma}\label{lemma:MASSIVELEMMA!}
	Let the event $E$ hold and $C$ exceed some global constant. Let $s$ and $e$ be such the fourth scenario applies, i.e.
	\begin{enumerate}
		\item $\nexists k': e_{k'} - \frac{10C}{\triangle_{k'}^2} \leq s+1 \leq e_{k'}$.
		\item $\nexists k'': s_{k''} +1 + \frac{10C}{\triangle_{k''}^2} \geq e \geq s_{k''}+1$
	\end{enumerate}	
	Then, the following holds true for all sparse $\textbf{J}$
	\begin{equation*}
	\mathcal{C} \left( \textbf{x}_{s,e}, \textbf{J}\right) \geq \frac{19}{20}C \left(\psi+|\textbf{J}|\log(p)\right) + \sum_{k \in \mathcal{D}_{s,e}} \left( \mathcal{C} \left( \textbf{x}_{(s_k+1):e_k},\textbf{J}_k\right)\right)
	\end{equation*}
	Moreover, the following statement is also true:
	\begin{equation*}
	\mathcal{C} \left( \textbf{x}_{s,e}, \textbf{1}\right) \geq \frac{19}{20}C \left(\psi+\sqrt{p\psi}\right) + \sum_{k \in \mathcal{D}_{s,e}} \left( \mathcal{C} \left( \textbf{x}_{(s_k+1):e_k},\textbf{J}_k\right)\right).
	\end{equation*}
\end{Lemma}

This Lemma can also be used to bound the increase in penalised cost obtained for the other three cases. The only difference is that $(s,e,\textbf{J})$ is first split up to twice in order to remove the short overlap with the true anomalous region at the beginning and/or the end. For the sake of brevity, we limit ourselves to write out the proof for the third case, for which the result is tightest. If, $\textbf{J}$ is sparse, we have that
\begin{align*}
&\mathcal{C}\left(x_{(s+1):e},\textbf{J}\right) \geq \mathcal{C}\left(x_{(s+1):e_{k'}},\textbf{J}\right) + \mathcal{C}\left(x_{(e_{k'}+1):s_{k''}},\textbf{J}\right) + 
\mathcal{C}\left(x_{(s_{k''}+1):e},\textbf{J}\right) - 2C\left(\psi+|\textbf{J}|\log(p) \right) \\
& > \mathcal{C}\left(x_{(s+1):e_{k'}},\textbf{J}\right) + 
\sum_{k \in \mathcal{D}_{s,e}} \left( \mathcal{C} \left( \textbf{x}_{(s_k+1):e_k},\textbf{J}_k\right)\right)
+ 
\mathcal{C}\left(x_{(s_{k''}+1):e},\textbf{J}\right) - \frac{22}{20}C\left(\psi+|\textbf{J}|\log(p) \right) ,
\end{align*}
where the inequality follows from Lemma \ref{lemma:MASSIVELEMMA!}. Similarly, if, $\textbf{J} = \textbf{1}$ is dense, we have that
\begin{align*}
&\mathcal{C}\left(x_{(s+1):e},\textbf{1}\right) \geq \mathcal{C}\left(x_{(s+1):e_{k'}},\textbf{1}\right) + \mathcal{C}\left(x_{(e_{k'}+1):s_{k''}},\textbf{1}\right) + 
\mathcal{C}\left(x_{(s_{k''}+1):e},\textbf{1}\right) - 2(C+1)\left(\psi+\sqrt{p\psi} \right) \\
&> \mathcal{C}\left(x_{(s+1):e_{k'}},\textbf{1}\right) + 
\sum_{k \in \mathcal{D}_{s,e}} \left( \mathcal{C} \left( \textbf{x}_{(s_k+1):e_k},\textbf{J}_k\right)\right)
+ \mathcal{C}\left(x_{(s_{k''}+1):e},\textbf{1}\right) - \frac{22}{20}C\left(\psi+\sqrt{p\psi} \right),
\end{align*}
where the inequalities follow from Lemma \ref{lemma:MASSIVELEMMA!}, $E_9$, and $C$ exceeding a global constant. This finishes the proof. \qed

\subsubsection{Proof of Proposition \ref{Prop:MAIN}}

\textbf{Proof of Proposition \ref{Prop:MAIN}}: Propositions \ref{Prop:BitStronger}, \ref{Prop:NOFPs}, \ref{Prop:PropertyI}, and \ref{Prop:PropertyI} give the result. 

\subsection{Proofs for Lemmata}

\subsubsection{Proof of Lemma \ref{lemma:Subgamma}}
The MGF of $Z=(x - a)^+$ is given by 
\begin{align*}
\mathbb{E}\left( e^{\lambda Z}\right) = \mathbb{P}\left(\chi_1^2 < a\right) + \int_{a}^{\infty} e^{\lambda (x-a)} \sqrt{\frac{2}{\pi x}} e^{-\frac{1}{2} x} dx =
\mathbb{P}\left(\chi_1^2 < a\right) + \frac{e^{-\lambda a}}{\sqrt{1-2 \lambda}} \mathbb{P}\left(\chi_1^2 > a(1-2\lambda)\right),
\end{align*}
for $0 \leq \lambda \leq 1/2$. Consequently, 
\begin{align*}
\frac{d}{d\lambda} \left( \mathbb{E}\left( e^{\lambda Z}\right)\right) &= 
\frac{2af(a)}{1-2\lambda} + \left( \frac{1}{1-2\lambda} - a\right) \frac{e^{-\lambda a}}{\sqrt{1-2\lambda}} \mathbb{P} \left( \chi_1^2 > a(1-2\lambda)
\right). 
\end{align*}
Evaluating the above at $\lambda = 0$ shows that the mean of $Z$ is indeed $\mu = 2af(a) + (1 - a)\mathbb{P}\left(\chi_1^2 > a \right)$. We therefore have 
\begin{align*}
& \frac{d}{d\lambda}\left( \log \left( \mathbb{E}\left( e^{\lambda Z}\right)\right) \right) -\mu 
= \frac{\frac{d}{d\lambda} \left( \mathbb{E}\left( e^{\lambda Z}\right)\right)}{\mathbb{E}\left( e^{\lambda Z}\right)} - \mu 
= \frac{\frac{2af(a)}{1-2\lambda} + \left( \frac{1}{1-2\lambda} - a\right) \frac{e^{-\lambda a}}{\sqrt{1-2\lambda}} \mathbb{P} \left( \chi_1^2 > a(1-2\lambda)
	\right)}{\mathbb{P}\left(\chi_1^2 < a\right) + \frac{e^{-\lambda a}}{\sqrt{1-2 \lambda}} \mathbb{P}\left(\chi_1^2 > a(1-2\lambda)\right)} - \mu \\
& = \frac{1}{1-2\lambda} \left[ 
\frac{2af(a) + \left( 1 - \left(1-2\lambda\right)a\right) \frac{e^{-\lambda a}}{\sqrt{1-2\lambda}} \mathbb{P} \left( \chi_1^2 > a(1-2\lambda)
	\right)}{\mathbb{P}\left(\chi_1^2 < a\right) + \frac{e^{-\lambda a}}{\sqrt{1-2 \lambda}} \mathbb{P}\left(\chi_1^2 > a(1-2\lambda)\right)} - \left(1-2\lambda\right)\mu 
\right] \\
&= \frac{1}{1-2\lambda} \left[ 
\frac{2af(a) - \left( 1 - a \right)\mathbb{P}\left(\chi_1^2 < a\right) - 2\lambda a \mathbb{P}\left(\chi_1^2 < a\right) }{\mathbb{P}\left(\chi_1^2 < a\right) + \frac{e^{-\lambda a}}{\sqrt{1-2 \lambda}} \mathbb{P}\left(\chi_1^2 > a(1-2\lambda)\right)} 
+ \left( 1 - \left(1-2\lambda\right)a\right)
-\left(1-2\lambda\right)\mu 
\right] \\
&= \frac{1}{1-2\lambda} \left[ 
\frac{\mu - \left( 1 - a \right) - 2\lambda a \mathbb{P}\left(\chi_1^2 < a\right) }{\mathbb{P}\left(\chi_1^2 < a\right) + \frac{e^{-\lambda a}}{\sqrt{1-2 \lambda}} \mathbb{P}\left(\chi_1^2 > a(1-2\lambda)\right)} 
+ \left( 1 - a - \mu \right)
+2(\mu  + a)\lambda
\right].
\end{align*}
Next note that
\begin{equation*}
\mathbb{P} \left( \chi_1^2 > a(1-2\lambda)\right) = \int_{a(1-2\lambda)}^{\infty}\sqrt{\frac{2}{\pi x}} e^{-x/2}dx = 
e^{\lambda a}\int_{a}^{\infty} \sqrt{\frac{y}{y-2\lambda a}}\sqrt{\frac{2}{\pi y}} e^{-y/2}dx.
\end{equation*}
Using the substitution $y = x - 2\lambda a$, shows that
\begin{equation}\label{eq:BoundsOnChisquareBounds}
\mathbb{P} \left( \chi_1^2 > a\right)<e^{-\lambda a}\mathbb{P} \left( \chi_1^2 > a(1-2\lambda)\right)<\frac{\mathbb{P} \left( \chi_1^2 > a\right)}{\sqrt{1-2\lambda}}.
\end{equation}
We can now use this result to further bound the MGF of the truncated $\chi^2_1$. We consider two cases separately: 

\textbf{Case 1}: $\mu - \left( 1 - a \right) - 2\lambda a \mathbb{P}\left(\chi_1^2 < a\right) \geq 0$. The lower bound in \ref{eq:BoundsOnChisquareBounds} shows that $\frac{d}{d\lambda}\left( \log \left( \mathbb{E}\left( e^{\lambda Z}\right)\right) \right) -\mu $ is bounded by 
\begin{align*}
& \frac{1}{1-2\lambda} \left[ 
\frac{\mu - \left( 1 - a \right) - 2\lambda a \mathbb{P}\left(\chi_1^2 < a\right) }{\mathbb{P}\left(\chi_1^2 < a\right) + \frac{1}{\sqrt{1-2 \lambda}} \mathbb{P}\left(\chi_1^2 > a\right)} 
+ \left( 1 - a - \mu \right)
+2(\mu  + a)\lambda 
\right] \\
& \leq \frac{1}{1-2\lambda} \left[ 
\mu - \left( 1 - a \right) - 2\lambda a \mathbb{P}\left(\chi_1^2 < a\right)
+ \left( 1 - a - \mu \right)
+2(\mu  + a)\lambda 
\right] 
= \frac{1}{1-2\lambda} \left[ 2(\mu  + a\mathbb{P}\left(\chi_1^2 > a\right))\lambda  \right] \\
&\leq \frac{2\lambda(1-\lambda)}{(1-2\lambda)^2}(\mu  + a\mathbb{P}\left(\chi_1^2 > a\right)) = \frac{2\lambda(1-\lambda)}{(1-2\lambda)^2}(2af(a)  + \mathbb{P}\left(\chi_1^2 > a\right))
\end{align*}
\textbf{Case 2}: $\mu - \left( 1 - a \right) - 2\lambda a \mathbb{P}\left(\chi_1^2 < a\right) < 0$. The upper bound in \ref{eq:BoundsOnChisquareBounds} shows that $\frac{d}{d\lambda}\left( \log \left( \mathbb{E}\left( e^{\lambda Z}\right)\right) \right) -\mu $ is bounded by \small
\begin{align*}
& \frac{1}{1-2\lambda} \left[ 
\frac{\mu + \left( 1 - a \right) - 2\lambda a \mathbb{P}\left(\chi_1^2 < a\right) }{\mathbb{P}\left(\chi_1^2 < a\right) + \frac{1}{1-2 \lambda} \mathbb{P}\left(\chi_1^2 > a\right)} 
+ \left( 1 - a - \mu \right)
+2(\mu  + a)\lambda 
\right] \\
& = \frac{1}{1-2\lambda} \left[ 
\left( 1 - a - \mu \right) \left( 1 - \frac{1}{\mathbb{P}\left(\chi_1^2 < a\right) + \frac{1}{1-2 \lambda} \mathbb{P}\left(\chi_1^2 > a\right)}\right) 
+2\lambda a \left( 1 - \frac{\mathbb{P}\left(\chi_1^2 < a\right)}{\mathbb{P}\left(\chi_1^2 < a\right) + \frac{1}{1-2 \lambda} \mathbb{P}\left(\chi_1^2 > a\right)}\right) 
+2\lambda \mu
\right] \\
& = \frac{1}{1-2\lambda} \left[ 
\left( 1 - a - \mu \right) \frac{\frac{1}{1-2 \lambda} \mathbb{P}\left(\chi_1^2 > a\right) - \mathbb{P}\left(\chi_1^2 > a\right)}{\mathbb{P}\left(\chi_1^2 < a\right) + \frac{1}{1-2 \lambda} \mathbb{P}\left(\chi_1^2 > a\right)} 
+2\lambda a \left( \frac{\frac{1}{1-2 \lambda} \mathbb{P}\left(\chi_1^2 > a\right)}{\mathbb{P}\left(\chi_1^2 < a\right) + \frac{1}{1-2 \lambda} \mathbb{P}\left(\chi_1^2 > a\right)}\right) 
+2\lambda \mu
\right] \\
& = \frac{1}{1-2\lambda} \left[ 
\left( 1 - a - \mu \right) \frac{2\lambda\mathbb{P}\left(\chi_1^2 > a\right)}{1 - 2\lambda\mathbb{P}\left(\chi_1^2 < a\right) } 
+2\lambda a \frac{ \mathbb{P}\left(\chi_1^2 > a\right)}{1-2\lambda\mathbb{P}\left(\chi_1^2 < a\right)}
+2\lambda \mu
\right]  = \frac{2\lambda}{1-2\lambda} \left[ \mu + (1-\mu) \frac{\mathbb{P}\left(\chi_1^2 > a\right)}{1 - 2\lambda\mathbb{P}\left(\chi_1^2 < a\right)}\right] \\
&= \frac{2\lambda}{(1-2\lambda)^2} \left[ \mu \left( 1- 2\lambda \right)+ (1-\mu)\mathbb{P}\left(\chi_1^2 > a\right) \frac{1-2\lambda}{1 - 2\lambda\mathbb{P}\left(\chi_1^2 < a\right)}\right] \\
&=  \frac{2\lambda}{(1-2\lambda)^2} \left[ \mu \left( 1- 2\lambda \right)+ (1-\mu)\mathbb{P}\left(\chi_1^2 > a\right) - (1-\mu)\mathbb{P}\left(\chi_1^2 > a\right)\frac{2\lambda\mathbb{P}\left(\chi_1^2 > a\right)}{1 - 2\lambda\mathbb{P}\left(\chi_1^2 < a\right)}\right]
\end{align*}
\normalsize
Using the fact that $\mu + \left( 1 - a \right) - 2\lambda a \mathbb{P}\left(\chi_1^2 < a\right) < 0$ and that $1-\mu  \geq 0$, we can bound this by
\begin{align*}
&=  \frac{2\lambda}{(1-2\lambda)^2} \left[ \mu \left( 1- \lambda \right)+ (1-\mu)\mathbb{P}\left(\chi_1^2 > a\right) - 2\lambda a\mathbb{P}\left(\chi_1^2 > a\right)^2\right] \\ 
& = \frac{2\lambda}{(1-2\lambda)^2} \left[ \left(\mu + a\mathbb{P}\left(\chi_1^2 > a\right)\right)\left( 1- \lambda \right)+ (1-\mu - a)\mathbb{P}\left(\chi_1^2 > a\right) - 2\lambda a\mathbb{P}\left(\chi_1^2 > a\right)^2 + a \lambda \mathbb{P}\left(\chi_1^2 > a\right)\right]
\end{align*}
Since $\lambda < \frac{1}{2}$ and $1-a-\mu \leq 0 $, we have that 
\begin{align*}
&(1-\mu - a)\mathbb{P}\left(\chi_1^2 > a\right) - 2\lambda a\mathbb{P}\left(\chi_1^2 > a\right)^2 + a \lambda \mathbb{P}\left(\chi_1^2 > a\right) \leq \lambda \mathbb{P}\left(\chi_1^2 > a\right) \left(    2 \left( 1-a-\mu \right) + a - 2 a\mathbb{P}\left(\chi_1^2 > a\right) \right) \\
&= \lambda \mathbb{P}\left(\chi_1^2 > a\right)  \left(
2 \left( \mathbb{E}\left(\chi_1^2 |\chi_1^2 < a\right)\mathbb{P}\left(\chi_1^2 < a\right) - a\mathbb{P}\left(\chi_1^2 < a\right) \right) + a - 2 a\mathbb{P}\left(\chi_1^2 > a\right)
\right)\\
&= \lambda \mathbb{P}\left(\chi_1^2 > a\right)  \left( 2  \mathbb{E}\left(\chi_1^2 |\chi_1^2 < a\right)\mathbb{P}\left(\chi_1^2 < a\right) - a \right) \leq 0 
\end{align*}
where the last inequality follows from the fact that $\mathbb{E}\left(\chi_1^2 |\chi_1^2 < a\right) \leq a/2$, which is due to the fact that the pdf of the $\chi_1^2$-distribution is decreasing. 

Consequently, 
\begin{equation*}
\frac{d}{d\lambda}\left( \log \left( \mathbb{E}\left( e^{\lambda Z}\right)\right) -\lambda\mu \right)  \leq \frac{2\lambda(1-\lambda)}{(1-2\lambda)^2}(2af(a)  + \mathbb{P}\left(\chi_1^2 > a\right)) = \frac{d}{d\lambda}\left( \frac{2(2af(a)  + \mathbb{P}\left(\chi_1^2 > a\right))\lambda^2}{2(1-2\lambda)} \right).
\end{equation*}
This shows that
\begin{equation*}
\log \left( \mathbb{E}\left( e^{\lambda (Z - \mu )}\right) \right) \leq \frac{2(2af(a)  + \mathbb{P}\left(\chi_1^2 > a\right))\lambda^2}{2(1-2\lambda)},
\end{equation*}
which finishes the proof. 

\subsubsection{Proof of Lemma \ref{lemma:Hazardrate}}
It is sufficient to show that 
\begin{equation*}
\mathbb{P}\left(Y_i \geq a +x | Y_i \geq a , v_i = 1 \right) \geq \mathbb{P} \left( Z > a+x | Z \geq a\right).
\end{equation*}
We have that 
\begin{equation*}
\mathbb{P}\left(Y_i \geq a +x | Y_i \geq a , v_i = 1 \right) = \frac{\mathbb{P}\left(\epsilon_i > \sqrt{a+x} - \mu \right)}{\mathbb{P}\left(\epsilon_i > \sqrt{a} - \mu \right)}
\end{equation*}
The derivative of left hand side with respect to $\mu$ is 
\begin{equation*} \frac{\mathbb{P}\left(\epsilon_i > \sqrt{a+x} - \mu \right)}{\mathbb{P}\left(\epsilon_i > \sqrt{a} - \mu \right)} \left(\frac{\phi(\sqrt{a+x} - \mu)}{\mathbb{P}\left(\epsilon_i > \sqrt{a+x} - \mu \right)} - \frac{\phi(\sqrt{a} - \mu)}{\mathbb{P}\left(\epsilon_i > \sqrt{a} - \mu \right)}\right) 
\end{equation*}
This is greater than 0, since the hazard rate of the Gaussian is increasing. Hence, 
\begin{equation*}
\mathbb{P}\left(Y_i \geq a +x | Y_i \geq a , v_i = 1 \right) = \frac{\mathbb{P}\left(\epsilon_i > \sqrt{a+x} - \mu \right)}{\mathbb{P}\left(\epsilon_i > \sqrt{a} - \mu \right)} \geq \frac{\mathbb{P}\left(\epsilon_i > \sqrt{a+x} \right)}{\mathbb{P}\left(\epsilon_i > \sqrt{a} \right)} = \mathbb{P} \left( Z > a+x | Z \geq a\right).
\end{equation*}

\subsubsection{Proof of Lemma \ref{lemma:LowerboundSubgaussian}}

Let $Z \sim \chi_1^2$ and write $\mu = \mathbb{E}\left(\left(Z-a\right)^+\right)$. The MGF $G(\lambda)$ of the random variable
\begin{equation*}
W = (a-Z)|( Z>a) + \frac{\mu }{\mathbb{P}\left(\chi_1^2 > a\right)} 
\end{equation*}
is then 
\small
\begin{align*}
G(\lambda) = \exp \left( \frac{\lambda\mu }{\mathbb{P}\left(\chi_1^2 > a\right)} \right)\frac{1 }{\mathbb{P}\left(\chi_1^2 > a\right)} \int_{0}^{\infty} \frac{1}{\sqrt{2\pi x}} e^{\lambda a - \lambda zx - \frac{1}{2}x }dx = \exp \left( \frac{\lambda\mu }{\mathbb{P}\left(\chi_1^2 > a\right)} + \lambda a \right) \frac{\mathbb{P}\left(\chi_1^2 > a(1+2\lambda)\right)}{\mathbb{P}\left(\chi_1^2 > a\right) \sqrt{1+2\lambda}}.
\end{align*}
Consequently, 
\begin{align*}
\frac{dG(\lambda)}{d\lambda} = \frac{1 }{\mathbb{P}\left(\chi_1^2 > a\right)}\left[ 
-\frac{2af(a)}{1+2\lambda}e^{-\lambda a} + \left( \frac{\mu }{\mathbb{P}\left(\chi_1^2 > a\right) } +  a - \frac{1}{1+2\lambda}\right)\frac{\mathbb{P}\left(\chi_1^2 > a(1+2\lambda)\right)}{\sqrt{1+2\lambda}}
\right]\exp \left( \frac{\lambda\mu }{\mathbb{P}\left(\chi_1^2 > a\right) } + \lambda a\right)
\end{align*}
and therefore, 
\begin{align*}
\frac{d\log \left(G(\lambda)\right)}{d\lambda} = \frac{\mu }{\mathbb{P}\left(\chi_1^2 > a\right) } +  a - \frac{1}{1+2\lambda} -\frac{2af(a)}{\sqrt{1+2\lambda}}\frac{e^{-\lambda a}}{\mathbb{P}\left(\chi_1^2 > a(1+2\lambda)\right)} 
\end{align*}
Since, 
\begin{align*}
\mathbb{P}\left(\chi_1^2 > a(1+2\lambda)\right) = \int_{ a(1+2\lambda)}^{\infty} \frac{1}{\sqrt{2\pi x}} e^{-\frac{x}{2}} dx =  e^{- \lambda a}\int_{ a}^{\infty} \frac{1}{\sqrt{2\pi y}} \sqrt{\frac{1}{1+2\lambda \frac{a}{y}}} e^{-\frac{y}{2} } dy \leq e^{- \lambda a}\mathbb{P}\left(\chi_1^2 > a\right),
\end{align*}
we must also have 
\begin{align*}
\frac{d\log \left(G(\lambda)\right)}{d\lambda} < \frac{2\lambda}{1+2\lambda} \left( 1+ \frac{2af(a)}{\mathbb{P}\left(\chi_1^2 > a\right)}\right) \leq 2\lambda \left( 1+ \frac{2af(a)}{\mathbb{P}\left(\chi_1^2 > a\right)}\right)
\end{align*}
and therefore
\begin{equation*}
G(\lambda) \leq \frac{\lambda^22\left( 1+ \frac{2af(a)}{\mathbb{P}\left(\chi_1^2 > a\right)}\right)}{2}.
\end{equation*}
This proves that $W$ is sub-Gaussian. Standard tail bounds for sub-Gaussian random variables then imply that independent random variables $W_1,...,W_k$ obeying the same law as $W$ satisfy
\begin{equation*}
\mathbb{P}\left(\sum_{i=1}^k > 2\sqrt{\left( 1+ \frac{2af(a)}{\mathbb{P}\left(\chi_1^2 > a\right)}\right)kt}\right) < e^{-t},
\end{equation*} 
for positive integers $k$ and all $t\in \mathbb{R}$. This finishes the proof. 
\subsubsection{Proof of Lemma \ref{lemma:Bounds_Regime3}}

The equality follows from Lemma \ref{lemma:Subgamma}. To prove the inequality, write $G(\tau) = \tau +2 a f(a) $, where $0 \leq \tau \leq 1$ and $a$ is defined by the equation $\mathbb{P}\left( \chi^2_1 > a\right) = \tau$. Note that $G(0) = 0$ and 
\begin{align*}
\frac{dG}{d\tau} = 1 + \frac{da}{d\tau} \left( f(a)-af(a) \right) = 1 - \frac{1}{f(a)} \left( f(a)-af(a) \right) = a > 0 
\end{align*}

Hence, $m +2paf(a)= pG(\frac{m}{p})$ is increasing in $m$. Moreover the following bounds hold on $a$:
\begin{equation*}
2\tau = 2\mathbb{P} \left( \chi^2_1 > a \right) < \mathbb{P}\left( \chi^2_2 > 2a \right) = \exp(-a).
\end{equation*}
Therefore, we have that
\begin{align*}
G(\tau) \leq \int_{0}^{\tau}-2\log(x) dx = -2\tau \log(\tau) + 2 \tau = 2\tau \log\left(\frac{1}{\tau}\right) + 2 \tau.
\end{align*}
Noting that $m +2paf(a)= pG(\frac{m}{p})$, finishes the proof.

\subsubsection{Proof of Lemma \ref{lemma:Mean_Increasing}}
We know from Lemma \ref{lemma:Subgamma}, that 
\begin{equation*}
\mathbb{E} \left((\chi_1^2 - b)^+|\chi_1^2 > b\right) = 1-b + 2bf(b)\mathbb{P}\left(\chi_1^2 > b\right)^{-1}.
\end{equation*}
Next note that 
\begin{equation*}
\mathbb{P}\left(\chi_1^2 > b\right) = \int_{b}^{\infty} \sqrt{\frac{2}{\pi x}}e^{-x/2} dx \leq \int_{b}^{\infty} \sqrt{\frac{2}{\pi b}}e^{-x/2} dx = 2f(b) 
\end{equation*}
Hence, 
\begin{equation*}
\mathbb{E} \left((\chi_1^2 - b)^+|\chi_1^2 > b\right) = 1-b + 2bf(b)\mathbb{P}\left(\chi_1^2 > b\right)^{-1} \geq 1-b + b = 1.
\end{equation*}
This finishes the proof. 
\subsubsection{Proof of Lemma \ref{lemma:Martingalebound}}

Let $\eta_1,...,\eta_{s+w} \stackrel{i.i.d.}{\sim} N(0,1)$ for some positive integer $s$. Define 
\begin{equation*}
Z_s := \max_{0 \leq a \leq w} (s+a)\left(\bar{\eta}_{1:(s+a)}\right)^2.
\end{equation*}
Write $T_a = \sum_{ t = 1}^a\eta_t$ and note that $e^{\lambda T_a}$ is a super-martingale for all $\lambda > 0$. The following holds: 
\begin{align*}
&\mathbb{P}\left( Z_s > u\right) \leq 
\sum_{i =\flooring{\log_b(s)}}^{\ceiling{\log_b(s+w)}} \mathbb{P} \left( \max_{b^i \leq s+a \leq b^{i+1}} (s+a)\left(\bar{\eta}_{1:(s+a)}\right)^2   > u \right) \leq 
\sum_{i=\flooring{\log_b(s)}}^{\ceiling{\log_b(s+w)}} \mathbb{P} \left( \max_{b^i \leq a' \leq b^{i+1}} \left(T_{a'}\right)^2   > b^{i}u \right) \\
&\leq 2\sum_{i=\flooring{\log_b(s)}}^{\ceiling{\log_b(s+w)}} \mathbb{P} \left( \max_{b^i \leq a' \leq b^{i+1}} T_{a'}   > \sqrt{b^{i}u} \right) = 2\sum_{i=\flooring{\log_b(s)}}^{\ceiling{\log_b(s+w)}} \min_{\lambda}\left[\mathbb{P} \left( \max_{b^i \leq a' \leq b^{i+1}}e^{ \lambda T_{a'} }  > e^{\sqrt{b^{i}u}\lambda} \right)\right] \\
&\leq  2\sum_{i=\flooring{\log_b(s)}}^{\ceiling{\log_b(s+w)}} \min_{\lambda}\left[\mathbb{E} \left(e^{ \lambda T_{\flooring{b^{i+1}}} } \right) e^{-\sqrt{b^{i}u}\lambda} \right] = 2\sum_{i=\flooring{\log_b(s)}}^{\ceiling{\log_b(s+w)}} \min_{\lambda}\left[e^{ \frac{\flooring{b^{i+1}}}{2} \lambda^2-\sqrt{b^{i}u}\lambda} \right] \\
&\leq 2\sum_{i=\flooring{\log_b(s)}}^{\ceiling{\log_b(s+w)}} \min_{\lambda}\left[e^{ \frac{b^{i+1}}{2} \lambda^2-\sqrt{b^{i}u}\lambda} \right] =  2\sum_{i=\flooring{\log_b(s)}}^{\ceiling{\log_b(s+w)}} e^{-\frac{u}{2b}} = 2(1+\ceiling{\log_b(s+w)} -\flooring{\log_b(s)} )e^{-\frac{u}{2b}} \\
&\leq 2(3+\log_b(s+w)-\log_b(s))e^{-\frac{u}{2b}} = 2(3+ \log_b(1+w/s))e^{-\frac{u}{2b}} \leq 2(3+ \log_b(w+1) )e^{-\frac{u}{2b}} \\
&\leq A \frac{1+\log(w+1)}{\log(b)}e^{-\frac{u}{2b}},
\end{align*}
for some global constant $A$. Here the fifth inequality follows from Doob's martingale inequality. 

Next note that 
\begin{align*}
&\mathbb{P} \left( \max_{0\leq f,d \leq w: j-f-d-i \geq 0} \left((j-f-d-i+1)\left(\bar{\bm{\eta}}_{(i+d):(j-f)}^{(c)}\right) ^2 \right)> u \right) \\
&\leq \sum_{ d = 0 } ^w \mathbb{P} \left( \max_{0\leq f \leq \min(w,j-i-d)} \left((j-f-d-i+1)\left(\bar{\bm{\eta}}_{(i+d):(j-f)}^{(c)}\right) ^2 \right)> u \right) \leq \sum_{ d = 0 } ^w \mathbb{P} \left( Z_{\max(1,j-i-2w)}> u \right) \\
& \leq  A(w+1)\frac{1+\log(w+1)}{\log(b)}e^{-\frac{u}{2b}}
\end{align*}

\subsubsection{Proof of Lemma \ref{lemma:Eventsets}}
In this section, we define the event that $E_a$ holds for a given set tuple $(i,j)$ to be $E_a^{(i,j)}$, for $a=1,...,11$. We know from the proof of Propositions \ref{prop:FPcontrol_mean_penalty_1} and \ref{prop:FPcontrol_mean_penalty_2} that 
\begin{equation*}
\mathbb{P}\left(E_1^{(i,j)}\right) > 1 - A_1e^{-\psi}
\end{equation*}
holds. A Bonferroni correction over all possible tuples $(i,j)$ then gives $\mathbb{P}\left(E_1\right)  > 1 - A_1n^2e^{-\psi}$. Furthermore, we have that 
\begin{equation*}
\mathbb{P}\left( E_2^{(i,j)}\right) = \mathbb{P}\left( \sum_{ c =1}^{p}(j-i+1)\left(\bar{\bm{\eta}}_{i:j}^{(c)}\right) ^2  < p + 2\psi + 2\sqrt{p\psi}\right) = \mathbb{P}\left( \chi^2_p < p + 2\psi + 2\sqrt{p\psi}\right) \geq 1 - e^{-\psi},
\end{equation*}
with the inequality following from the tail bounds proven in \cite{laurent2000adaptive}. A Bonferroni correction then gives $\mathbb{P}\left(E_2\right)  > 1 - n^2e^{-\psi}$. Next note that for any fixed fixed $i$, $j$, and $c$
\begin{equation*}
\mathbb{P} \left( (j-i+1)\left(\bar{\bm{\eta}}_{i:j}^{(c)} + \bar{\bm{\mu}}_{i:j}^{(c)} \right) ^2 > s \right) \geq \mathbb{P} \left( (j-i+1)\left(\bar{\bm{\eta}}_{i:j}^{(c)}\right) ^2 > s \right)
\end{equation*}
holds for all $s \geq 0$. Therefore
\begin{equation*}
\mathbb{P} \left( \sum_{ c = 1}^p(j-i+1)\left(\bar{\bm{\eta}}_{i:j}^{(c)} + \bar{\bm{\mu}}_{i:j}^{(c)} \right) ^2 > p - 2\sqrt{p\psi} \right) \geq \mathbb{P} \left(  \sum_{ c = 1}^p(j-i+1)\left(\bar{\bm{\eta}}_{i:j}^{(c)}\right) ^2 > p - 2\sqrt{p\psi} \right) \geq 1 -e^{-\psi},
\end{equation*}
with the last inequality again flowing from \cite{laurent2000adaptive}.  A Bonferroni correction then gives $\mathbb{P}\left(E_3\right)  > 1 - n^2e^{-\psi}$. Next note that 
\begin{equation*}
\frac{1}{\sqrt{|S|}}\sum_{c \in S}\sqrt{j-i+1}\bar{\bm{\eta}}_{i:j} \sim N(0,1)
\end{equation*}
We can then use the well known tail bounds on the Normal distribution to show that
\begin{equation*}
\mathbb{P}\left(\left|\frac{1}{\sqrt{|S|}}\sum_{c \in S}\sqrt{j-i+1}\bar{\bm{\eta}}_{i:j}\right| < \sqrt{2\psi + 2|S|\log(p)}\right) \geq 1 - A_4p^{-|S|}e^{-\psi},
\end{equation*}
for a constant $A_4$. A Bonferroni correction over all possible sets $S$ then shows that 
\begin{align*}
&\mathbb{P}\left(\left|\frac{1}{\sqrt{|S|}}\sum_{c \in S}\sqrt{j-i+1}\bar{\bm{\eta}}_{i:j}\right| < \sqrt{2\psi + 2|S|\log(p)} \; \; \; \forall S \subset \{1,...,p\}\right) \\
&\geq 1 - \sum_{m=1}^p|\{S|S \subset \{1,...,p\}, |S|=m\}|A_4p^{-|m|}e^{-\psi} \geq 1 - \sum_{m=1}^p\frac{p!}{(p-m)!m!}A_4p^{-|m|}e^{-\psi} \\
&\geq 1 - \sum_{m=1}^p\frac{1}{m!}p^mA_4p^{-|m|}e^{-\psi}\geq 1 - \left(A_4e\right)e^{-\psi}.
\end{align*}
A Bonferroni correction over the indices $i$ and $j$ then proves that $\mathbb{P}\left(E_4\right)  > 1 - \left(A_4e\right)n^2e^{-\psi}$. Next, for fixed $i$ and $j$,
\begin{align*}
&\mathbb{P} \left(\sum_{ c \notin S }(j-i+1)\left(\bar{\bm{\eta}}_{i:j}^{(c)} + \bar{\bm{\mu}}_{i:j}^{(c)} \right) ^2  > p - 2\sqrt{p\psi} - 2\psi -2|S|\log(p) \right)\\
& \geq \mathbb{P} \left(\sum_{ c \notin S }(j-i+1)\left(\bar{\bm{\eta}}_{i:j}^{(c)} \right) ^2  > p - 2\sqrt{p\psi} - 2\psi -2|S|\log(p) \right) \\
& \geq 1 -\mathbb{P} \left(\sum_{ i=1}^p(j-i+1)\left(\bar{\bm{\eta}}_{i:j}^{(c)} \right) ^2  \leq p - 2\sqrt{p\psi} \right) - \mathbb{P} \left(\sum_{ c \in S }(j-i+1)\left(\bar{\bm{\eta}}_{i:j}^{(c)} \right) ^2  >   2\psi -2|S|\log(p) \right)\\
& \geq 1 - (1+A_1)e^{-\psi}
\end{align*}
A Bonferroni correction over all indices $i$ and $j$ then gives $\mathbb{P}\left(E_5\right)> 1-(1+A_1)n^2e^{-\psi}$. Next we note that 
\begin{equation*}
\left(\sum_{ c \in S }\left( \sum_{t=i}^j \left(\bm{\mu}_t^{(c)} - \bar{\bm{\mu}}_{i:j}^{(c)}  \right)\bm{\eta}_t^{(c)} \right) \right) \left(\sqrt{\sum_{ c \in S } \sum_{t=i}^j \left(\bm{\mu}_t^{(c)} - \bar{\bm{\mu}}_{i:j}^{(c)}  \right)^2}\right)^{-1} \sim N(0,1). 
\end{equation*}
Consequently, \small
\begin{equation*}
\mathbb{P}\left( \left|\left(\sum_{ c \in S }\left( \sum_{t=i}^j \left(\bm{\mu}_t^{(c)} - \bar{\bm{\mu}}_{i:j}^{(c)}  \right)\bm{\eta}_t^{(c)} \right) \right) \left(\sqrt{\sum_{ c \in S } \sum_{t=i}^j \left(\bm{\mu}_t^{(c)} - \bar{\bm{\mu}}_{i:j}^{(c)}  \right)^2}\right)^{-1}  \right|   > \sqrt{2\psi+2 \left| S \cap W_{i,j} \right| \log(p)}  \right) \leq A_4p^{-|S \cap W_{i,j}|}e^{-\psi},
\end{equation*}
\normalsize for some constant $A_4$. A Bonferroni correction over the sets $S$ then gives that \small
\begin{align*}
&\mathbb{P}\left( \left|\left(\sum_{ c \in S }\left( \sum_{t=i}^j \left(\bm{\mu}_t^{(c)} - \bar{\bm{\mu}}_{i:j}^{(c)}  \right)\bm{\eta}_t^{(c)} \right) \right)   \right|   \leq  \sqrt{\sum_{ c \in S } \sum_{t=i}^j \left(\bm{\mu}_t^{(c)} - \bar{\bm{\mu}}_{i:j}^{(c)}  \right)^2} \sqrt{2\psi+2 \left| S \cap W_{i,j} \right| \log(p)}  \; \; \; \forall S \subset \{1,...,p\}\right) \\
&= \mathbb{P}\left( \left|\left(\sum_{ c \in S\cap W_{i,j} }\left( \sum_{t=i}^j \left(\bm{\mu}_t^{(c)} - \bar{\bm{\mu}}_{i:j}^{(c)}  \right)\bm{\eta}_t^{(c)} \right) \right)   \right|   \leq  \sqrt{\sum_{ c \in S \cap W_{i,j}} \sum_{t=i}^j \left(\bm{\mu}_t^{(c)} - \bar{\bm{\mu}}_{i:j}^{(c)}  \right)^2} \sqrt{2\psi+2 \left| S \cap W_{i,j} \right| \log(p)}  \; \; \; \forall S \subset \{1,...,p\}\right) \\
&= \mathbb{P}\left( \left|\left(\sum_{ c \in W}\left( \sum_{t=i}^j \left(\bm{\mu}_t^{(c)} - \bar{\bm{\mu}}_{i:j}^{(c)}  \right)\bm{\eta}_t^{(c)} \right) \right)   \right|   \leq  \sqrt{\sum_{ c \in W} \sum_{t=i}^j \left(\bm{\mu}_t^{(c)} - \bar{\bm{\mu}}_{i:j}^{(c)}  \right)^2} \sqrt{2\psi+2 \left|W \right| \log(p)}  \; \; \; \forall W \subset W_{i,j}\right) \\
&\leq 1 - \sum_{W \subset W_{i,j}} \mathbb{P}\left( \left|\left(\sum_{ c \in W}\left( \sum_{t=i}^j \left(\bm{\mu}_t^{(c)} - \bar{\bm{\mu}}_{i:j}^{(c)}  \right)\bm{\eta}_t^{(c)} \right) \right)   \right|   >  \sqrt{\sum_{ c \in W} \sum_{t=i}^j \left(\bm{\mu}_t^{(c)} - \bar{\bm{\mu}}_{i:j}^{(c)}  \right)^2} \sqrt{2\psi+2 \left|W \right| \log(p)}  \; \; \; \forall W \subset W_{i,j}\right) \\
& \leq 1 - \sum_{ |W| = 1}^{|W_{i:j}|} \frac{p!}{(p-|W|)!(|W|)!} A_4p^{-|W|} e^{-\psi} \leq 1 - (A_4e)e^{-\psi}
\end{align*} \normalsize
We note that
\begin{equation*}
\frac{(j-j')(j'-i+1)}{j-i+1}\left(\bar{\bm{\eta}}_{i:j'}^{(c)} - \bar{\bm{\eta}}_{(j'+1):j}^{(c)}\right) ^2 \sim \chi^2_1,
\end{equation*}
and 
\begin{equation*}
\mathbb{P}\left(\frac{(j-j')(j'-i+1)}{j-i+1}\left(\bar{\textbf{x}}_{i:j'}^{(c)} - \bar{\textbf{x}}_{(j'+1):j}^{(c)}\right) ^2 > t \right)  \geq \mathbb{P}\left(\chi^2_1 >t\right) \; \; \; \forall t>0.
\end{equation*}
Therefore, proving that constants $A_7$, $A_8$, and $A_9$ exist such that $\mathbb{P}\left(E_7^{(i,j',j)}\right) > 1 - A_7e^{-\psi}$,  $\mathbb{P}\left(E_8^{(i,j',j)}\right) > 1 - A_8e^{-\psi}$, and $\mathbb{P}\left(E_9^{(i,j',j)}\right) > 1 - A_9e^{-\psi}$ hold for fixed $i$, $j'$, and $j$ is equivalent to proving the existence of constants $A_1$, $A_2$, and $A_3$ such that $\mathbb{P}\left(E_1^{(i,j)}\right) > 1 - A_1e^{-\psi}$,  $\mathbb{P}\left(E_2^{(i,j)}\right) > 1 - A_2e^{-\psi}$, and $\mathbb{P}\left(E_3^{(i,j)}\right) > 1 - A_3e^{-\psi}$ hold. This was already done earlier in the proof. A Bonferroni correction over all possible $i$, $j'$, and $j$ then yields $\mathbb{P}\left(E_7\right) > 1 - A_7n^3e^{-\psi}$,  $\mathbb{P}\left(E_8\right) > 1 - A_8n^3e^{-\psi}$, and $\mathbb{P}\left(E_9\right) > 1 - A_9n^3e^{-\psi}$. 

The fact that
\begin{equation*}
\left(\sum_{ c \in \textbf{J}_k}\sqrt{j-i+1}\bar{\bm{\eta}}_{i:j}^{(c)}\right) \left(\sqrt{2|\textbf{J}_k|\psi} \right)^{-1} \sim N(0,1)
\end{equation*}
shows that $\mathbb{P}\left(E_{10}^{(i,j)}\right) > 1 - A_{10}e^{-\psi}$ for some constant $ A_{10}$. The cardinality of the set of allowed tuples $(i,j)$ is strictly less than $n^2$. Consequently $\mathbb{P}\left(E_{10}\right) > 1 - A_{10}n^2e^{-\psi}$. The same argument can be used to show that $\mathbb{P}\left(E_{11}^{(i,e_k,j)}\right) > 1 - A_{11}e^{-\psi}$. A Bonferroni correction over all triplets $(i,e_k,j)$ then proves that $\mathbb{P}\left(E_{11}\right) > 1 - A_{11}n^3e^{-\psi}$

\subsubsection{Proof of Lemma \ref{lemma:splitsaving}}
This Lemma can be proven using straightforward algebra. 
\begin{align*}
\mathcal{S}\left(\textbf{x}_{i:j},\textbf{J}\right) = \sum_{ c \in \textbf{J} } (j-i+1) \left( \bar{\textbf{x}}_{i:j}^{(c)}\right)^2 = \sum_{ c \in \textbf{J} } (j-i+1)^{-1} \left( \bar{\textbf{x}}_{i:j}^{(c)}\right)^2 = \sum_{ c \in \textbf{J} }\left[ \frac{\left((j'+1-i)\bar{\textbf{x}}_{i:j'}^{(c)} + (j-j')\bar{\textbf{x}}_{(j'+1):j}^{(c)}\right)^2}{j-i+1}\right].
\end{align*}
Next we note that the following holds for all $a$, $b$, $y$, and $z$:
\begin{align*}
\frac{(ay+bz)^2}{a+b}= \frac{a^2y^2+2abyz+b^2z^2}{a+b} = ay^2 + bz^2 + \frac{-aby^2+2abyz-baz^2}{a+b} = ay^2 + bz^2 - \frac{ab}{a+b}(y-z)^2.
\end{align*}
Thus,
\begin{equation*}
\mathcal{S}\left(\textbf{x}_{i:j},\textbf{J}\right) = \sum_{ c \in S } (j'+1-i)\left(\bar{\textbf{x}}_{i:j'}^{(c)}\right)^2 + \sum_{ c \in S } (j-j')\left(\bar{\textbf{x}}_{(j'+1):j}^{(c)}\right)^2  -  \sum_{ c \in S } \frac{(j-j')(j'-i+1)}{j-i+1} \left(\bar{\textbf{x}}_{i:j'}^{(c)} - \bar{\textbf{x}}_{(j'+1):j}^{(c)} \right)^2,
\end{equation*}
which finishes the proof. \qed 

\subsubsection{Proof of Lemma \ref{lemma:splitlemma}}
This result deals with the $p$ term of the penalty incurred for splitting a sparse fitted segment into two and follows directly from $E_9$ and Lemma \ref{lemma:splitsaving}. Indeed, by Lemma \ref{lemma:splitsaving} implies that 
\begin{equation*}
\mathcal{C}\left(\textbf{x}_{i:j'},\textbf{1}\right) +\mathcal{C}\left(\textbf{x}_{(j'+1):j},\textbf{1}\right) -  \mathcal{C}\left(\textbf{x}_{i:j},\textbf{1}\right) = p + C\psi + C\sqrt{p\psi} - \sum_{ c = 1}^p\frac{(j-j')(j'-i+1)}{j-i+1}\left(\bar{\textbf{x}}_{i:j'}^{(c)} - \bar{\textbf{x}}_{(j'+1):j}^{(c)}\right) ^2.
\end{equation*}
Given $E_9$, the above is bounded by 
\begin{equation*}
p + C\psi + C\sqrt{p\psi} - p + 2\sqrt{p\psi}  = C\psi + C\sqrt{p\psi} + 2\sqrt{p\psi}.
\end{equation*} 
This finishes the proof.\qed 

\subsubsection{Proof of Lemma \ref{lemma:merging reduces cost}}

This lemma shows that merging two neighbouring fitted segments reduces the penalised cost by $O(C)$ and follows almost immediately from Lemma \ref{lemma:splitsaving}. We consider the cases $|\textbf{J}_k| \leq k^*$ and $|\textbf{J}_k| > k^*$ separately. Let $|\textbf{J}_k| \leq k^*$. Then
\begin{align*}
&\mathcal{ C} \left(x_{i:j'},\textbf{J}_k\right) + \mathcal{ C} \left(x_{(j'+1):j},\textbf{J}_k\right) - \mathcal{ C} \left(x_{i:j},\textbf{J}_k\right) = C\psi + C|\textbf{J}_k|\log(p) - \sum_{ c \in \textbf{J}_k } \frac{(j-j')(j'-i+1)}{j-i+1}\left(\bar{\bm{\eta}}_{i:j'}^{(c)} - \bar{\bm{\eta}}_{(j'+1):j}^{(c)}\right) ^2 \\
& \geq C\psi + C|\textbf{J}_k|\log(p) - 2\psi - 2|\textbf{J}_k|\log(p) \geq \frac{79}{80}C \left( \psi+|\textbf{J}_k|\log(p)\right),
\end{align*}
where the first inequality follows from $E_7$ and the second one holds if $C$ exceeds some global constant. Now let $|\textbf{J}_k| \geq k^*$
\begin{align*}
&\mathcal{ C} \left(x_{i:j'},\textbf{1}\right) + \mathcal{ C} \left(x_{(j'+1):j},\textbf{1}\right) - \mathcal{ C} \left(x_{i:j},\textbf{1}\right) = p + C\psi + C\sqrt{p\psi} - \sum_{ c =1 }^p \frac{(j-j')(j'-i+1)}{j-i+1}\left(\bar{\bm{\eta}}_{i:j'}^{(c)} - \bar{\bm{\eta}}_{(j'+1):j}^{(c)}\right) ^2 \\
& \geq p + C\psi + C\sqrt{p\psi} - 2\psi - 2\sqrt{p\psi} - p \geq  \frac{79}{80}C \left( \psi+\sqrt{p\psi}\right),
\end{align*}
where the first inequality follows from $E_8$ and the second one holds if $C$ exceeds some global constant. \qed

\subsubsection{Proof of Lemma \ref{lemma:NonVSsome}}
This Lemma proves MVCAPA has power at detecting anomalous regions. We begin by considering the case in which $J_k$ is dense. We have: \small
\begin{align*}
\mathcal{C}\left(\textbf{x}_{i:j} , \textbf{1}\right) &= p + C\psi + C\sqrt{p\psi} - \sum_{ c = 1}^p \left(j-i+1\right)\left(\textbf{x}_{i:j}^{(c)}\right)^2 \\
&=  p + C\psi + C\sqrt{p\psi} - \sum_{ c = 1}^p (j-i+1)\left(\bar{\bm{\eta}}_{i:j}^{(c)}\right)^2 - |\textbf{J}_k|\bm{\mu}_k^2(j-i+1) - 2\bm{\mu}_k \sqrt{j-i+1}\sum_{ c \in \textbf{J}_k }\left( \sqrt{j-i+1} \bar{\bm{\eta}}_{i:j}\right) \\
&\leq C\psi + (C+2)\sqrt{p\psi} - |\textbf{J}_k|\bm{\mu}_k^2(j-i+1) + 2\sqrt{(j-i+1)\bm{\mu}_k^2}
\sqrt{2|\textbf{J}_k|\psi} \\
&\leq  C\psi + (C+2)\sqrt{p\psi} - \frac{1}{2}|\textbf{J}_k|\bm{\mu}_k^2(j-i+1) + 4\psi \leq  C\psi + (C+2)\sqrt{p\psi} - \frac{1}{2}|\textbf{J}_k|\bm{\mu}_k^2\frac{4C}{\triangle_k^2} + 4\psi \\
&=(C+4)\psi + (C+2)\sqrt{p\psi} - 2C(\psi+\sqrt{p\psi}) \leq 0 
\end{align*} \normalsize
with the first inequality following form $E_{10}$ and $E_3$, the second from the AM-GM inequality,  the third from the condition on $j-i+1$, and the last one holds if $C$ exceeds a global constant. 

The proof for when $J_k$ is sparse is almost identical. We have that: 
\small
\begin{align*}
&\mathcal{C}\left(\textbf{x}_{i:j} , \textbf{J}_k\right) = C\psi + C|\textbf{J}_k|\log(p) - \sum_{ c \in \textbf{J}_k }\left(j-i+1\right)\left(\bm{\mu}_k + \bar{\bm{\eta}}_{i:j}^{(c)}\right)^2 \\
&=  C\psi + C|\textbf{J}_k|\log(p) - \sum_{ c \in \textbf{J}_k } (j-i+1)\left(\bar{\bm{\eta}}_{i:j}^{(c)}\right)^2 - |\textbf{J}_k|\bm{\mu}_k^2(j-i+1) - 2\bm{\mu}_k \sqrt{j-i+1}\sum_{ c \in \textbf{J}_k }\left( \sqrt{j-i+1} \bar{\bm{\eta}}_{i:j}\right) \\
&\leq C\psi + C|\textbf{J}_k|\log(p) - |\textbf{J}_k|\bm{\mu}_k^2(j-i+1) + 2\sqrt{(j-i+1)\bm{\mu}_k^2}
\sqrt{2|\textbf{J}_k|\psi + 2|\textbf{J}_k|^2\log(p)} \\
&\leq   C\psi + C|\textbf{J}_k|\log(p) - \frac{1}{2}|\textbf{J}_k|\bm{\mu}_k^2(j-i+1) + 4\psi +4|\textbf{J}_k|\log(p) \leq   (C+4)\psi + (C+4)|\textbf{J}_k|\log(p) - \frac{1}{2}|\textbf{J}_k|\bm{\mu}_k^2\frac{4C}{\triangle_k^2} \\
& =(C+4)\psi + (C+4)|\textbf{J}_k|\log(p) - 2C(\psi+|\textbf{J}_k|\log(p)) \leq 0 ,
\end{align*} \normalsize
where the first inequality follows from $E_4$, the second from the AM-GM inequality,  the third from the condition on $j-i+1$, and the last one holds if $C$ exceeds a global constant. \qed

\subsubsection{Proof of Lemma \ref{lemma:SplitCOSTSAVING}}
This Lemma prevents fitted changes from containing too many observations belonging to the typical distribution. We limit ourselves to proving the result for the first case, since the proof of the second case is symmetrical. We begin by proving the result for the case in which $|\textbf{J}_k| \leq k^*$. Writing, $e' = e_{k} + \lceil 10\frac{C}{\triangle_{k}^2} \rceil $  and $s' = e_{k} - \lceil 10 \frac{C}{\triangle_{k}^2} \rceil $
\begin{align*}
\mathcal{C}\left(x_{i:j} , \textbf{J}_k\right) &\geq \mathcal{C}\left(x_{i:s'} , \textbf{J}_k\right) + \mathcal{C}\left(x_{(s'+1):e'} , \textbf{J}_k\right) + \mathcal{C}\left(x_{(e'+1):j} , \textbf{J}_k\right) - 2C\psi - 2C|\textbf{J}_k|\log(p) \\
&\geq \mathcal{C}\left(x_{i:s'} , \textbf{J}_k\right) + \mathcal{C}\left(x_{(s'+1):e'} , \textbf{J}_k\right) - (C+2) \left( \psi + |\textbf{J}_k|\log(p) \right)
\end{align*}
Next, note that Lemma \ref{lemma:splitsaving} implies that \small
\begin{equation*}
\mathcal{C}\left( x_{(s'+1):e'}, \textbf{J}_k \right) = \mathcal{C}\left( x_{(s'+1):e_k}, \textbf{J}_k \right) + \mathcal{C}\left( x_{(e_k+1):e'}, \textbf{J}_k \right) - C\left( \psi + |\textbf{J}_k|\log(p) \right) + \sum_{ c \in \textbf{J}_k }  \frac{e'-s'}{2} \left( \bm{\mu}_k + \bar{\bm{\eta}}_{(s'+1):e_k} - \bar{\bm{\eta}}_{(e_k+1):e'} \right)^2
\end{equation*} \normalsize
Moreover, we have that
\begin{align*}
&\sum_{ c \in \textbf{J}_k }  \frac{e'-s'}{2} \left( \bm{\mu}_k + \bar{\bm{\eta}}_{(s'+1):e_k} - \bar{\bm{\eta}}_{(e_k+1):e'} \right)^2 \\
&= \frac{e'-s'}{2}|\textbf{J}_k|\bm{\mu}_k^2 - \bm{\mu}_k(e'-s')\sum_{ c \in \textbf{J}_k } \left(\bar{\bm{\eta}}_{(s'+1):e_k} - \bar{\bm{\eta}}_{(e_k+1):e'}\right) +  \frac{e'-s'}{2}\sum_{ c \in \textbf{J}_k } \left(\bar{\bm{\eta}}_{(s'+1):e_k} - \bar{\bm{\eta}}_{(e_k+1):e'}\right)^2 \\
&\geq \frac{e'-s'}{2}|\textbf{J}_k|\bm{\mu}_k^2 - 2\sqrt{(e'-s')\bm{\mu}_k^2} \sqrt{2|\textbf{J}_k|\psi}  \geq \frac{e'-s'}{3}|\textbf{J}_k|\bm{\mu}_k^2 - 12\psi = \frac{20}{3}C\left( \psi + |\textbf{J}_k| \log(p)\right) -12\psi,
\end{align*}
where the first inequality follows from $E_{11}$ and the second inequality from the AM-GM inequality. Combining all the above, we obtain that \small
\begin{align*}
\mathcal{C}\left(x_{i:j} , \textbf{J}_k\right) &\geq \mathcal{C}\left(x_{i:s'} , \textbf{J}_k\right) + \mathcal{C}\left( x_{(s'+1):e_k}, \textbf{J}_k \right) + \mathcal{C}\left( x_{(e_k+1):e'}, \textbf{J}_k \right) + \left(\frac{14}{3}C - 14 \right) \psi + \left(\frac{14}{3}C - 2 \right)|\textbf{J}_k|\log(p)   \\
&\geq \mathcal{C}\left(x_{i:e_k}, \textbf{J}_k\right) + \frac{19}{20} C\left(\psi + |\textbf{J}_k|\log(p)\right) + (C-2)\left(\psi + |\textbf{J}_k|\log(p)\right) + \left(\frac{14}{3}C - 14 \right) \psi + \left(\frac{14}{3}C - 2 \right)|\textbf{J}_k|\log(p) \\
&\geq 6C \left( \psi + |\textbf{J}_k|\log(p) \right),
\end{align*} \normalsize
where the second inequality follows from Lemma \ref{lemma:merging reduces cost} and $E_2$. The proof for the case in which $|\textbf{J}_k| > k^*$ is very similar. We the have that 
\begin{align*}
\mathcal{C}\left(x_{i:j} , \textbf{1}\right) &\geq \mathcal{C}\left(x_{i:s'} , \textbf{1}\right) + \mathcal{C}\left(x_{(s'+1):e'} , \textbf{1}\right) + \mathcal{C}\left(x_{(e'+1):j} , \textbf{1}\right) - 2C\psi - 2(C+2)\sqrt{p\psi}\\
&\geq \mathcal{C}\left(x_{i:s'} , \textbf{1}\right) + \mathcal{C}\left(x_{(s'+1):e'} , \textbf{1}\right) - (C+6) \left( \psi + \sqrt{p\psi} \right),
\end{align*}
with the first inequality following from Lemma \ref{lemma:splitsaving} and the event $E_3$ the second being due to $E_2$. The remainder of the proof of the Lemma is  very similar to the sparse case and has therefore been omitted. 

\subsubsection{Proof of Lemma \ref{lemma:SplitCOSTSAVINGDENSE}}

This Lemma shows that the optimal partition can not contain fitted segments containing more than $10 \frac{C}{\triangle_{k}^2}$ observations from both the typical distribution and a dense anomalous region. If $\textbf{J}=\textbf{1}$ the result follows a fortiori from Lemma \ref{lemma:SplitCOSTSAVING}. Assume now that $|\textbf{J}| \leq k^*$. As in the proof of Lemma \ref{lemma:SplitCOSTSAVING}, we limit ourselves to proving the first case, the proof of the other one being symmetrical. The following holds:
\begin{align*}
&\mathcal{C}\left(x_{i:j},\textbf{J}\right) = \mathcal{C}\left(x_{i:j},\textbf{1}\right)  -(p+C\psi+C\sqrt{p\psi}) + \sum_{ c \notin \textbf{J} }(j-i+1)\left(\bar{\textbf{x}}_{i:j}\right)^2 + C\psi+C|\textbf{J}|\log(p)
\\
&\geq \mathcal{C}\left(x_{i:j},\textbf{1}\right)  - C(\psi+\sqrt{p\psi}) -  2\sqrt{p\psi} -2\psi - 2|\textbf{J}|\log(p) + C\psi+C|\textbf{J}|\log(p) \geq \mathcal{C}\left(x_{i:e_k},\textbf{1}\right)  + 4C(\psi+\sqrt{p\psi}),
\end{align*}
where the first inequality follows from the event $E_5$ and the second one from Lemma \ref{lemma:SplitCOSTSAVING} and a choice of $C$ exceeding some global constant. \qed

\subsubsection{Proof of Lemma \ref{lemma:merging twice reduces cost}}

This lemma shows that merging two neighbouring fitted segments reduces penalised cost by $O(C)$ -- even when they are separated by a gap. The proof is very similar to that of Lemma \ref{lemma:merging reduces cost}. In fact, Lemma \ref{lemma:merging twice reduces cost} follows a fortiori from Lemma \ref{lemma:merging reduces cost} when $j' = j''$. When $j' \neq j''$ we consider the $|\textbf{J}_k| \leq k^*$ and $|\textbf{J}_k| > k^*$ separately. Let $|\textbf{J}_k| \leq k^*$. Then, 
\begin{align*}
&\mathcal{ C} \left(x_{i:j'},\textbf{J}_k\right) + \mathcal{ C} \left(x_{(j''+1):j},\textbf{J}_k\right) - \mathcal{ C} \left(x_{i:j},\textbf{J}_k\right) \\
&\geq \mathcal{ C} \left(x_{i:j'},\textbf{J}_k\right) + \left[\mathcal{ C} \left(x_{(j'+1):j''},\textbf{J}_k\right) - C\psi - C|\textbf{J}_k|\log(p) \right] + \mathcal{ C} \left(x_{(j''+1):j},\textbf{J}_k\right) - \mathcal{ C} \left(x_{i:j},\textbf{J}_k\right) \\
&\geq  - C\psi -  C|\textbf{J}_k|\log(p) + \frac{79}{80}C \left( \psi+|\textbf{J}_k|\log(p)\right) + \frac{79}{80}C \left( \psi+|\textbf{J}_k|\log(p)\right) \geq \frac{19}{20}C \left( \psi+|\textbf{J}_k|\log(p)\right),
\end{align*} 
where the second inequality follows from applying Lemma \ref{lemma:merging reduces cost} twice. The proof for the case in which $|\textbf{J}_k| > k^*$ is very similar. We have that 
\begin{align*}
&\mathcal{ C} \left(x_{i:j'},\textbf{1}\right) + \mathcal{ C} \left(x_{(j''+1):j},\textbf{1}\right) - \mathcal{ C} \left(x_{i:j},\textbf{1}\right) \\
&\geq \mathcal{ C} \left(x_{i:j'},\textbf{1}\right) + \left[ \mathcal{ C} \left(x_{(j'+1):j''},\textbf{1}\right) - C\psi - (C+2)\sqrt{p\psi} \right] + \mathcal{ C} \left(x_{(j''+1):j},\textbf{1}\right) - \mathcal{ C} \left(x_{i:j},\textbf{1}\right) \\
&\geq - C\psi - (C+2)\sqrt{p\psi}  + \frac{79}{80}C \left( \psi + \sqrt{p\psi}\right) + \frac{79}{80}C \left( \psi + \sqrt{p\psi}\right) \geq \frac{19}{20}C\left(\psi+\sqrt{p\psi}\right),
\end{align*}
where the first inequality follows from $E_3$, the third inequality follows from applying Lemma  \ref{lemma:merging reduces cost} twice, and the third holds if $C$ exceeds a global constant. \qed

\subsubsection{Proof of Lemma \ref{lemma:trimming(h)edges}}

This Lemma shows that if a fitted segment contains observations belonging to the typical distribution it can be trimmed to containing only anomalous observations without increasing the penalised cost by more than $O(1)$ . We begin by proving the sparse case
\begin{align*}
&\mathcal{C}\left(\textbf{x}_{i:j} , \textbf{J}\right) \geq \mathcal{C}\left(\textbf{x}_{i:j'} , \textbf{J}\right) + \left(\mathcal{C}\left(\textbf{x}_{(j'+1):j} , \textbf{J}\right) - C\psi - C|\textbf{J}|\log(p)\right) \geq \mathcal{C}\left(\textbf{x}_{i:j'} , \textbf{J}\right) -2\psi-2|\textbf{J}|\log(p) \\ 
&\geq  (\mathcal{C}\left(\textbf{x}_{i:(i'-1)} , \textbf{J}\right)  - C\psi - C|\textbf{J}|\log(p)  ) + \mathcal{C}\left(\textbf{x}_{i':j'} , \textbf{J}\right) -2\psi-2|\textbf{J}|\log(p) \geq \mathcal{C}\left(\textbf{x}_{i':j'} , \textbf{J}\right) -4\psi-4|\textbf{J}|\log(p),
\end{align*}
where the first and third inequality follows from the fact that introducing free splits reduces un-penalised cost whilst the second and third inequality follows from $E_1$. Note that if $j'=j$ and/or $i'=i$ the first and second and/or the third and forth step are not necessary. The result nevertheless holds. A similar proof can be derived for the dense case:
\begin{align*}
&\mathcal{C}\left(\textbf{x}_{i:j} , \textbf{J}\right) \geq \mathcal{C}\left(\textbf{x}_{i:j'} , \textbf{J}\right) + \left(\mathcal{C}\left(\textbf{x}_{(j'+1):j} , \textbf{J}\right) - C\psi - C\sqrt{p\psi}\right) - 2 \sqrt{p\psi} \geq \mathcal{C}\left(\textbf{x}_{i:j'} , \textbf{J}\right) -2\psi-4\sqrt{p\psi} \\ 
&\geq  (\mathcal{C}\left(\textbf{x}_{i:(i'-1)} , \textbf{J}\right)  - C\psi - C\sqrt{p\psi}  ) + \mathcal{C}\left(\textbf{x}_{i':j'} , \textbf{J}\right) -2\psi-6\sqrt{p\psi} \geq \mathcal{C}\left(\textbf{x}_{i':j'} , \textbf{J}\right) -4\psi-8\sqrt{p\psi},
\end{align*}
with the first and third inequalities following form Lemma \ref{lemma:splitlemma}, and the second and fourth from $E_2$. \qed
\normalsize

\subsubsection{Proof of Lemma \ref{lemma:changeproperty}}

This Lemma links the savings of a fitted segment to the signal strength of the corresponding segment. We have that 
\begin{align*}
&\alpha\left(C\psi + C|\textbf{J}|\log(p)\right) \leq \sum_{c \in \textbf{J}}\left(\bm{\mu}_k + \bar{\bm{\eta}}_{i:j}^{(c)} \right)^2(j-i+1) \\
&= |\textbf{J}\cap\textbf{J}_k|(j-i+1)\bm{\mu}_k^2 + 2\sqrt{j-i+1}\bm{\mu}_k \sum_{c \in \textbf{J}\cap\textbf{J}_k} \sqrt{j-i+1}\bar{\bm{\eta}}_{i:j}^{(c)} + \sum_{c \in \textbf{J}} \left(\sqrt{j-i+1}\bar{\bm{\eta}}_{i:j}^{(c)}\right)^2 \\
& \leq |\textbf{J}\cap\textbf{J}_k|(j-i+1)\bm{\mu}_k^2 + 2\sqrt{j-i+1}|\bm{\mu}_k|\sqrt{2\psi|\textbf{J}\cap\textbf{J}_k| + 2|\textbf{J}\cap\textbf{J}_k|^2 \log(p)} + 2\psi + 2|\textbf{J}|\log(p) \\
& \leq |\textbf{J}|(j-i+1)\bm{\mu}_k^2 + 2\sqrt{j-i+1}|\bm{\mu}_k|\sqrt{2\psi|\textbf{J}| + 2|\textbf{J}|^2 \log(p)} + 2\psi + 2|\textbf{J}|\log(p) \\
& = \left( \sqrt{ |\textbf{J}|(j-i+1)\bm{\mu}_k^2} + \sqrt{2\psi + 2|\textbf{J}|\log(p)}\right)^2,
\end{align*}
with the first inequality following from $E_1$ and $E_4$ and th second from the fact that $|\textbf{J}\cap\textbf{J}_k| \leq |\textbf{J}|$.This therefore implies that 
\begin{equation*}
\sqrt{|\textbf{J}|(j-i+1)\bm{\mu}_k^2 } \geq \left( \sqrt{\alpha C} -\sqrt{2}\right)\sqrt{\psi+|\textbf{J}| \log(p)}
\end{equation*}

\subsubsection{Proof of Lemma \ref{lemma:Sparse_To_Sparse_CONDITION1}}
This Lemma shows that if removing a fitted sparse segment does not result in a reduction in penalised cost of $O(\frac{1}{20}C)$, the increase in penalised cost incurred for replacing it with the sparse ground truth is $O(\frac{1}{20}C)$. We will use a very similar strategy to the one we used to prove Lemma \ref{lemma:Sparse_To_Dense_CONDITION1}. We begin by noting that \small
\begin{equation}\label{eq:sparsetosparse}
\mathcal{C}\left(\textbf{x}_{i:j},\textbf{J}_k\right) - \mathcal{C}\left(\textbf{x}_{i:j},\textbf{J}\right) =  C \left(|\textbf{J}_k| - |\textbf{J}| \right)\log(p) - \sum_{ c \in \textbf{J}_k \setminus \textbf{J} }(j-i+1) \left( \mu + \bar{\bm{\eta}}_{i:j}^{(c)}\right)^2 + \sum_{ c \in \textbf{J} \setminus \textbf{J}_k }(j-i+1) \left(\bar{\bm{\eta}}_{i:j}^{(c)}\right)^2
\end{equation}
\normalsize If $|\textbf{J}|> \frac{19}{20} |\textbf{J}_k|$, $E_1$ bounds (\ref{eq:sparsetosparse}) by 
\begin{equation*}
C \left(|\textbf{J}_k| - |\textbf{J}| \right)\log(p) + 2\psi + 2|\textbf{J}|\log(p) \leq \frac{1}{10}C|\textbf{J}_k|\log(p) + 2\psi + \left(2-\frac{1}{20}C\right)|\textbf{J}|\log(p) \leq \frac{1}{10}C|\textbf{J}_k|\log(p) + 2\psi,
\end{equation*}
with the last inequality holding if $C$ exceeds some global constant. If $|\textbf{J}|\leq \frac{19}{20} |\textbf{J}_k|$ we write $\textbf{A} = \textbf{J}_k \setminus \textbf{J}$ and bound (\ref{eq:sparsetosparse}) by 
\begin{equation*}
C \left(|\textbf{J}_k| - |\textbf{J}| \right)\log(p) +2\psi + 2|\textbf{J}| \log(p) - |\textbf{A}| \bm{\mu}_k^2(j-i+1) + 2\sqrt{\bm{\mu}_k^2(j-i+1)} \sqrt{|\textbf{A}|\psi + |\textbf{A}|^2\log(p)}
\end{equation*}
using $E_1$ and $E_4$. Lemma \ref{lemma:changeproperty} implies that
\begin{align*}
\sqrt{(j-i+1)\bm{\mu}_k^2} \geq \frac{1}{\sqrt{|\textbf{J}|}}\left(\sqrt{\frac{19}{20}C}-2\right) \sqrt{\psi + |\textbf{J}|\log(p)}. 
\end{align*}
Consequently, copying parts of the proof of Lemma \ref{lemma:Sparse_To_Dense_CONDITION1} , we have that
\begin{equation*}
|\textbf{A}| \bm{\mu}_k^2(j-i+1) - 2\sqrt{\bm{\mu}_k^2(j-i+1)} \sqrt{|\textbf{A}|\psi + |\textbf{A}|^2\log(p)} > \frac{37}{40}C|\textbf{A}|\log(p),
\end{equation*}
which shows that (\ref{eq:sparsetosparse}) is bounded by \footnotesize
\begin{align*}
C \left(|\textbf{J}_k| - |\textbf{J}| \right)\log(p) +2\psi + 2|\textbf{J}| \log(p) - \frac{37}{40}C|\textbf{A}|\log(p) & \leq \frac{1}{10}C|\textbf{J}_k|\log(p) +2\psi +(2 - \frac{1}{40}C)|\textbf{J}|\log(p) \\
& \leq \frac{1}{10}C|\textbf{J}_k|\log(p) +2\psi,
\end{align*} \normalsize
where the first inequality follows from the fact that $|\textbf{J}_k| < |\textbf{J}| + |\textbf{A}|$ and the second one holds if $C$ exceeds a global constant. This finishes the proof. 

\subsubsection{Proof of Lemma \ref{lemma:Sparse_To_Dense_CONDITION1}}
This Lemma shows that if removing a fitted sparse segment does not result in a reduction in penalised cost of $O(\frac{1}{20}C)$, the increase in penalised cost incurred for replacing it with the dense ground truth is $O(\frac{1}{20}C)$. We have that
\begin{equation}\label{eq:Boundme}
\mathcal{C}\left( \textbf{x}_{i:j},\textbf{1}\right) - \mathcal{C}\left( \textbf{x}_{i:j},\textbf{J}\right) = p + C\sqrt{p\psi} -C|\textbf{J}|\log(p) - \sum_{ c \notin \textbf{J}}(j-i+1)\left(\bar{\textbf{x}}_{i:j}^{(c)}\right)^2
\end{equation}
We consider 2 cases separately. If $|\textbf{J}|>\frac{19}{20}k^*$, the event $E_5$ implies that the above can be bounded by 
\begin{equation*}
p + C\sqrt{p\psi} -C|\textbf{J}|\log(p) - \left(p-2\sqrt{p\psi} - 2\psi - 2|\textbf{J}|\log(p)\right) \geq 2\psi+ (C+2)\sqrt{p\psi} - (C-2)\frac{19}{20}\sqrt{p\psi} \geq 
\frac{1}{10}C\sqrt{p\psi} + 2\psi,
\end{equation*}
provided $C$ exceeds some global constant. If $|\textbf{J}| \leq \frac{19}{20}k^*$, we introduce the set $\textbf{A} = \textbf{J}_k\setminus \textbf{J}$. The quantity in (\ref{eq:Boundme}) is then equal to 
\begin{align*}
&p + C\sqrt{p\psi} -C|\textbf{J}|\log(p) - |\textbf{A}|(j-i+1)\bm{\mu}_k^2 + 2 \sqrt{(j-i+1)}\bm{\mu}_k\sum_{ c \in \textbf{A}} \sqrt{(j-i+1)} \bar{\bm{\eta}}_{i:j}^{(c)} - \sum_{ c \notin \textbf{J}}(j-i+1)\left(\bar{\bm{\eta}}_{i:j}^{(c)}\right)^2 \\
&\leq (C+2)\sqrt{p\psi} - (C-2) |\textbf{J}|\log(p) + 2\psi - |\textbf{A}|(j-i+1)\bm{\mu}_k^2 + 2 \sqrt{(j-i+1)\bm{\mu}_k^2} \sqrt{2|\textbf{A}|\psi+2|\textbf{A}|^2\log(p)},
\end{align*}
where the inequality flows from $E_1$, $E_4$, and $E_5$. If $C$ exceeds a fixed constant, the above is less than 
\begin{equation}\label{eq:BOUNDme}
\frac{41}{40}C\sqrt{p\psi} - \frac{37}{40}C |\textbf{J}|\log(p) + 2\psi - |\textbf{A}|(j-i+1)\bm{\mu}_k^2 + 2 \sqrt{(j-i+1)\bm{\mu}_k^2} \sqrt{2|\textbf{A}|\psi+2|\textbf{A}|^2\log(p)}
\end{equation}
Lemma \ref{lemma:changeproperty} now implies that
\begin{align*}
\sqrt{(j-i+1)\bm{\mu}_k^2} \geq \frac{1}{\sqrt{|\textbf{J}|}}\left(\sqrt{\frac{19}{20}C}-2\right) \sqrt{\psi + |\textbf{J}|\log(p)}. 
\end{align*}
Therefore \small
\begin{align*}
&|\textbf{A}|\sqrt{(j-i+1)\bm{\mu}_k^2} - 2\sqrt{2|\textbf{A}|\psi+2|\textbf{A}|^2\log(p)} \geq  \left(\sqrt{\frac{19}{20}C}-2\right)\sqrt{\frac{|\textbf{A}|}{|\textbf{J}|}|\textbf{A}|\psi + |\textbf{A}|^2\log(p)} - 2\sqrt{2|\textbf{A}|\psi+2|\textbf{A}|^2\log(p)} \\
&\geq \left(\sqrt{\frac{19}{20}C}-2\right)\sqrt{\frac{1}{20}|\textbf{A}|\psi + |\textbf{A}|^2\log(p)} - 2\sqrt{2|\textbf{A}|\psi+2|\textbf{A}|^2\log(p)},
\end{align*}
\normalsize which exceeds 0 if $C$ exceeds a global constant. Therefore 
\begin{align*}
&|\textbf{A}|^2(j-i+1)\bm{\mu}_k^2 - 2|\textbf{A}|\sqrt{(j-i+1)\bm{\mu}_k^2}\sqrt{2|\textbf{A}|\psi+2|\textbf{A}|^2\log(p)} \\
&\geq \frac{|\textbf{A}|}{|\textbf{J}|}\left(\sqrt{\frac{19}{20}C}-2\right)^2\left(\psi + |\textbf{J}|\log(p) \right) - 2 \frac{\sqrt{\frac{19}{20}C}-2}{|\textbf{J}|}\sqrt{2|\textbf{A}|\psi+2|\textbf{A}|^2\log(p)} \sqrt{\psi + |\textbf{J}|\log(p)} \\
&\geq \left(\sqrt{\frac{19}{20}C}-2\right)^2\left(\frac{|\textbf{A}|}{|\textbf{J}|}\psi + |\textbf{A}|\log(p) \right) - 2 \sqrt{\frac{19}{20}C}\sqrt{2\psi+2|\textbf{A}|\log(p)} \sqrt{\frac{|\textbf{A}|}{|\textbf{J}|}\psi + |\textbf{A}|\log(p)}\\
&\geq \left(\sqrt{\frac{19}{20}C}-2\right)^2\left(\frac{|\textbf{A}|}{|\textbf{J}|}\psi + |\textbf{A}|\log(p) \right) -  \sqrt{\frac{19}{20}C} \left( \left(2+\frac{|\textbf{A}|}{|\textbf{J}|}\right)\psi + 3|\textbf{A}|\log(p)\right) \\
& = \left(\left(\sqrt{\frac{19}{20}C}-2\right)^2 - 3\sqrt{\frac{19}{20}C} \right) |\textbf{A}|\log(p) + \left(\left(\left(\sqrt{\frac{19}{20}C}-2\right)^2 - \sqrt{\frac{19}{20}C}\right) \frac{|\textbf{A}|}{|\textbf{J}|} - 2\right)\psi, 
\end{align*}
where the third inequality follows from the AM-GM-inequality. If $C$ exceeds a fixed constant this will exceed
\begin{equation*}
\frac{37}{40}C|\textbf{A}|\log(p),
\end{equation*}
Hence the quantity in (\ref{eq:BOUNDme}) is bounded by 
\begin{equation*}
\frac{41}{40}C\sqrt{p\psi} - \frac{37}{40}C \left(|\textbf{J}|+|\textbf{A}|\right)\log(p) + 2\psi \leq  \frac{41}{40}C\sqrt{p\psi} -  \frac{37}{40}Ck^*\log(p) + 2\psi = \frac{1}{10}C\sqrt{p\psi} + 2\psi.
\end{equation*}
This finishes the proof. \qed

\subsubsection{Proof of Lemma \ref{lemma:Dense_to_sparse}}

This Lemma bounds the increase in penalised cost incurred when transitioning from a fitted dense segment to the sparse ground truth. We have that
\begin{align*}
& \mathcal{C}\left( \textbf{x}_{i:j},\textbf{J}_k\right) - \mathcal{C}\left( \textbf{x}_{i:j},\textbf{1}\right) = C|\textbf{J}_k|\log(p) - \left(C\sqrt{p\psi} + p \right) + \sum_{ c \notin \textbf{J}_k}(j-i+1)\left(\bar{\bm{\eta}}_{i:j}^c\right)^2 \\
&\leq C|\textbf{J}_k|\log(p) - \left(C\sqrt{p\psi} + p \right) + \left(p + 2\psi + 2\sqrt{p\psi} \right) = C|\textbf{J}_k|\log(p) - C\sqrt{p\psi} + 2\sqrt{p\psi} + 2\psi\\
&\leq  \frac{13}{20} C|\textbf{J}_k|\log(p) - \frac{6}{10} C\sqrt{p\psi} + 2\psi \leq \frac{1}{10} C|\textbf{J}_k|\log(p) - \frac{1}{20} C\sqrt{p\psi} + 2\psi,
\end{align*}
for large enough $C$. Here the first inequality follows from $E_2$ and the second inequality holds because $|\textbf{J}_k| \leq k^*$. \qed

\subsubsection{Proof of Lemma \ref{lemma:Sparse_To_Dense_CONDITION2}}
The proof is very similar to that of Lemma \ref{lemma:Sparse_To_Dense_CONDITION1} and has therefore been omitted.\\

\subsubsection{Proof of Lemma \ref{lemma:Sparse_To_Sparse_CONDITION2}}
The proof is very similar to that of Lemma \ref{lemma:Sparse_To_Sparse_CONDITION1} and has therefore been omitted.\\

\subsubsection{Proof of Lemma \ref{lemma:MASSIVELEMMA!}}

This Lemma shows that splitting up long fitted changes containing multiple sparse anomalous regions along the ground truth reduces the penalised cost by $O(C)$ We begin by considering 
\begin{align*}
&\mathcal{C} \left( \textbf{x}_{s,e}, \textbf{1}\right) - \sum_{k \in \mathcal{D}_{s,e}} \left( \mathcal{C} \left( \textbf{x}_{(s_k+1):e_k},\textbf{J}_k\right)\right) = p+C\psi+ C\sqrt{p\psi} + \sum_{ c = 1}^p \left( \sum_{t=s}^{e} \left(\textbf{x}_t^{(c)} - \bar{\textbf{x}}_{s:e}^{(c)}\right)^2 \right) \\
&- \sum_{k \in \mathcal{D}_{s,e}}\left( \sum_{c \in \textbf{J}_k}\left( \sum_{t=s_k+1 }^{e_k}  \left( \textbf{x}_t^{(c)} - \bar{\textbf{x}}_{(s_k+1):e_k}^{(c)} \right)^2   \right)   +   C\psi + C|\textbf{J}_k|\log(p) \right) - \sum_{ c = 1}^p\sum_{t: \nexists k:  \; c \in \textbf{J}_k  \land t \in [s_k+1,e_k] }\left(\bm{\eta}^{(c)}_t\right)^2 \\
&\geq p+C\psi+ C\sqrt{p\psi} + \sum_{ c = 1}^p \left( \sum_{t=s}^{e} \left(\bm{\mu}_t^{(c)} - \bar{\bm{\mu}}_{s:e}^{(c)} + \bm{\eta}_t^{(c)} - \bar{\bm{\eta}}_{s:e}^{(c)}\right)^2 \right) \\ 
&- \sum_{k \in \mathcal{D}_{s,e}}\left( \sum_{c \in \textbf{J}_k}\left( \sum_{t=s_k+1 }^{e_k}  \left( \bm{\eta}_t^{(c)}  \right)^2   \right)   +   C\psi + C|\textbf{J}_k|\log(p) \right) - \sum_{ c = 1}^p\sum_{t: \nexists k:  \; c \in \textbf{J}_k  \land t \in [s_k+1,e_k] }\left(\bm{\eta}^{(c)}_t\right)^2 \\
& = p+C\psi+ C\sqrt{p\psi} + \sum_{ c = 1}^p \left( \sum_{t=s}^{e} \left( \bm{\eta}_t^{(c)} - \bar{\bm{\eta}}_{s:e}^{(c)}\right)^2 \right)  + \sum_{ c = 1}^p \left( \sum_{t=s}^{e} \left(\bm{\mu}_t^{(c)} - \bar{\bm{\mu}}_{s:e}^{(c)} \right)^2 \right)  \\ 
&+2\sum_{ c = 1}^p \left( \sum_{t=s}^{e} \left(\bm{\mu}_t^{(c)} - \bar{\bm{\mu}}_{s:e}^{(c)} \right) \left(\bm{\eta}_t^{(c)} - \bar{\bm{\eta}}_{s:e}^{(c)}\right) \right)  -  \sum_{ c = 1}^p \left( \sum_{t=s}^{e} \left(\bm{\eta}_t^{(c)} \right)^2 \right) - \sum_{k \in \mathcal{D}_{s,e}}\left( C\psi + C|\textbf{J}_k|\log(p) \right) \\
& =  p + C\psi + C \sqrt{p\psi}  - \sum_{ c = 1}^p \left( (e-s+1)\left( \bar{\bm{\eta}}_{s:e}^{(c)}\right)^2 \right) + \sum_{ c = 1}^p \left( \sum_{t=s}^{e} \left(\bm{\mu}_t^{(c)} - \bar{\bm{\mu}}_{s:e}^{(c)} \right)^2 \right) - \sum_{k \in \mathcal{D}_{s,e}}\left( C\psi + C|\textbf{J}_k|\log(p) \right) \\
& +2\sum_{ c = 1}^p \left( \sum_{t=s}^{e} \left(\bm{\mu}_t^{(c)} - \bar{\bm{\mu}}_{s:e}^{(c)} \right) \left(\bm{\eta}_t^{(c)} \right) \right) \\
&\geq \frac{19}{20}C\left( \psi + \sqrt{p\psi}\right) + \sum_{ c = 1}^p \left( \sum_{t=s}^{e} \left(\bm{\mu}_t^{(c)} - \bar{\bm{\mu}}_{s:e}^{(c)} \right)^2 \right) - \sum_{k \in \mathcal{D}_{s,e}}\left( C\psi + C|\textbf{J}_k|\log(p) \right) \\
&-2\sqrt{\sum_{ c = 1}^p \sum_{t=s}^e \left(\bm{\mu}_t^{(c)} - \bar{\bm{\mu}}_{s:e}^{(c)}  \right)^2}\sqrt{2\psi+2 \left|W_{s,e} \right| \log(p)} \\
& \geq \frac{19}{20}C\left( \psi + \sqrt{p\psi}\right) + \frac{1}{2}\sum_{ c = 1}^p \left( \sum_{t=s}^{e} \left(\bm{\mu}_t^{(c)} - \bar{\bm{\mu}}_{s:e}^{(c)} \right)^2 \right) - \sum_{k \in \mathcal{D}_{s,e}}\left( C\psi + C|\textbf{J}_k|\log(p) \right)  - 8\psi - 8  \left|W_{s,e} \right| \log(p),
\end{align*}
where the first inequality follows from the fact that the residual sum of squares is minimised at the mean, the second inequality follows from $E_2$ and $E_6$, and the last inequality follows from the AM-GM inequality. 

Next note that 
\begin{equation*}
\sum_{t=s}^{e} \left(\bm{\mu}_t^{(c)} - \bar{\bm{\mu}}_{s:e}^{(c)} \right)^2 
\end{equation*}
corresponds to the residual sum of squares obtained by fitting $\bm{\mu}_e^{(c)},...,\bm{\mu}_s^{(c)}$ as a single segment. Consequently, breaking it up into smaller segments does not increase un-penalised cost. More precisely, for any partition $\tau_{s:e}=\{s,\tau_1,...,\tau_m,e\} $ of the segment $(s+1,e)$, 
\begin{equation*}
\sum_{t=s+1}^{e} \left(\bm{\mu}_t^{(c)} - \bar{\bm{\mu}}_{(s+1):e}^{(c)} \right)^2 \geq \sum_{k=0}^{m} \left( \sum_{t=\tau_m+1}^{\tau_{m+1}} \left(\bm{\mu}_t^{(c)} - \bar{\bm{\mu}}_{(\tau_m+1):\tau_{m+1}}^{(c)} \right)^2 \right)
\end{equation*}
holds. In particular, we therefore have that 
\begin{align*}
&\frac{1}{2}\sum_{ c = 1}^p \left( \sum_{t=s}^{e} \left(\bm{\mu}_t^{(c)} - \bar{\bm{\mu}}_{s:e}^{(c)} \right)^2 \right) \geq \sum_{k: e_k \in [s,e]}\frac{1}{2}\left( \sum_{ c \in \textbf{J}_k} \left(\sum_{e_k-\lceil\frac{10C}{\triangle_{k}^2}\rceil}^{e_k+\lfloor\frac{10C}{\triangle_{k}^2}\rfloor} \left(\bm{\mu}_t^{(c)} - \bar{\bm{\mu}}_{\left(e_k-\lceil\frac{10C}{\triangle_{k}^2}\rceil\right):\left(e_k+\lfloor\frac{10C}{\triangle_{k}^2}\rfloor\right)}^{(c)} \right)^2\right) \right)  \\
&+ \sum_{k: s_k \in [s,e]}\frac{1}{2}\left( \sum_{ c \in \textbf{J}_k} \left(\sum_{s_k- \lfloor \frac{10C}{\triangle_{k}^2} \rfloor }^{s_k+\lceil \frac{10C}{\triangle_{k}^2} \rceil} \left(\bm{\mu}_t^{(c)} - \bar{\bm{\mu}}_{\left(s_k-\lfloor\frac{10C}{\triangle_{k}^2}\rfloor\right):\left(s_k+\lceil\frac{10C}{\triangle_{k}^2}\rceil\right)}^{(c)} \right)^2\right) \right)  \\
& = \frac{1}{2}\sum_{k: e_k \in [s,e]}\left( |\textbf{J}_k| 2 \left \lceil \frac{10C}{\triangle_{k}^2} \right \rceil \frac{\bm{\mu}_k^2}{4}\right) + \frac{1}{2}\sum_{k: s_k \in [s,e]}\left( |\textbf{J}_k| \frac{20C}{\triangle_{k}^2} \frac{\bm{\mu}_k^2}{4}\right) \geq \frac{1}{2}\sum_{k \in \mathcal{D}_{s,e}}\left( |\textbf{J}_k| \frac{20C}{\triangle_{k}^2} \frac{\bm{\mu}_k^2}{4}\right) \\
& =  \sum_{k \in \mathcal{D}_{s,e}} \frac{5}{2}C \left( \psi + |\textbf{J}_k| \log(p)\right),
\end{align*}
where the first inequality follows from using a partition which cuts $\frac{10C}{\triangle_{k}^2}$ either side of the starting points and end points of true anomalous regions contained in $[s,e]$ and the second inequality follows from the definition of $\mathcal{D}_{s,e}$. Consequently, we have that
\begin{align*}
&\mathcal{C} \left( \textbf{x}_{s,e}, \textbf{1}\right) - \sum_{k \in \mathcal{D}_{s,e}} \left( \mathcal{C} \left( \textbf{x}_{(s_k+1):e_k},\textbf{J}_k\right)\right) \\ 
&\geq 
\frac{19}{20}C\left( \psi + \sqrt{p\psi}\right) +  \sum_{k \in \mathcal{D}_{s,e}} \frac{5}{2}C \left( \psi + |\textbf{J}_k| \log(p)\right) - \sum_{k \in \mathcal{D}_{s,e}}\left( C\psi + C|\textbf{J}_k|\log(p) \right)  - 8\psi - 8  \left|W_{s,e} \right| \log(p) \\
&\geq  \frac{19}{20}C\left( \psi + \sqrt{p\psi}\right),
\end{align*}
where the first inequality follows from assembling the previous two results, and the second one holds if $C$ exceeds a global constant. We also have that: \small
\begin{align*} 
&\mathcal{C} \left( \textbf{x}_{s,e}, \textbf{J}\right) - \sum_{k \in \mathcal{D}_{s,e}} \left( \mathcal{C} \left( \textbf{x}_{(s_k+1):e_k},\textbf{J}_k\right)\right) = C\psi+ C|\bm{J}|\log(p) + \sum_{ c \in \textbf{J}} \left( \sum_{t=s}^{e} \left(\textbf{x}_t^{(c)} - \bar{\textbf{x}}_{s:e}^{(c)}\right)^2 \right) \\
&- \sum_{k \in \mathcal{D}_{s,e}}\left( \sum_{c \in \textbf{J}_k \cap \textbf{J}}\left( \sum_{t=s_k+1 }^{e_k}  \left( \textbf{x}_t^{(c)} - \bar{\textbf{x}}_{(s_k+1):e_k}^{(c)} \right)^2   \right)   +   C\psi + C|\textbf{J}_k|\log(p) \right) - \sum_{ c \in \textbf{J}} \sum_{t: \nexists k:  \; c \in \textbf{J}_k  \land t \in [s_k+1,e_k] }\left(\bm{\eta}^{(c)}_t\right)^2 \\
& + \sum_{k \in \mathcal{D}_{s,e}}\left( \sum_{c \in \textbf{J}_k \setminus \textbf{J}}\left( (e_k-s_k)  \left( \bar{\textbf{x}}_{(s_k+1):e_k}^{(c)} \right)^2   \right)  \right) \\
&\geq C\psi+ C|\bm{J}|\log(p) + \sum_{ c \in \textbf{J}} \left( \sum_{t=s}^{e} \left(\bm{\mu}_t^{(c)} - \bar{\bm{\mu}}_{s:e}^{(c)}\right)^2 \right) + \sum_{ c \in \textbf{J}} \left( \sum_{t=s}^{e} \left(\bm{\eta}_t^{(c)} - \bar{\bm{\eta}}_{s:e}^{(c)}\right)^2 \right) + 2 \sum_{ c \in \textbf{J}} \left( \sum_{t=s}^{e} \left(\bm{\eta}_t^{(c)} - \bar{\bm{\eta}}_{s:e}^{(c)}\right)\left(\bm{\mu}_t^{(c)} - \bar{\bm{\mu}}_{s:e}^{(c)}\right)\right) \\ 
&-\sum_{ c \in \textbf{J}} \left( \sum_{t=s}^{e} \left(\bm{\eta}_t^{(c)} \right)^2 \right) -\sum_{k \in \mathcal{D}_{s,e}}  \left( C\psi + C|\textbf{J}_k|\log(p) \right) + \sum_{k \in \mathcal{D}_{s,e}}\left( \sum_{c \in \textbf{J}_k \setminus \textbf{J}} (e_k-s_k)  \bm{\mu}_k ^2  + 2(e_k-s_k)\bm{\mu}_k \sum_{c \in \textbf{J}_k \setminus \textbf{J}} \left( \bar{\bm{\eta}}_{(s_k+1):e_k}^{(c)} \right)   \right) \\
&\geq( C-2)(\psi+ |\bm{J}|\log(p)) - \sum_{k \in \mathcal{D}_{s,e}}  \left( C\psi + C|\textbf{J}_k|\log(p) \right)+\sum_{ c \in \textbf{J}} \left( \sum_{t=s}^{e} \left(\bm{\mu}_t^{(c)} - \bar{\bm{\mu}}_{s:e}^{(c)}\right)^2 \right)  + 2 \sum_{ c \in \textbf{J}} \left( \sum_{t=s}^{e} \left(\bm{\mu}_t^{(c)} - \bar{\bm{\mu}}_{s:e}^{(c)}\right)\bm{\eta}_t^{(c)}\right) \\
&+\sum_{k \in \mathcal{D}_{s,e}}\left(  (e_k-s_k) |\textbf{J}_k \setminus \textbf{J}| \bm{\mu}_k ^2  - 2\sqrt{(e_k-s_k)\bm{\mu}_k^2|\textbf{J}_k \setminus \textbf{J}|(2\psi+2|\textbf{J}_k \setminus \textbf{J}|\log(p))} \right) \\
&\geq ( C-2)(\psi+ |\bm{J}|\log(p)) - \sum_{k \in \mathcal{D}_{s,e}}  \left( C\psi + C|\textbf{J}_k|\log(p) \right)+\sum_{ c \in \textbf{J}} \left( \sum_{t=s}^{e} \left(\bm{\mu}_t^{(c)} - \bar{\bm{\mu}}_{s:e}^{(c)}\right)^2 \right)  \\
&-2\sqrt{\sum_{ c \in \textbf{J}} \left( \sum_{t=s}^{e} \left(\bm{\mu}_t^{(c)} - \bar{\bm{\mu}}_{s:e}^{(c)}\right)^2 \right)} \sqrt{2\psi+2|W_{s:e}|\log(p)} +\sum_{k \in \mathcal{D}_{s,e}}\left( \frac{1}{2} (e_k-s_k) |\textbf{J}_k \setminus \textbf{J}| \bm{\mu}_k ^2  - 8(\psi+|\textbf{J}_k \setminus \textbf{J}|\log(p)) \right) \\
&\geq ( C-2)(\psi+ |\bm{J}|\log(p)) - \sum_{k \in \mathcal{D}_{s,e}} (C+8) \left( \psi + |\textbf{J}_k|\log(p) \right)+\sum_{ c \in \textbf{J}} \left( \sum_{t=s}^{e} \left(\bm{\mu}_t^{(c)} - \bar{\bm{\mu}}_{s:e}^{(c)}\right)^2 \right) + \frac{1}{2}\sum_{k \in \mathcal{D}_{s,e}}\left(  (e_k-s_k) |\textbf{J}_k \setminus \textbf{J}| \bm{\mu}_k ^2  \right) \\
& - 8(\psi+|W_{s:e}|\log(p)) - \frac{1}{2}\sum_{ c \in \textbf{J}} \left( \sum_{t=s}^{e} \left(\bm{\mu}_t^{(c)} - \bar{\bm{\mu}}_{s:e}^{(c)}\right)^2 \right) \\
& \geq \frac{19}{20}C(\psi+|\textbf{J}|\log(p)) - \sum_{k \in \mathcal{D}_{s,e}} 2C \left( \psi + |\textbf{J}_k|\log(p) \right) + \frac{1}{2} \left(\sum_{ c \in \textbf{J}} \left( \sum_{t=s}^{e} \left(\bm{\mu}_t^{(c)} - \bar{\bm{\mu}}_{s:e}^{(c)}\right)^2 \right) + \sum_{k \in \mathcal{D}_{s,e}}\left(  (e_k-s_k) |\textbf{J}_k \setminus \textbf{J}| \bm{\mu}_k ^2  \right) \right)
\end{align*} \normalsize
A very similar argument as the one used for the dense case can be used to show that
\begin{equation*}
\frac{1}{2} \sum_{ c \in \textbf{J}} \left( \sum_{t=s}^{e} \left(\bm{\mu}_t^{(c)} - \bar{\bm{\mu}}_{s:e}^{(c)}\right)^2 \right) \geq \sum_{k \in \mathcal{D}_{s,e}} \frac{5}{2}C |\textbf{J}_k \cap \textbf{J}|\frac{\bm{\mu}_k^2}{\triangle_{k}^2}.
\end{equation*}
Consequently, 
\begin{align*}
&\mathcal{C} \left( \textbf{x}_{s,e}, \textbf{J}\right) - \sum_{k \in \mathcal{D}_{s,e}} \left( \mathcal{C} \left( \textbf{x}_{(s_k+1):e_k},\textbf{J}_k\right)\right) \\
&\geq \frac{19}{20}C(\psi+|\textbf{J}|\log(p)) - \sum_{k \in \mathcal{D}_{s,e}} 2C \left( \psi + |\textbf{J}_k|\log(p) \right) + \sum_{k \in \mathcal{D}_{s,e}} \left[\frac{5}{2}C |\textbf{J}_k \cap \textbf{J}|\frac{\bm{\mu}_k^2}{\triangle_{k}^2}  + \frac{5C\bm{\mu}_k^2}{2\triangle_{k}^2}|\textbf{J}_k \setminus \textbf{J}| \right] \\
&= \frac{19}{20}C(\psi+|\textbf{J}|\log(p)) - \sum_{k \in \mathcal{D}_{s,e}} 2C \left( \psi + |\textbf{J}_k|\log(p) \right) + \sum_{k \in \mathcal{D}_{s,e}} \frac{5}{2}C \left(\psi + |\textbf{J}_k|\log(p)\right) \geq \frac{19}{20}C(\psi+|\textbf{J}|\log(p)),
\end{align*}
where the first inequality follows from the condition on the segment length $e_k-s_k$. \qed

\newpage

\subsection{Additional Simulation Results}

\vspace{-20pt}

\begin{figure} 
	\begin{subfigure}[b]{0.5\linewidth}
		\centering
		\includegraphics[width=\linewidth]{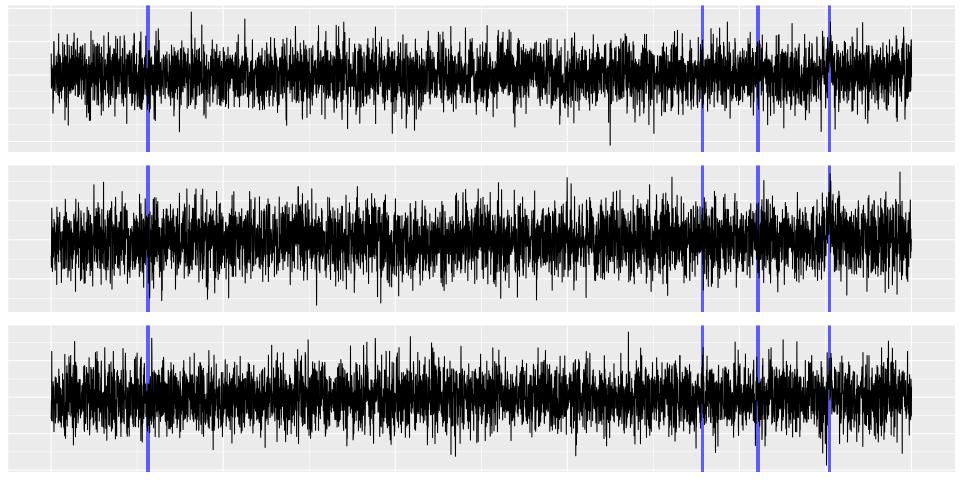} 
		\caption{Example}
		\label{fig:all_component_example} 
		\vspace{4ex}
	\end{subfigure} 
	\begin{subfigure}[b]{0.5\linewidth}
		\centering
		\includegraphics[width=\linewidth]{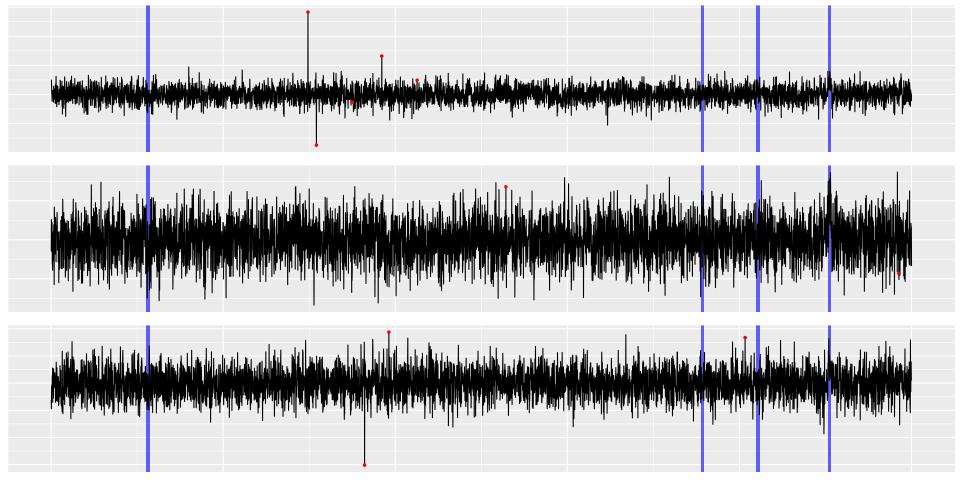} 
		\caption{Example, with pt. anomalies}
		\label{fig:all_component_ANOM_example} 
		\vspace{4ex}
	\end{subfigure}
	\vspace{-20pt}
	\begin{subfigure}[b]{0.5\linewidth}
		\centering
		\includegraphics[width=0.9\linewidth]{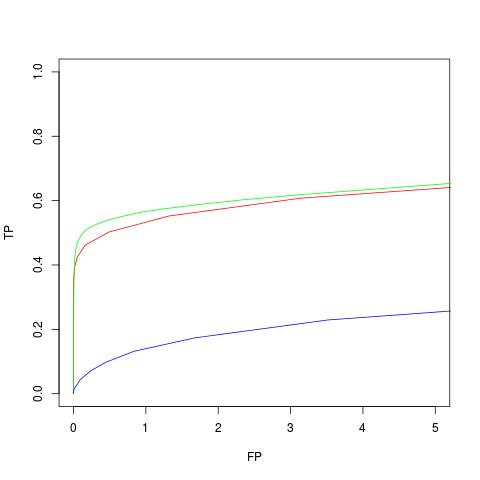} 
		\caption{p=10} 
		\label{fig:all_component_small} 
		\vspace{4ex}
	\end{subfigure} 
	\begin{subfigure}[b]{0.5\linewidth}
		\centering
		\includegraphics[width=0.9\linewidth]{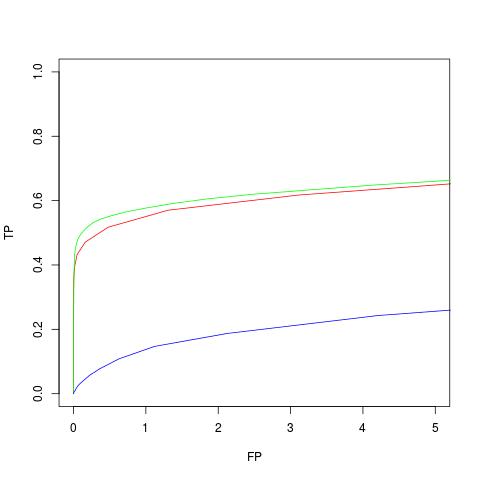} 
		\caption{p=10, with pt. anomalies} 
		\label{fig:all_component_ANOM_small} 
		\vspace{4ex}
	\end{subfigure} 
	\vspace{-20pt}
	\begin{subfigure}[b]{0.5\linewidth}
		\centering
		\includegraphics[width=0.9\linewidth]{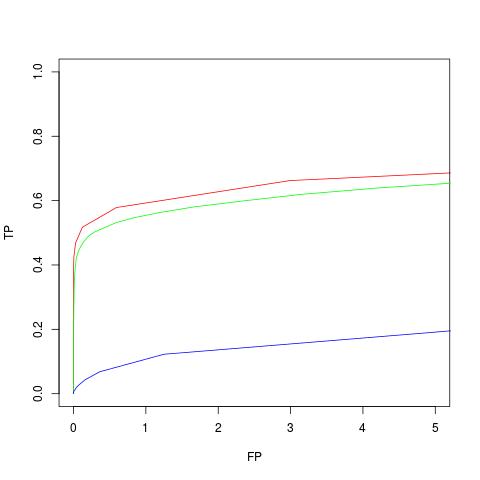} 
		\caption{p=100} 
		\label{fig:all_component_large} 
		\vspace{4ex}
	\end{subfigure}
	\begin{subfigure}[b]{0.5\linewidth}
		\centering
		\includegraphics[width=0.9\linewidth]{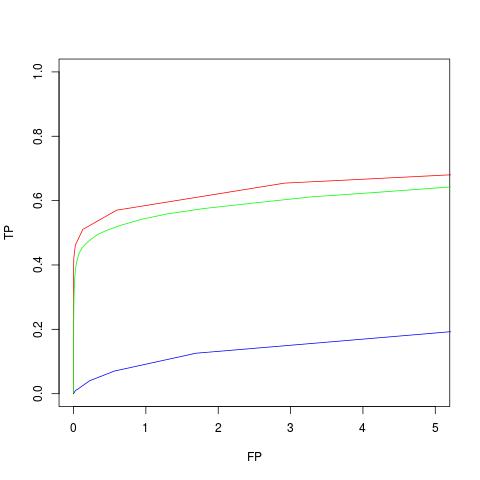} 
		\caption{p=100, with pt. anomalies} 
		\label{fig:all_component_ANOM_larg} 
		\vspace{4ex}
	\end{subfigure}
	\caption{Example series and ROC curves for setting 2. MVCAPA is in red, PASS in green, and Inspect in blue.}
	\label{fig:all_component} 
\end{figure}

\end{document}